\documentclass[]{JHEP3}

\JHEPspecialurl{http://jhep.sissa.it/JOURNAL/JHEP3.tar.gz}

\usepackage{amsthm,latexsym,amssymb,amsfonts,array}
\usepackage[nosort]{cite}
\usepackage[centertags,intlimits]{amsmath}
\usepackage{graphicx}
\usepackage{epsf}
\usepackage{longtable}
\usepackage{epsfig}

\makeatletter

\@addtoreset{equation}{section}
\makeatother
\newcommand{\al}{\alpha}

\newcommand{\bas}{\backslash}
\newcommand{\be}{\beta}
\newcommand{\beq}{\begin{equation}}
\newcommand{\beqa}{\begin{eqnarray}}

\newcommand{\eeq}{\end{equation}}
\newcommand{\eeqa}{\end{eqnarray}}

\newcommand{\f}{\frac}

\newcommand{\hs}{\hspace{0.1 cm}}

\newcommand{\ga}{\gamma}

\newcommand{\ka}{\ka}

\newcommand{\mbb}{\mathbb}
\newcommand{\mc}{\mathcal}
\newcommand{\mf}{\mathfrak}
\newcommand{\nn}{\nonumber}

\newcommand{\om}{\omega}
\newcommand{\pa}{\partial}

\newcommand{\tchi}{\tilde{\chi}}
\newcommand{\tm}{\tilde{m}}
\newcommand{\tn}{\tilde{n}}
\newcommand{\Zn}{{\mathbb Z}}
\newcommand{\pg}{SU(2,1;\mbb{Z}[i])}
\newcommand{\xit}{\tilde{\xi}}
\newcommand{\xib}{\bar{\xi}}
\newcommand{\xitb}{\bar{\xit}}

\newcommand{\cx}{{\mathbb C}}
\newcommand{\tr}{\text{tr}}
\newcommand{\cxh}{\mathbb{CH}}
\newcommand{\cF}{{\mathcal F}}
\newcommand{\cZ}{{\mathcal Z}}

\newcommand{\cV}{{\mathcal{V}}}
\newcommand{\cW}{{\mathcal{W}}}
\newcommand{\cosm}{{\mathcal{K}}}
\newcommand{\tcosm}{{\tilde{\mathcal{K}}}}
\newcommand{\IZ}{\mbb{Z}}
\newcommand{\IR}{\mbb{R}}

\preprint{ULB-TH/09-32}

\title{Instanton Corrections to the Universal Hypermultiplet and Automorphic Forms on {\boldmath $SU(2,1)$}}

\author{ Ling Bao${}^{\spadesuit}$, Axel Kleinschmidt${}^{\diamondsuit}$, Bengt E. W. Nilsson${}^{\spadesuit}$, Daniel Persson${}^{\spadesuit}$${}^{\diamondsuit}$ \phantom{andand} and Boris Pioline${}^{\clubsuit}$\\

${}^{\spadesuit}$ Fundamental Physics, 
Chalmers University of Technology, \\
SE-412 96, G\"oteborg, Sweden\\

${}^{\diamondsuit}$ Physique Th\'eorique et Math\'ematique, \\
Universit\'e Libre
de Bruxelles \& International Solvay Institutes, \\
ULB-Campus Plaine C.P. 231, B-1050 Bruxelles, Belgium\\

${}^\clubsuit$ Laboratoire de Physique Th\'eorique et Hautes
Energies, \\ CNRS UMR 7589 and  
Universit\'e Pierre et Marie Curie - Paris 6,\\
4 place Jussieu, 75252 Paris cedex 05, France \\

E-mail: \email{ling.bao@chalmers.se}, \email{axel.kleinschmidt@ulb.ac.be}, \email{tfebn@chalmers.se}, \email{daniel.persson@chalmers.se}, \email{pioline@lpthe.jussieu.fr}}

\abstract{
The hypermultiplet moduli space in Type IIA string theory compactified on a rigid Calabi-Yau threefold $\mc{X}$,  corresponding to the ``universal hypermultiplet'', is described at tree-level by the symmetric space $SU(2,1)/(SU(2)\times U(1))$. To determine the quantum corrections to this metric, we posit
that a discrete subgroup of the continuous tree-level isometry group $SU(2,1)$,
namely  the Picard modular group $SU(2,1;\mbb{Z}[i])$,  must remain unbroken in the exact metric -- including all perturbative and non-perturbative quantum corrections. This assumption is expected to be valid when $\mc{X}$ admits complex multiplication by $\mbb{Z}[i]$. Based on this hypothesis, we construct an $SU(2,1;\mbb{Z}[i])$-invariant, non-holomorphic Eisenstein series,
and tentatively propose that this Eisenstein series provides the exact contact potential on the twistor space over the universal hypermultiplet moduli space. We analyze its non-Abelian Fourier expansion, and show that the Abelian and non-Abelian Fourier coefficients take the required form for instanton corrections due to Euclidean D2-branes wrapping special Lagrangian submanifolds, and to Euclidean NS5-branes wrapping the entire Calabi-Yau threefold, respectively. While this tentative proposal
fails to reproduce the correct one-loop correction, the consistency of the Fourier expansion with 
physics expectations provides strong support for the usefulness of the Picard modular group in constraining the quantum moduli space. }
 

\begin{document}

\section{Introduction and Summary}
\label{Section:Introduction}

String theory compactified on a manifold $\mc{X}$  typically leads to 
a low energy effective action with an often large number of massless scalar fields
valued in a moduli space $\mc{M}$. 
In general, the Riemannian metric on $\mc{M}$ is deformed by perturbative
and non-perturbative quantum corrections, making it very difficult to
determine the exact form of the quantum effective action. In this paper we study the particular case of compactifications of type IIA string theory on a rigid Calabi-Yau threefold $\mc{X}$ (i.e. with Betti number $h_{2,1}(\mc{X})=0$). In this case, the hypermultiplet part of the moduli space $\mc{M}$
is known to be described, at tree-level in the string perturbative expansion, by
the symmetric space $\mc{M}_{\rm UH}=SU(2,1)/(SU(2)\times U(1))$.  We analyze the quantum corrections to this classical geometry, and in particular conjecture the form of D2-brane and 
NS5-brane instanton contributions. Before entering into the details, 
we begin by discussing some of the  ideas 
leading up to our proposal. 

\subsection{Rigid Moduli Spaces for $\mc{N}\geq 4$ and Eisenstein series}

For compactifications preserving $\mc{N}\ge 4$ supersymmetry in $D=4$, 
the moduli space is always locally a symmetric space $\mc{M}=G/K$, with 
$G$ being a global symmetry and $K$, the maximal compact
subgroup of $G$, being a local R-symmetry. 
In particular,  $\mc{M}$ has restricted holonomy group $K$ and is rigid
(see, e.g., \cite{Aspinwall} for a nice discussion). Quantum corrections 
are encoded in the global structure of $\mc{M}$, given by a double coset 
\beq
\mc{M}_{\mathrm{exact}}=G(\mbb{Z})\backslash G/K
\eeq 
where $G(\mbb{Z})$ is typically an arithmetic subgroup of $G$, known  
as the $S$-, $T$- or $U$-duality
group, depending on the context \cite{Hull:1994ys,Witten:1995ex,Obers:1998fb}. 
For example, M-theory compactified on $T^7$ (or type IIA/B
on $T^6$),  gives rise to $\mc{N} =8$ supergravity in four dimensions,
whose  exact moduli space is conjectured to be
$E_{7(7)}(\mbb{Z})\backslash E_{7(7)}/(SU(8)/\mbb{Z}_2)$ \cite{Hull:1994ys}.  
In such cases, the quantum effective action is expected to be invariant under
$G(\mbb{Z})$, which gives a powerful constraint on possible quantum corrections.

This idea was exploited with great success in the seminal work \cite{Green:1997tv} 
in the context of  type IIB supergravity in ten dimensions, 
where the exact coefficient of the higher-derivative $\mc{R}^4$-type corrections were proposed to be given by a
non-holomorphic Eisenstein series $\mc{E}_{3/2}^{SL(2, \mbb{Z})}$
as a function of the ``axio-dilaton'' $C_{(0)}+ie^{-\phi}$, valued on the fundamental domain 
 $\mc{M}=SL(2, \mbb{Z})\backslash SL(2,\mbb{R})/SO(2)$ of the Poincar\'e upper half plane.
This proposal reproduced the known tree-level and one-loop 
corrections \cite{Green:1981yb,Gross:1986iv}, predicted 
the absence of higher loop corrections, later verified by an explicit two-loop computation
\cite{D'Hoker:2005ht}, and suggested the exact form of D(-1)-instanton contributions, later 
corroborated by explicit matrix model computations \cite{Kostov:1998pg,Moore:1998et}. 
From the mathematical point of view,
perturbative corrections and instanton contributions correspond, respectively, to the constant terms and
Fourier coefficients of the automorphic form $\mc{E}_{3/2}^{SL(2, \mbb{Z})}$.
This work was extended to toroidal compactifications
of M-theory, where the $\mc{R}^4$-type corrections were argued to be given by 
Eisenstein series of the respective U-duality group \cite{PiolineKiritsis,Green:1997di,ObersPioline},
predicting the contributions of Euclidean D$p$-brane instantons,
and, when $n\geq 6$, NS5-branes. Unfortunately, extracting the constant
terms and  Fourier coefficients of Eisenstein series is not an easy task, and  
it has been difficult to put the conjecture to the test.  Part of our motivation 
is to develop the understanding of Eisenstein series beyond the relatively well
understood case of $G( \mbb{Z})=SL(n, \mbb{Z})$.

\subsection{The Hypermultiplet Moduli Space of $ \mc{N} =2$ Supergravity}

Compactifications with fewer unbroken supersymmetries ($ \mc{N} \leq
2$ in $D=4$)  lead to moduli spaces which are generically not 
symmetric spaces. An interesting example is type IIA string
theory compactified on a Calabi-Yau threefold $\mc{X}$, leading to $ \mc{N} =2$ supergravity in four
dimensions coupled to $h_{1,1}$ vector multiplets and $h_{2,1}+1$ hypermultiplets.
The moduli space locally splits into a direct
product $\mc{M}=\mc{M}_{\mathrm{V}}\times \mc{M}_{\mathrm{H}}$, where
$\mc{M}_{\mathrm{V}}$ is a $2h_{1,1}$-dimensional special K\"ahler manifold, and
$\mc{M}_{\mathrm{H}}$ a $4(h_{2,1}+1)$-dimensional quaternion-K\"ahler manifold, respectively.
 $\mc{M}_{\mathrm{V}}$
encodes the (complexified) K\"ahler structure of $\mc{X}$, while
$\mc{M}_{\mathrm{H}}$ encodes deformations of the complex structure. 
$\mc{M}_{\mathrm{V}}$ is exact at tree-level in the perturbative string expansion,
and well understood thanks to classical mirror symmetry (see e.g.
\cite{MirrorSymmetry} for an extensive introduction). 
In this paper we focus on the less understood hypermultiplet moduli space $\mc{M}_{\mathrm{H}}$.
Note however that upon further compactification on a circle, $\mc{M}_{\mathrm{V}}$
is extended to a $4(h_{1,1}+1)$-dimensional quaternion-K\"ahler manifold by the
$c$-map, and the vector and hypermultiplet moduli spaces become equally
complicated, being exchanged under T-duality along the circle \cite{Cecotti:1988qn}.

Contrary to $\mc{M}_{\mathrm{V}}$, the hypermultiplet moduli space $\mc{M}_{\mathrm{H}}$ receives perturbative and non-perturbative corrections in the string coupling\cite{BeckerBeckerStrominger,AntoniadisMinasian1,Strominger,Gunther:1998sc,AntoniadisMinasian2}. The non-perturbative corrections are due to Euclidean D2-branes wrapping special
Lagrangian submanifolds in $\mc{X}$, as well as to Euclidean NS5-branes wrapping the
entire Calabi-Yau threefold \cite{BeckerBeckerStrominger}.\footnote{See \cite{Nick} for a recent analysis of these effects in heterotic compactifications.} It has been an
outstanding problem to understand how these effects modify the geometry of the
moduli space $\mc{M}_{\mathrm{H}}$, mainly due to the fact that quaternion-K\"ahler
geometry is much more complicated than special K\"ahler geometry. 
Recently, however, it has become apparent that twistor techniques can be
efficiently applied to quaternion-K\"ahler geometry.  
In particular, deformations of the
quaternion-K\"ahler geometry of $\mc{M}_{\mathrm{H}}$ are in one-to-one correspondance
with deformations of its twistor space $\mc{Z}_{\mc{M}_{\mathrm{H}}}$,  a $\mbb{C}P^1$ bundle over
$\mc{M}_{\mathrm{H}}$ \cite{Salamon,Swann,LeBrun} (see 
\cite{deWitVandoren1,deWitVandoren2,Alexandrov:2008ds,Alexandrov:2008nk,Alexandrov:2008gh} 
for a physics realization of this equivalence). One virtue of this
approach is that, contrary to $\mc{M}_{\mathrm{H}}$, the twistor space
$\mc{Z}_{\mc{M}_{\mathrm{H}}}$ is K\"ahler, and therefore quantum corrections
to $\mc{M}_{\mathrm{H}}$ can in principle be described in terms of the
K\"ahler potential on its twistor space $\mc{Z}_{\mc{M}_{\mathrm{H}}}$. 
Furthermore, $\mc{Z}_{\mc{M}_{\mathrm{H}}}$ being a complex contact manifold,
it can be described by holomorphic data, namely a set of complex
symplectomorphisms between complex Darboux coordinate patches.

Using these techniques, much headway has been made in summing up part of
the instanton corrections to hypermultiplet moduli spaces in both  type IIA and
IIB string theory \cite{Vandoren1,Vandoren2,Alexandrov:2009zh,Alexandrov:2009qq}. These
techniques were combined with the 
$SL(2,\mbb{Z})$-invariance of the four-dimensional effective action  in \cite{Vandoren1}, to obtain
the quantum corrections to 
$\mc{M}_{\mathrm{H}}^{\mathrm{IIB}}$ due to  F1, D$(-1)$ and D1 instantons\footnote{The 
holomorphic action of $SL(2,\mbb{Z})$ on the twistor space $\mc{Z}_{\mc{M}_{\mathrm{H}}}$
has been recently clarified in \cite{Alexandrov:2009qq}.}. In
this context, the Eisenstein series $\mc{E}_{3/2}^{SL(2, \mbb{Z})}$, discussed
above, reappears as the D(-1) instanton contribution to the K\"ahler potential
on the twistor space of  $\mc{M}_{\mathrm{H}}^{\mathrm{IIB}}$.
This result was then mapped over to the IIA side using mirror symmetry
\cite{Vandoren2}, providing the quantum corrections to
$\mc{M}_{\mathrm{H}}^{\mathrm{IIA}}$ from Euclidean D2-branes wrapping $A$-cycles in
$\mc{X}$. Subsequently, the contribution from
D2-branes wrapping arbitrary combinations of $A$ and $B$-cycles was obtained 
to linear order by covariantizing the result of
\cite{Vandoren2} under ``electric-magnetic duality'' between $A$- and
$B$-cycles in the Calabi-Yau \cite{Alexandrov:2008nk,Alexandrov:2009zh}. 
By the T-duality argument mentioned above, this
also provides the contributions of 4D BPS black holes to the vector multiplet
moduli space in type IIA or IIB string theory compactified on $\mc{X}\times S_1$.
However, the  NS5-brane contributions  (or,  on the vector multiplet side, 
the Kaluza-Klein monopole contributions)
have so far proven to be considerably more elusive, although they can be in principle reached 
following the  ``roadmap''  proposed in \cite{Vandoren2}.  By postulating invariance
under a larger discrete group $SL(3,\mbb{Z})$, a subset of the NS5-brane contributions
corresponding to the ``extended universal hypermultiplet" was 
conjectured in \cite{AutomorphicNS5}. This analysis (and presumably also the analysis
in \cite{Vandoren1}) breaks down for rigid Calabi-Yau
threefolds, the sector which we address in this work.

\subsection{Rigid Calabi-Yau Threefolds  and the Picard Modular Group}
\label{Section:RigidCY3}
In the present paper, we study the hypermultiplet moduli space  $\mc{M}_{\mathrm{H}}$
in a restricted setting, 
namely for type IIA string theory compactified on a {rigid} Calabi-Yau 
threefold $\mc{X}$ (i.e. with $h_{2,1}(\mc{X})=0$).
By the T-duality argument indicated above, our analysis applies equally well to the vector multiplet
moduli space in type IIB string theory compactified on  $\mc{X} \times S_1$. 
Rigid Calabi-Yau threefolds  are  rare, but examples can be found in the 
mathematics (see, e.g., \cite{Noriko}) and the physics literature (see, e.g.,
\cite{Candelas:1985en,Strominger:1985it,Candelas:1993nd,Bershadsky:1993cx}).
One of their peculiarities is that they do not admit a mirror  in the usual
sense, since
$h_{1,1}\geq 1$ for any Calabi-Yau threefold\footnote{It is possible that the superconformal
field theory on $\mc{X}$ admits a mirror description as a Landau-Ginzburg model $LG$, but 
it is not obvious that this equivalence should extend at the non-perturbative level. Put differently, 
it is unclear whether type IIA on 
$LG$ can still be lifted to M-theory, and whether type IIB on $LG$ still exhibits $SL(2,\IZ)$
symmetry.}. Thus, it is no longer clear that $\mc{M}_{\mathrm{H}}$ should admit 
an isometric action of $SL(2,\mbb{Z})$. Moreover, rigid 
Calabi-Yau threefolds do not admit a K3 fibration, so are not amenable to heterotic/type II 
duality  \cite{Aspinwall:1995vk}.

For such rigid Calabi-Yau threefolds then, the hypermultiplet sector consists solely of the ``universal hypermultiplet", given at tree-level by the quaternion-K\"ahler
symmetric space $\mc{M}_{\mathrm{UH}}(\mc{X})=SU(2,1)/\big(SU(2)\times U(1)\big)$. The metric on $\mc{M}_{\mathrm{UH}}(\mc{X})$ is obtained via the $c$-map procedure
from the complex structure moduli space of $\mc{X}$  \cite{Cecotti:1988qn,Ferrara:1988ff}.
Since $\mc{X}$ has no complex structure deformations, its prepotential $F(X)$ is determined
from the special geometry relations
\beq
X=\int_{\mc{A}} \Omega , \qquad \qquad \f{\partial F}{\partial X} = \int_{\mc{B}} \Omega\ 
\label{Periods}
\eeq
to be quadratic, namely 
\beq
F(X)=\tau X^2 /2\ ,\qquad \tau := \f{\int_{\mc{B}} \Omega}{\int_{\mc{A}}\Omega}
\label{tau}
\eeq
where $\Omega\in H^{3,0}(\mc{X})$ is the holomorphic 3-form,  
$(\mc{A}, \mc{B})$ is an integral symplectic basis of  $H_{3}(\mc{X},\mbb{Z})$,
and $\tau$ is a fixed complex number, the period matrix.  
The $c$-map then leads to the metric\footnote{Concretely, this metric may be obtained from the general $c$-map metric given in Eq. 4.31 of \cite{Alexandrov:2008nk} by setting $h_{2,1}=0$, $F(X)=\tau X^2/2$ and implementing the change of variables: $r=e^{-2\phi}$, $\zeta^{0}=-2\sqrt{2} \chi$, $\tilde\zeta_0 = 2\sqrt{2} \tilde\chi$, $\sigma =8\psi$.}
\beq ds^2_{\mc{M}_{\mathrm{UH}}}(\tau)= d\phi^2+e^{2\phi} \f{|d\tilde\chi+\tau d\chi|^2}{\Im \tau} + e^{4\phi} \big(d\psi +\chi
d\tilde{\chi}-\tilde{\chi}d\chi\big)^2.
\label{dsuh}
\eeq
In type IIA string theory compactified on $\mc{X}$, $e^{\phi}$ is the four-dimensional string coupling, $\psi$ is the NS-NS axion, dual to the 2-form $B_{(2)}$ in $D=4$,
and $(\chi, \tilde\chi)$ are the periods of the Ramond-Ramond 3-form $C_{(3)}$:
\beq
\chi = \int_{\mc{A}} C_{(3)}, \qquad \qquad \tilde\chi = \int_{\mc{B}} C_{(3)}.
\label{RRperiods}
\eeq
In the dual type IIB string theory on $\mc{X} \times S_1$, $e^{\phi}$ is instead the inverse radius of the circle in 4D Planck units, while $\chi,\tilde{\chi}$ are the components of the ten-dimensional Ramond-Ramond 4-form $C_{(4)}$ on $H^{3}(\mc{X},\IR)\times S^1$ and $\psi$ is
the NUT potential (the dual of the Kaluza-Klein gauge field).
 
Classically, the family of metrics \eqref{dsuh}, parametrized by $\tau$, are all locally isometric
to the symmetric space $SU(2,1)/(SU(2)\times U(1))=\mbb{CH}^2$ 
(see Section \ref{Section:SU(2,1)} for details). A standard choice is  to take $\tau=i$, leading to the familiar form of the left-invariant metric on $\mbb{CH}^2$. However, at the quantum level the choice of $\tau$ is not inoccuous. Indeed 
the Ramond-Ramond scalars $(\chi, \tilde\chi)$ are known \cite{Morrison:1995yi} to parametrize 
the intermediate Jacobian
\beq
J(\mc{X})=\f{H^{3}(\mc{X}, \mbb{C})}{H^{3}(\mc{X}, \mbb{Z})} \ .
\eeq
Equivalently they are subject to discrete identifications 
\beq
\label{chiid}
(\chi, \tilde\chi)\rightarrow (\chi+a, \tilde\chi+b), \hs a,b\in \mbb{Z}\ .
\eeq
If $\mc{X}$ is a rigid Calabi-Yau threefold, its intermediate Jacobian is an elliptic curve 
\beq
J(\mc{X}) = \mbb{C}/ ( \IZ + \tau \IZ) 
\eeq
where $\tau$ is the period matrix defined in (\ref{tau}). Thus, different choices of $\tau$
lead to locally isometric but globally inequivalent metrics. In the present work we shall restrict to the particular choice\footnote{We are grateful to Jan Stienstra for pointing out this assumption,
which was implicit in the first version of this work. We comment on the possible duality symmetries
for rigid Calabi-Yau compactifications 
with $\tau\neq i$ in Section \ref{Conclusions}, see also \cite{Bao:2010cc}.}  $\tau=i$,
 corresponding to rigid Calabi-Yau threefolds for which the intermediate Jacobian is a square torus $\mbb{C}/\mbb{Z}[i]$, where $\mbb{Z}[i]$  denotes the {Gaussian integers} $\{z\in\mbb{C}\ |\ \Re(z), \Im(z)\in \mbb{Z}\}$, corresponding to the ring of integers in the imaginary quadratic number field $\mbb{Q}(i)$. Mathematically, this choice implies  in particular  that $J(\mc{X})$
admits ``complex multiplication'', a notion that originates from the study of elliptic curves $\mbb{C}/(\mbb{Z}+\tau\mbb{Z})$, which are said to admit complex multiplication (or are of ``CM-type'') if and only if the modular parameter $\tau$ takes values in an imaginary quadratic extension of $\mbb{Q}$.  
Many but not all rigid Calabi-Yau threefolds admit complex multiplication; a necessary and sufficient criterion is that the intermediate Jacobian of the Calabi-Yau threefold is of CM-type (see, e.g., \cite{Noriko} for a review). An example of a rigid Calabi-Yau threefold that does not admit complex multiplication is provided by  the hypersurface constructed in \cite{Schoen}.\footnote{We also note that complex multiplication has appeared previously in the physics literature in \cite{Moore:1998pn,Gukov:2002nw}.} For examples of rigid Calabi-Yau threefolds that admit complex multiplication by $\mbb{\mbb{Z}}[i]$, as is relevant in the present work, see \cite{Noriko}.

While quantum corrections are bound to break the continuous
isometric action of $SU(2,1)$ on the hypermultiplet moduli space 
$\mc{M}_{\mathrm{UH}}$, 
we posit that they preserve a discrete arithmetic subgroup 
$G(\mbb{Z})\subset SU(2,1)$ (note however that we do not assume that $ \mc{M}^{\mathrm{exact}}_{\mathrm{UH}}$
is a double coset $G(\mbb{Z})\backslash G/K$).\footnote{The fate of the 
continuous symmetry $SU(2,1)$  at the level of higher derivative corrections to the 
low energy effective action 
will be analyzed in a follow-up paper \cite{CompactificationPaper}.}
On physical grounds, the subgroup $G(\mbb{Z})$ should contain the following action on the moduli:
\begin{enumerate}
\item  A discrete Heisenberg group $ N (\mbb{Z})$, acting by discrete (Peccei-Quinn) shift symmetries on the axions $\chi, \tilde{\chi}$ and $\psi$:
\beqa\label{heisshiftsscalars2}
\chi & \longmapsto& \chi+a\,,\nn\\
\tilde\chi & \longmapsto& \tilde\chi +b\,,\nn\\
\psi & \longmapsto&\psi +\frac12c -a\tilde\chi +b \chi,
\eeqa
where $a, b\in \mbb{Z}$ and $c\in 4\mbb{Z}$, while leaving the dilaton $\phi$ invariant. 
In the type IIA setting, the breaking of the continuous shifts of $\chi$ and $\tilde{\chi}$ are due to D2-brane instantons, while the breaking of the continuous shift of $\psi$ is due to the 
NS5-brane instantons. In the following, we shall make the stronger assumption that $\psi$ 
is in fact periodic with period $1/2$ (i.e. $c\in \mbb{Z}$). While this half-integer periodicity is 
predicted by the discrete symmetry that we are about to propose, it implies that only
NS5-branes in multiples of four\footnote{Thus, we find disagreement with~\cite{BeckerBecker},
who found that single NS5-brane instantons break continous Peccei-Quinn symmetries to 
\eqref{heisshiftsscalars2} with $c\in \mbb{Z}$.}
 can correct the metric. 

\item The ``electric-magnetic duality''   $R$ which interchanges the R-R scalars $\chi$ and $\tilde{\chi}$~\cite{BeckerBecker}:
\beq
R\hs :\hs (\chi, \tilde{\chi}) \hs \longmapsto \hs (-\tilde{\chi}, \chi).
\label{emduality}
\eeq
This symmetry is only expected to hold for rigid Calabi-Yau compactifications with $\tau=i$.
In that  case it amounts to a change of symplectic basis for $H_3(\mc{X}, \mbb{Z})$.

\item We further assume that a discrete subgroup $SL(2,\IZ)$ of the four-dimensional 
$S$-duality (or, on the type IIB side, Ehlers symmetry), acting 
in the standard non-linear way on the complex parameter $\chi+i e^{-\phi}$
on the slice $\tilde\chi=\psi=0$, is left unbroken by quantum corrections.  
As in earlier endeavours \cite{LambertWest2,Bao:2007er,Michel,Colonnello:2007qy,Bao:2007fx},
it is difficult to justify this assumption rigorously, but the fact, demonstrated herein,
that it leads to physically sensible results can be taken as support for this assumption.\footnote{On non-rigid Calabi-Yau manifolds one can use mirror symmetry to obtain S-duality transformations in IIA from the $SL(2,\mbb{Z})$ symmetry of type IIB, as has been advocated in literature for example in~\cite{Gunther:1998sc,deWit:1996ix,Bohm:1999uk,Manschot:2009ia}.}
\end{enumerate}
Based on these assumptions, it follows that, when $\tau=i$, $G(\mbb{Z})$ must be the {\em Picard modular group}\footnote{The nomenclature ``Picard group'' is not unique, in fact our Picard group is a member of a family of similar groups $PSU(1,n+1;\mbb{Z}[i])$ of which the case $n=0$, corresponding to $PSL(2,\mbb{Z}[i])$ is also often called {the} Picard group. In this paper we will always mean $SU(2,1;\mbb{Z}[i])$ when speaking of the Picard group.}
$SU(2,1;\mbb{Z}[i])$, defined as the intersection (see e.g. \cite{FaFrLaPa,FrancsicsLax})
\beq
\label{defpic}
SU(2, 1;\mbb{Z}[i]) := SU(2,1)\cap SL(3,\mbb{Z}[i]).
\eeq
Indeed, the symmetries 1,2,3 above reproduce the list of generators of the Picard modular group
obtained in  \cite{FaFrLaPa}.  We find it remarkable that adjoining electric-magnetic duality and S-duality to 
the physically well-established Peccei-Quinn shift symmetries  
generates an interesting discrete subgroup of $SU(2,1)$.

 We conclude this section by emphasizing that although the moduli $\{\phi, \chi, \tilde\chi, \psi\}\in \mc{M}_{\text{UH}}$
occur universally for any Calabi-Yau threefold, the ``universal hypermultiplet
moduli space" \eqref{dsuh} is not a universal subsector of the hypermultiplet moduli space 
$\mc{M}_{\mathrm{H}}(\mc{X})$ for non-rigid $\mc{X}$ \cite{Aspinwall}. (This is
is in contrast to the 
``extended universal hypermultiplet sector " introduced in \cite{AutomorphicNS5}.)
However, in cases where  $\mc{M}$ is a symmetric space, it can often be written 
as a fiber bundle over $\mc{M}_{\mathrm{UH}}$. One example is type II string theory
compactified on $T^7$, where the moduli space can be written as the fiber bundle
\cite{Cecotti:1988qn,Aspinwall}
\beq
\f{E_{7(7)}}{SU(8)/\mbb{Z}_2} \hs \to \hs
 \f{SU(2,1)}{SU(2)\times U(1)} \times \f{SL(2, \mbb{R})}{SO(2)} \ .
 \label{E7Bundle}
\eeq
In this context, it would be interesting to investigate which discrete subgroup of $SU(2,1)$ is singled out by the intersection with the U-duality group $E_{7(7)}(\mbb{Z})$ \cite{Hull:1994ys}. A similar decomposition as (\ref{E7Bundle}) also occurs for very special $\mc{N}=2$ supergravity theories,
where the second factor on the r.h.s. is replaced by a non-compact version of the
5-dimensional U-duality group \cite{Gunaydin:2001bt,Gunaydin:2004md}. Thus, it is
plausible that the considerations in this paper may have bearing on the more general case 
of non-rigid Calabi-Yau threefolds.

\subsection{Automorphic Forms for the Picard Modular Group}

Having identified \eqref{defpic} as a candidate symmetry group, we apply standard
machinery to construct a simple automorphic form $\mc{E}_s(\mc{K})$ of $\pg$, which 
satisfies 
\beq
\Delta_{\mbb{CH}^2}\, \mc{E}_s(\phi, \lambda, \ga)=s(s-2)\, \mc{E}_s(\phi, \lambda, \ga)\,,
\label{EigenvalueEquation}
\eeq
where $\Delta_{\mbb{CH}^2}$ is the Laplace-Beltrami operator on $\mbb{CH}^2$, given in 
\eqref{Laplacian} below.
For the sake of completeness, and because they mutually enlighten each other, we shall present
three equivalent constructions of the Eisenstein series for the Picard modular group:
\begin{itemize}
\item 
First,  we generalize 
the construction of non-holomorphic Eisenstein series for 
real classical groups over the integers  in \cite{ObersPioline} 
to unitary groups over the Gaussian integers,
and consider the constrained lattice sum
\beq\label{eisenintro}
\mc{E}_s(\mc{K}):= \sideset{}{'}\sum_{\vec{\om}\in
  \mbb{Z}[i]^3\atop \vec{\om}^{\dagger}\cdot\eta\cdot
\vec{\om}=0} \Big[\vec{\om}^{\dagger}\cdot \mc{K}\cdot \vec{\om}\Big]^{-s}, 
\eeq
where $\mc{K}=\mc{V}\mc{V}^\dagger $ is a Hermitian matrix parametrizing the coset space
$SU(2,1)/\big(SU(2)\times U(1)\big)$, and the sum runs over non-zero triplets
of Gaussian integers $\vec{\om}$ subject to a certain quadratic constraint \eqref{constraint}. 
The prime indicates that $\vec \om=(0,0,0)$ is excluded from the sum. More generally,
throughout this paper a prime on a summation symbol will indicate that the zero value is 
excluded from the sum.

\item 
Secondly, we consider the $\pg $-invariant Poincar\'e series 
\beq
\mc{P}_s(\mc{Z}) := \sum_{\ga\in {N}(\mbb{Z})\backslash
  \pg } \mc{F}(\ga\cdot \mc{Z})^s, 
\eeq
where $\mc{F}(\mc{Z})$ is a function on $SU(2,1)/\big(SU(2)\times U(1)\big)$ which
is manifestly invariant under the Heisenberg subgroup ${N}(\mbb{Z})\subset
\pg$.  As shown in Section \ref{Section:Eisenstein}, 
this construction yields the same automorphic form as before up to an $s$-dependent
factor,
\beq
\mc{E}_s(\mc{K}) =4\zeta_{\mbb{Q}(i)}(s) \, \mc{P}_s(\mc{Z}) \ ,
\eeq
where $\zeta_{\mbb{Q}(i)}(s) $ is the Dedekind zeta function for the Gauss field $\mbb{Q}(i)$,
defined in \eqref{dedezeta}.
  
\item 
Thirdly, in Appendix \ref{pAdicConstruction} we show that the same 
automorphic form can be obtained by the general adelic method explained 
 e.g. in  \cite{PiolineWaldron1,Kazhdan,PiolineWaldron2},
\beq\label{eqn:padicintro}
\mc{E}_s(\mc{V}) =\sideset{}{'}\sum_{(C_1,C_2)\in P\mbb{Q}(i)^2\atop |C_1|^2-2\Im(C_2)=0}
  \rho(\mc{V})\cdot   \prod_{g\, \text{prime} }f_g(C_1,C_2)\ ,
\eeq
where  the infinite product runs over all Gaussian prime numbers including $g=\infty$. Here, 
$\mc{V}$ is a coset representative of
$SU(2,1)/\big(SU(2)\times U(1)\big)$, $P\mbb{Q}(i)^2$ is the projective space
$\mbb{Q}(i)^3/\mbb{Q}(i)^*$, $\rho$ is the principal continuous series representation of
$SU(2,1)$ on functions on $P\mbb{Q}(i)^2$,  $f_{\infty}:=f_{{K}}$ is the 
$SU(2)\times U(1)$-invariant spherical vector
from \cite{PiolineGunaydin}, and $f_g$ ($g< \infty$) its $p$-adic 
counterpart.
\end{itemize}
Generally, any function $f$ on $\mbb{CH}^2$ which obeys the following three conditions:
\begin{enumerate}
\item $f$ satisfies the Laplace equation $\Delta_{\mbb{CH}^2}\, f ={s(s-2)}\, f$,  
\item $f$ is invariant under the Heisenberg group $N(\mbb{Z})$ (see \eqref{heisshiftsscalars2}), 
\item $f$ has at most polynomial growth in $e^\phi$ as $e^{\phi}\rightarrow 0$,
\end{enumerate}
admits a non-Abelian Fourier expansion of the form\footnote{A similar  expansion 
was given in \cite{Ishikawa} but we disagree with the details in this reference. For instance, the second constant term appears to be missing in Thm 5.3.1ii) of \cite{Ishikawa}. Such a term is required on general grounds \cite{Langlands}.}
\beq
\begin{split}
f(\phi, \chi, \tilde\chi, \psi)&=A(s)\, e^{-2s\phi} + B(s)\, e^{-2(2-s)\phi} \\
&+ e^{-2\phi}\sideset{}{'}\sum_{(\ell_1, \ell_2)\in\mbb{Z}^2}
C^{(\text{A})}_{\ell_1, \ell_2}(s) \, 
K_{2s-2}\Big(2\pi e^{-\phi}\sqrt{\ell_1^2+\ell_2^2}\Big) e^{-2\pi i (\ell_1\chi+\ell_2\tilde\chi)} \\
&+ e^{-\phi} \sideset{}{'}\sum_{k\in\mbb{Z}} \sum_{\ell=0}^{4|k|-1}\sum_{n\in\mbb{Z}+\f{\ell}{4|k|}}\sum_{r=0}^{\infty} 
C^{\text{(NA)}}_{r,k, \ell}(s) \hs |k|^{1/2-s} e^{-4\pi |k| (\tilde\chi-n)^2} \\
&  \quad \times \,H_{r}\Big(\sqrt{8\pi |k|} (\tilde\chi-n)\Big) W_{-r-\f{1}{2}, s-1} \Big(4\pi |k| e^{-2\phi}\Big) e^{8\pi i k n \chi-4\pi i k(\psi+\chi\tilde\chi)},
\end{split}
\label{FirstFourier}
\eeq
where $K_s$, $H_{r}$ and $W_{-r-\f{1}{2}, s-1}$  denote the modified Bessel function, Hermite polynomial  and Whittaker function, respectively. The first line in \eqref{FirstFourier} corresponds to the ``constant terms", i.e.  the leading terms in an expansion at the cusp $e^{\phi}\rightarrow 0$. The second line is the Abelian contribution, corresponding to an ordinary Fourier expansion with respect to the Abelianized Heisenberg group 
$N/Z$, where $N$ is the Heisenberg subgroup of  upper triangular matrices in $SU(2,1)$ and $Z$ is the center of $N$ (in this particular case it coincides with the commutator subgroup $[N,N]$). 
Finally, the last two lines represent the non-Abelian contribution, i.e. 
the part of $f$ which transforms non-trivially under the
action of the center $Z$ of the Heisenberg group. We have chosen to write them in the
``$\tilde\chi$-polarization'', where the $a$ and $c$ shifts in \eqref{heisshiftsscalars2}
are diagonalized, but it is also possible to use the ``$\chi$-polarization'', where $b$ and $c$
shifts are diagonalized (see Section \ref{section:generalFourierexpansion} for a detailed discussion of the choice of polarization).
The constant terms 
$A(s),B(s)$ and the numerical Fourier coefficients 
$C^{(\text{A})}_{\ell_1, \ell_2}(s)$ will be derived in Section \ref{Section:FourierExpansion}.

Our main mathematical results may now be summarized in the following propositions and conjecture:

\vspace{.3cm}

\noindent {\bf Proposition 1 (Constant Terms):} {\it The constant terms in the Fourier expansion of the Eisenstein series $\mc{E}_s(\phi, \chi, \tilde\chi, \psi)$ in Eq. (\ref{FirstFourier}) are given by 
\beq
\mc{E}_s^{(\text{const})}= 4\zeta_{\mbb{Q}(i)}(s) \left\{ e^{-2s\phi} + \frac{\mathfrak{Z}(2-s)}{\mathfrak{Z}(s)} e^{-2(2-s)\phi}\right\}.
\label{ConstantTermFormula}
\eeq
 Here we have defined the ``Picard zeta function'' $\mathfrak{Z}(s)$ through
\beq 
\mathfrak{Z}(s) := \zeta_{\mbb{Q}(i)*}(s) \, \beta_*(2s-1),
\eeq
 where $ \zeta_{\mbb{Q}(i)*}(s)$ and $\beta_*(2s-1)$ are the completed Dedekind zeta function (\ref{completedDedekind}) and Dirichlet beta function (\ref{completedBeta}), respectively. }
 This proposition is proven in Sections \ref{Section:FirstConstantTerm} and \ref{Section:SecondConstantTerm}.\\

\noindent {\bf Proposition 2 (Abelian Fourier Coefficients):} {\it The numerical Fourier coefficients $C^{(\text{A})}_{\ell_1, \ell_2}(s)$ in the expansion of $\mc{E}_s(\phi, \chi, \tilde\chi, \psi)$ with respect to the abelianized Heisenberg group $N/[N,N]$ are given by
\beq
C^{(\text{A})}_{\ell_1, \ell_2}(s)= \frac{2\zeta_{\mbb{Q}(i)}(s)}{\mathfrak{Z}(s)}\big[\ell_1^2+\ell_2^2\big]^{s-1} \sum_{\omega'|\Lambda} |\omega^{\prime}|^{2-2s} 
\sum_{z|\frac{\Lambda}{\omega^{\prime}}} |z|^{4-4s},
\eeq
where $\sum_{\omega^{\prime} | \Lambda}$ denotes a sum over Gaussian divisors of $\Lambda=\ell_2-i\ell_1\in \mbb{Z}[i]$, and $\omega^{\prime}$ is a Gaussian integer such that $\gcd\big(\Re(\omega^{\prime}), \Im(\omega^{\prime})\big)=1$. } This  proposition is proven in Section \ref{AbelianCoefficients} and Appendix \ref{AbelianMeasureDetails}.

\vspace{.3cm}

\noindent  To have a complete understanding of the Fourier expansion one must also extract the non-Abelian numerical coefficients $C^{\text{(NA)}}_{r,k, \ell}(s)$, corresponding to the Fourier expansion of $\mc{E}_s$ with respect to the center $Z=[N,N]$ of the Heisenberg group $N$. Some preliminary results in this direction are presented in Section \ref{NonAbelianCoefficients}. 

\vspace{.3cm}

\noindent {\bf Conjecture (Functional Relation):}  {\it The Eisenstein series $\mc{P}_s=(4\zeta_{\mbb{Q}(i)}(s))^{-1}\,  \mc{E}_s$ satisfies the following functional relation:}
\beq
\label{conj}
\mathfrak{Z}(s) \, \mc{P}_s = \mathfrak{Z}(2-s) \, \mc{P}_{2-s}\ .
\eeq

\noindent The map $s\mapsto 2-s$ corresponds to the action of the restricted Weyl group of $\mf{su}(2,1)$. The
validity of \eqref{conj} at the level of the constant and Abelian terms holds true by the above propositions, but our incomplete understanding of the non-Abelian terms prevents us from stating it as a theorem. We note that a different functional relation for $\mc{P}_s$ was stated in \cite{Orloff}  but it appears to be inconsistent with the constant terms (\ref{ConstantTermFormula}).
 
\subsection{Eisenstein Series and the Exact Universal Hypermultiplet Geometry}

As a test of the idea that the Picard modular group should control the quantum corrections
to the universal hypermultiplet moduli space, we tentatively propose that the $\pg$-invariant 
Eisenstein series $\mc{E}_s(\phi, \chi, \tilde\chi, \psi)$, for the specific value $s=3/2$, gives a non-perturbative completion of the contact potential $e^{\Phi(x^{\mu}, z)}$ restricted to a certain holomorphic 
section $z=z(x^{\mu})$ of the twistor space $\mc{Z}_{\mc{M}_{\text{UH}}}$ of the universal hypermultiplet moduli space.  In this context, the constant terms of the Fourier expansion (\ref{FirstFourier}) represent the classical and one-loop contributions  to the moduli space metric, respectively. Matching the power of the dilaton $e^\phi$ is what fixes the value of $s$ above.  This proposal is tentative however,
as it turns out  that the numerical coefficient of the one-loop term predicted by (\ref{FirstFourier}) 
is inconsistent with its known value in string theory \cite{AntoniadisMinasian1} 
(in particular, it has the wrong sign). We proceed nevertheless, since the structure
of the non-Abelian Fourier expansion \eqref{FirstFourier} is largely independent of the
automorphic form under consideration.

In this scenario, the Abelian terms in (\ref{FirstFourier}) have the suitable form to 
describe the effects of D2-brane instantons wrapping supersymmetric 3-cycles in the homology class
$\ell_1\mc{A}+\ell_2\mc{B}\in H_3(\mc{X},\mbb{Z})$, where $\mc{A}$ and $\mc{B}$ provide a symplectic basis for $H_3(\mc{X}, \mbb{Z})$. The  Abelian Fourier coefficients $C^{(\text{A})}_{\ell_1, \ell_2}(3/2)$, related to the instanton measure $\mu_{3/2}(\ell_1, \ell_2)$ via (\ref{NumericalAbelianFourierCoefficients}), should count supersymmetric cycles in the above
holomogy class. The  instanton measure $\mu_{3/2}(\ell_1, \ell_2)$ generalizes the familiar D$(-1)$ instanton measure $\mu_{3/2}(N)$ of \cite{Green:1997tv} which is also known to capture the effects of pure charge $N$ $A$-type D2-brane instantons \cite{Vandoren2} (corresponding to $\ell_2=0$). 
On the other hand non-Abelian terms 
have the suitable form to represent the
effects of charge $k$ NS5-brane instantons, possibly bound to D2-branes.
We take these encouraging facts as evidence that the Picard modular group 
should be a powerful principle in constraining the exact metric on the 
hypermultiplet moduli space in rigid Calabi-Yau compactifications, and in particular
in determining the NS5-brane instantons which have  remained elusive for generic 
Calabi-Yau compactifications. We comment on alternative choices of automorphic form for $SU(2,1;\mbb{Z}[i])$ which may alleviate the shortcomings of the Eisenstein series $\mc{E}_{3/2}$ 
in Section~\ref{Conclusions}.

\subsection{Outline}

In Section \ref{Section:SU(2,1)} we give a detailed description of the group $SU(2,1)$,
the symmetric space $SU(2,1)/(SU(2)\times U(1))$ and the Picard modular group $\pg$.
In Section \ref{Section:Eisenstein} we  construct  an $\pg$-invariant Eisenstein series $\mc{E}_s(\phi, \chi, \tilde\chi, \psi)$ in the principal continuous series of $SU(2,1)$ by two different methods. We proceed in Section  \ref{Section:FourierExpansion} to compute the Fourier expansion of $\mc{E}_s$, extracting explicit forms for the constant terms as well as the Abelian and non-Abelian Fourier coefficients. Finally, in Section~\ref{Section:InstantonCorrections}, we use the automorphic form  $\mc{E}_s(\phi, \chi, \tilde\chi, \psi)$ at order $s=3/2$ to conjecture the exact form of the D2-brane and NS5-brane instanton corrections to the universal hypermultiplet moduli space $\mc{M}_{\mathrm{UH}}$. In Appendix~\ref{Gaussians}, we review some basic facts about Gaussian integers and Gaussian primes as well as a reminder of Dirichlet series.
The third construction of the Eisenstein series $\mc{E}_s(\phi, \chi, \tilde\chi, \psi)$ using an adelic construction is given for completeness in Appendix \ref{pAdicConstruction}. This construction can be viewed as an extension of the analysis in Section 2 of \cite{PiolineGunaydin} to the automorphic setting. In Appendix~\ref{AbelianMeasureDetails} we provide details on the derivation of the Abelian Fourier coefficients $C_{\ell_1, \ell_2}^{(\text{A})}(s)$.

\section{On the Picard Modular Group $\pg$}
\label{Section:SU(2,1)}

As indicated in the introduction, a key character in this paper is
the symmetric space $SU(2,1)/(SU(2)\times U(1))$. This space describes
the tree-level moduli space of the universal hypermultiplet in type IIA string theory
compactified on a rigid Calabi-Yau threefold.\footnote{We note that the coset space $SU(2,1)/(SU(2)\times U(1))$ also appears as the moduli space of the Einstein--Maxwell system when dimensionally
reduced from $D=4$ to $D=3$ on a spacelike circle.} In this section,
we set up notations for the group $SU(2,1)$, give two equivalent descriptions of 
the symmetric space $SU(2,1)/(SU(2)\times U(1))$,  and introduce the Picard modular group $SU(2,1;\Zn[i])$.

\subsection{The Group $SU(2,1)$ and its Lie Algebra $\mf{su}(2,1)$}
\label{sec:su21sec}

The Lie group $SU(2,1)$ is defined as a subgroup of the group $GL(3,\cx)$ of
invertible $(3\times 3)$ complex matrices via
\beq\label{SU21def}
SU(2,1) = \left\{ g \in GL(3,\cx)\,:\, g^\dagger \eta g = \eta\,\,
  \,\text{and}\, \det(g)=1\right\}\,.
\eeq
Here, the defining metric $\eta$ is given by
\beq\label{invariantmetric}
\eta = \left(\begin{array}{ccc}
0 & 0 &-i\\
0 & 1 & 0 \\
i & 0 & 0
\end{array}\right)
\eeq
and has signature $(++-)$. The condition $g^\dagger \eta g=\eta$
already implies $|\det(g)|=1$ and so we can also think of $SU(2,1)$ as the set
of unitary matrices $U(2,1)$ modulo a pure phase, $SU(2,1)\cong PU(2,1)$, with the
projectivization $P$ referring to the equivalence relation $g\sim g
e^{i\alpha}$ for $\alpha \in [0,2\pi)$. The diagonal matrices
$e^{i\alpha}\text{diag}(1,1,1)$ form the center of the group $U(2,1)$.

The Lie group $SU(2,1)$ as defined in (\ref{SU21def}) has as Lie algebra
of real dimension 8
\beq\label{su21def}
\mf{su}(2,1) =\left\{ X \in \mf{gl}(3,\cx) \,:\, X^\dagger \eta + \eta X =0
  \,\,\text{and}\,\, \tr(X)=0\right\}\,.
\eeq
It consists of four compact and four non-compact generators, and
the maximal real torus is one-dimensional. We define the non-compact and
compact Cartan generators
\beq\label{cartans}
H = \left(\begin{array}{ccc}1&0&0\\0&0&0\\0&0&-1\end{array}\right)\,,\quad
J = \left(\begin{array}{ccc}i&0&0\\0&-2i&0\\0&0&i\end{array}\right)\,,
\eeq
the positive step operators
\beq\label{posstep}
X_1  = \left(\begin{array}{ccc}0&-1+i&0\\0&0&1-i\\0&0&0\end{array}\right)\,,\quad
\tilde{X}_1
=\left(\begin{array}{ccc}0&1+i&0\\0&0&1+i\\0&0&0\end{array}\right)\,,\quad
X_2 =  \left(\begin{array}{ccc}0&0&1\\0&0&0\\0&0&0\end{array}\right)\,,
\eeq\label{negstep}
and the negative step operators
\beq
Y_{-1}
=\left(\begin{array}{ccc}0&0&0\\1+i&0&0\\0&-1-i&0\end{array}\right)\,,\quad
\tilde{Y}_{-1}
=\left(\begin{array}{ccc}0&0&0\\-1+i&0&0\\0&-1+i&0\end{array}\right)\,,\quad
Y_{-2} = \left(\begin{array}{ccc}0&0&0\\0&0&0\\-1&0&0\end{array}\right)\,.
\eeq
The subscript refers to the eigenvalue under the adjoint action of the
non-compact Cartan generator $H$, e.g. $[H,X_1]=X_1$; the adjoint action of
the compact Cartan generator $J$ is not diagonalisable over the real
numbers. Furthermore, the generators satisfy
\beq\label{HeisenbergAlgebra}
\left[X_1,\tilde{X}_1\right]=-4 X_2\,,
\eeq
such that the positive step operators form a Heisenberg algebra.
Furthermore, the negative step operators $Y$ are minus the Hermitian conjugate
of the positive step operators $X$.

The Lie algebra $\mf{su}(2,1)$ has a natural five grading by the generator $H$
as a direct sum of vector spaces
\beq\label{5grading}
\mf{su}(2,1)=\mf{g}_{-2}\oplus \mf{g}_{-1} \oplus \mf{g}_0 \oplus \mf{g}_1
\oplus \mf{g}_2\,,
\eeq
with
\beqa
\mf{g}_{-2} = \mbb{R}Y_{-2}\,,\,
\mf{g}_{-1} = \mbb{R}Y_{-1}\oplus \mbb{R}\tilde{Y}_{-1}\,,\,
\mf{g}_0 = \mbb{R}H\oplus \mbb{R} J \,,\,
\mf{g}_1  = \mbb{R}X_{1}\oplus \mbb{R}\tilde{X}_{1}\,,\,
\mf{g}_{2} = \mbb{R}X_{2}\,.
\eeqa
One sees that the $H$-eigenspaces with eigenvalue $\pm 1$ are degenerate. This
is a characteristic feature of the reduced root system $BC_1$ underlying the
real form $\mf{su}(2,1)$ of $\mf{sl}(3,\cx)$. There is a single simple root $\alpha$
since the real rank of $\mf{su}(2,1)$ is one, and there are non-trivial root
spaces $\mf{g}_1$ and $\mf{g}_2$ corresponding to $\alpha$ and $2\alpha$,
respectively.\footnote{A discussion of the restricted root
  system can for example be found  in \cite{LivRel}.}
The $\mf{sl}(2,\mbb{R})$ subalgebra associated with the
$2\alpha$ root space is  canonically normalised and can be given a standard
basis for example with $H$, $E=X_2$ and $F=-Y_{-2}$, so that
$[E,F]=H$. The corresponding $SL(2,\IR)$ subgroup of $SU(2,1)$ is given by
matrices of the form
\beq\label{sl2sub}
\left\{ \left(\begin{array}{ccc}
a & 0 & b\\
0 & 1 & 0\\
c & 0 & d\end{array}\right) \,:\,
a,b,c,d\in \mbb{R}\,\,\text{and}\,\, ad-bc=1\right\}
\subset SU(2,1)\,.
\eeq
Under this embedding, the fundamental representation of $SU(2,1)$ decomposes as $3=2\oplus 1$.
There exists a second, non-regular embedding of $SL(2,\mbb{R})$ inside $SU(2,1)$,  consisting
of matrices of the form
\beq\label{sl2sub2}
SL(2,\mbb{R}) = \left\{ \left(\begin{array}{ccc}
a^2 & (-1+i)a b & i b^2\\
(-1-i)a c & ad+bc & (1-i)b d\\
-i c^2 & (1+i)c d & d^2\end{array}\right) \,:\,
a,b,c,d\in \mbb{R}\,\,\text{and}\,\, ad-bc=1\right\}\,.
\eeq
Under this embedding, the fundamental representation of $SU(2,1)$ remains irreducible.
The two subgroups (\ref{sl2sub}) and (\ref{sl2sub2}) together generate the whole of $SU(2,1)$.

The Iwasawa decomposition of the Lie algebra $\mf{su}(2,1)$ reads
\beq \label{su21iwa}
\mf{su}(2,1)=\mf{n}_+ \oplus \mf{a} \oplus \mf{k} ,
\eeq
where the non-compact (Abelian) Cartan subalgebra is defined as $\mf{a}=\mbb{R}H$,
while the nilpotent subspace
$\mf{n}_+=\mbb{R}X_1\oplus\mbb{R}\tilde{X}_1\oplus \mbb{R}X_2$ is spanned by the
positive step operators. The compact subalgebra of
$\mf{su}(2,1)$ is $\mf{k}=\mf{su}(2)\oplus\mf{u}(1)$ as a direct sum of Lie
algebras.\footnote{By contrast, the Iwasawa decomposition (\ref{su21iwa}) is
  only a direct sum of vector spaces and not of Lie algebras.}
The generators of $\mf{su}(2)$ and $\mf{u}(1)$ are given explicitly by the anti-Hermitian matrices
\beqa
\hat{K}_1 &=& \frac14\left(X_{1} + Y_{-1}\right)\,,\quad
\hat{K}_2 = \frac14\left(\tilde{X}_{1} + \tilde{Y}_{-1}\right) \,,\quad
\hat{K}_3 =  \frac14\left(X_{2} + Y_{-2}  + J\right) \,,\nn \\
\hat{J} &=& \frac{3}{4}\left( X_{2} + Y_{-2}\right)-\frac14 J\,.
\eeqa
These satisfy $\left[\hat{J} ,\hat{K}_i\right] =0$ and
$\left[\hat{K}_i,\hat{K}_j\right] = -\epsilon_{ijk} \hat{K}_k$.
The Weyl group  of the  reduced root system $BC_1$ is
\beq
\label{wgrp}
\cW (\mf{su}(2,1)) = \cW(BC_1) \cong \Zn_2\,,
\eeq
corresponding to the Weyl reflection with respect to $\alpha$.

\subsection{Complex Hyperbolic Space}
\label{ComplexHyperbolicSpace}

The group $SU(2,1)$ acts transitively and isometrically on the complex two-dimensional 
space\footnote{This is referred to as the ``unbounded hyperquadric model'' in
\cite{FrancsicsLax}.}
\beq\label{chs}
\cxh^2 = \left\{ \cZ=(z_1,z_2)\in \cx^2 \,:\,\cF(\cZ)>0\right\}\,,
\eeq
equipped with the K\"ahler metric 
\beq\label{metricZ}
ds^2 = \frac14 \cF^{-2}\left[dz_1d\bar{z}_1 + iz_2dz_1d\bar{z}_2
-i \bar{z}_2 dz_2d\bar{z}_1 +2\Im(z_1)dz_2d\bar{z}_2\right]\,.
\eeq
The ``height function"  $\cF:\cx^2\to \mbb{R}$  is defined by
\beq\label{KaehlerArgument}
\cF(\cZ) := \Im(z_1) - \frac12 |z_2|^2 \,,
\eeq
and provides a K\"ahler potential for the metric \eqref{metricZ},
\beq\label{kahlerpot}
K_{\mbb{CH}^2}(\cZ) = -\log \cF(\cZ)\,.
\eeq
The action of $SU(2,1)$ on $\cZ\in\cxh^2$ is via fractional linear transformations
\beq\label{su21act}
g\cdot \cZ = \frac{A\cZ+B}{C\cZ+D}\quad \quad \text{for}\quad
g=\left(\begin{array}{cc}A&B\\C&D\end{array}\right)\,,
\eeq
where the blocks $A$, $B$, $C$ and $D$ have the sizes $(2\times 2)$, $(2\times 1)$, $(1\times
2)$ and $(1\times 1)$, respectively, so that the denominator is
a complex number. Since the height function transforms as 
\beq\label{trmheight}
\cF(g\cdot \cZ) = \frac{\cF(\cZ)}{|C\cZ+D|^2}\,,
\eeq
the condition $\cF(\cZ)>0$ is preserved and the action is isometric. In fact, when verifying (\ref{trmheight})
one only requires the condition $g^\dagger\eta g =\eta$ so that
(\ref{su21act}) defines an action of all of $U(2,1)$ on complex hyperbolic
two-space. Since elements from the center act trivially, one can restrict to
$PU(2,1)\cong SU(2,1)$.
We will refer to the space $\cxh^2$ defined in (\ref{chs}) as the {\em complex hyperbolic space}, or the {\em complex upper half plane}. The slice $z_2=0, \Im(z_1)>0$ inside $\cxh^2$
is preserved by the action of the $SL(2,\IR)$ subgroup in  \eqref{sl2sub}, and gives an embedding of the standard Poincar\'e upper half plane inside $\cxh^2$.

\subsection{Relation to the Scalar Coset Manifold $SU(2,1)/(SU(2)\times U(1))$}
\label{sec:su21coset}

The complex hyperbolic upper half plane is isomorphic to the Hermitian symmetric space
\beq\label{su21cos}
\cxh^2 \cong SU(2,1)/(SU(2)\times U(1))\,,
\eeq
where the right hand side should properly be restricted to the connected
component of the identity. This space can be parametrized  by four real variables 
$\{\phi, \chi, \tilde\chi, \psi\}$, using the coset representative in the 
Iwasawa gauge (\ref{su21iwa}), 
\beqa\label{cosetel}
\cV =  e^{ \chi X_1 + \tilde\chi\tilde{X}_1 +2\psi X_2} e^{-\phi  H}
= \left(\begin{array}{ccc}
e^{-\phi} & \tilde\chi-\chi +i (\chi+\tilde\chi) &
e^{\phi}\left(2\psi+i(\chi^2+\tilde\chi^2)\right)\\
0 & 1 & e^\phi \left(\chi+\tilde\chi+i(\tilde\chi-\chi)\right)\\
0 & 0 & e^{\phi}
\end{array}\right)\,.
\eeqa
The symmetric space is a right coset in our conventions, the coset element $\cV$
transforming as
$\cV\to g\cV k^{-1}$ with $g\in SU(2,1)$ and $k\in SU(2)\times U(1)$.  The
four scalar fields can take arbitrary real values.

It is convenient to define the Hermitian matrix
\beq\label{kdef}
\cosm = \cV\cV^\dagger
\eeq
that transforms as $\cosm\to g\cosm g^\dagger$ under the action of $g\in
SU(2,1)$. Explicitly, this matrix reads
\beq
\mc{K}=\left( \begin{array}{ccc}
e^{-2\phi}+|\lambda|^2+e^{2\phi}|\ga|^2 & i\bar\lambda+e^{2\phi} \bar\lambda\ga & e^{2\phi} \ga \\
-i{\lambda}+e^{2\phi} {\lambda}\bar{\ga} & 1+e^{2\phi} |\lambda|^2 & e^{2\phi}{\lambda}\\
e^{2\phi}\bar{\ga} & e^{2\phi} \bar\lambda & e^{2\phi} \\
\end{array} \right),
\label{generalizedmetric}
\eeq
where, for later convenience, we have defined the complex variables
\beq\label{eisenvars}
\lambda:=\chi+\tilde\chi+i(\tilde\chi-\chi), \qquad \ga:= 2\psi+\f{i}{2}|\lambda|^2.
\eeq
From $\mc{K}$ one can obtain the metric on the symmetric space via
\beq
ds^2 = -\frac18\text{tr} \left(d\cosm \,d(\cosm^{-1})\right)
= \frac18\text{tr}\left( \cV^{-1}d\cV + (\cV^{-1}d\cV)^\dagger\right)^2\,.
\eeq
Working this out for the coset element (\ref{cosetel}) one finds the following
$SU(2,1)$ invariant metric
\beq\label{metricscalars}
ds^2 = d\phi^2 + e^{2\phi}(d\chi^2+d\tilde\chi^2)
  + e^{4\phi}(d\psi+\chi d\tilde\chi-\tilde\chi d\chi)^2\,.
\eeq
Comparing (\ref{metricscalars}) to (\ref{metricZ}) leads to the identification
\beqa\label{scalarsZ}
z_1 &=& 2\psi + i \left(e^{-2\phi}+\frac12 |z_2|^2\right) = 2\psi+i\left(e^{-2\phi}+\chi^2+\tilde\chi^2\right)\,,\nn\\
z_2 &=& \chi + \tilde\chi + i(\tilde\chi-\chi)\,.
\eeqa
Note that $z_1=\ga +ie^{-2\phi}, z_2=\lambda$, and the condition $0<\cF(\cZ)=e^{-2\phi}$ 
is automatically satisfied. Note that in terms of the real variables $\{\phi, \chi, \tilde\chi, \psi\}$ the K\"ahler potential (\ref{kahlerpot}) simply becomes $K_{\mbb{CH}^2}=2\phi$.

In the variables $\cZ=(z_1,z_2)$ given by (\ref{scalarsZ}), the matrix $\cosm$
of (\ref{kdef}) takes the simple form
\beq\label{cosetmatrix}
\cosm  = \tcosm+\eta \,,
\eeq
where $\eta$ is the defining matrix of $SU(2,1)$ given in
(\ref{invariantmetric}) and
\beq\label{ktildematrix}
\tcosm =  e^{2\phi}\left(\begin{array}{ccc}
|z_1|^2 & z_1\bar{z}_2 & z_1\\
\bar{z}_1 z_2 & |z_2|^2&z_2 \\
\bar{z}_1& \bar{z}_2 & 1
\end{array}\right)\,,
\eeq
where one should bear in mind that $e^{2\phi}=1/\cF(\cZ)$.
The relations \eqref{scalarsZ} together with \eqref{su21act} allow one to determine the action of 
an element of $SU(2,1)$ in the real coordinates $\phi,\chi,\tilde\chi,\psi$.
In particular, one may check that on the slice $\tilde\chi=\psi=0$, the $SL(2,\IR)$
subgroup  (\ref{sl2sub2})  acts by fractional linear transformations on the
complex modulus $\chi + i e^{-\phi}$. This action may be considered as
a remnant of the $SL(2,\IR)$ S-duality in ten-dimensional type IIB string theory.
Similarly, on the slice $z_2=0$ the $SL(2,\IR)$ subgroup  (\ref{sl2sub})  acts by fractional linear transformations on the complex modulus $2\psi + i e^{-2\phi}$, which realizes
 four-dimensional S-duality.

\subsection{Coset Transformations and Subgroups of $SU(2,1)$}
\label{subsec:CosetTransf}

We now study the effect of some particular elements of $SU(2,1)$ on complex hyperbolic two-space
which have an immediate physical interpretation.

\subsubsection*{Heisenberg Translations}
 Let $ N  $ denote the exponential of the nilpotent algebra of positive step operators $\mf{n}_+$. We define the following elements of $ N  $
\beq\label{heisgen}
T_1 = \left(\begin{array}{ccc}
1 & -1+i & i \\
0 & 1 & 1-i\\
0 &0 & 1\end{array}\right)\,,
\quad
\tilde{T}_1  = \left(\begin{array}{ccc}
1 & 1+i & i \\
0 & 1 & 1+i\\
0 &0 & 1\end{array}\right)\,,
\quad
T_2 = \left(\begin{array}{ccc}
1 & 0 & 1 \\
0 & 1 & 0\\
0 &0 & 1\end{array}\right)\,.
\eeq
These are defined such that $T_1=\exp(X_1)$ etc. Any element  $n\in N  $ can be written as
\beqa\label{GeneralHeisenberg}
n &=& (T_1)^a (\tilde{T}_1)^b (T_2)^{c+2ab} = e^{a X_1 + b\tilde{X}_1 +c X_2} \nn\\
&=& \left(\begin{array}{ccc}
1 & a(-1+i) +b(1+i) & c +i (a^2+b^2)\\
0 &1 & a(1-i) +b(1+i) \\
0 &0 &1\end{array}\right)
\eeqa
for $a,b,c\in\mbb{R}$. The effect of this transformation on $\cZ=(z_1,z_2)$ is
\beqa\label{HeisenbergAction}
z_1 & \longmapsto& z_1 + \big[a(-1+i)+b(1+i)\big] z_2 +c+i(a^2+b^2)\,,\nn\\
z_2 & \longmapsto& z_2 + a(1-i)+b(1+i)\,,
\eeqa
or in terms of the four scalars fields of (\ref{cosetel})
\beqa\label{heisshiftsscalars}
\phi & \longmapsto& \phi\,,\nn\\
\chi & \longmapsto& \chi+a\,,\nn\\
\tilde\chi & \longmapsto& \tilde\chi +b\,,\nn\\
\psi & \longmapsto&\psi +\frac12c -a\tilde\chi +b \chi\,.
\eeqa
The appearance of the shift parameters $a$ and $b$ in the transformation of
$\psi$ is due to the non-Abelian structure of $\mf{n}_+$ given by the
Heisenberg algebra (\ref{HeisenbergAlgebra}). This effect is also evident in
the first line of the expression (\ref{GeneralHeisenberg}) for the general
element of $ N  $. From the point of view of the coset, the Heisenberg
translations do not require any compensating transformation as they preserve
the Iwasawa gauge.

\subsubsection*{Rotations}
Rotations are generated by the compact Cartan element $J$ of
$\mf{su}(2,1)$ given in (\ref{cartans}). Let
\beq\label{rot}
R = \exp(\pi J/2) = \left(\begin{array}{ccc}
i & 0 & 0\\
0 & -1& 0\\
0 & 0 & i \end{array}\right),
\eeq
then the most general transformation of this type is given by $R^\sigma:= e^{\sigma \pi J/2}$, for
$\sigma =0,1,2,3$, and acts on $\cZ=(z_1,z_2)$ via
\beq
z_1 \to  z_1\,,\quad
z_2 \to  e^{i\pi\sigma/2} z_2\,.
\eeq
In terms of the four scalar fields this transformation reads
\beqa
\phi & \longmapsto& \phi\,,\nn\\
\chi & \longmapsto& \cos(\pi\sigma/2)\chi-\sin(\pi\sigma/2)\tilde\chi\,,\nn\\
\tilde\chi & \longmapsto& \sin(\pi\sigma/2)\chi+\cos(\pi\sigma/2)\tilde\chi\,,\nn\\
\psi & \longmapsto&\psi\,
\eeqa
and so rotates the two scalars $\chi$ and $\tilde\chi$ among each other while
leaving the other two invariant.  The compensating transformation to restore
the Iwasawa gauge for the coset element (\ref{cosetel}) is $k=R^\sigma$.

\subsubsection*{Involution}

The last transformation of interest is the involution
\beq\label{inv}
S = \left(
\begin{array}{ccc}
0 & 0 & i \\
0 & -1 & 0 \\
-i & 0 & 0
\end{array}
\right),
\eeq
which acts on $\cZ=(z_1, z_2)$ according to
\beq
z_1  \mapsto  -\frac{1}{z_1}\,,\quad
z_2  \mapsto -i\frac{z_2}{z_1}\,,
\eeq
corresponding to the non-trivial generator in the Weyl group \eqref{wgrp}. For the real scalars
themselves we find the following transformation
\beqa
\phi & \longmapsto & -\frac{1}{2}\ln\left[
  \frac{e^{-2\phi}}{4\psi^2 +[e^{-2\phi}+ \chi^2 + \tchi^2]^2 }
\right]\,, \nn \\
\chi &  \longmapsto & \phantom{-}\frac{2\psi\tchi - (e^{-2\phi}+\chi^2 +
  \tchi^2 )\chi}{4\psi^2+ [e^{-2\phi}+\chi^2 + \tchi^2]^2}\,,\nn\\
\tchi &  \longmapsto & \phantom{-}\frac{2\psi\chi + (e^{-2\phi}+\chi^2 +
  \tchi^2 )\tchi}{4\psi^2+ [e^{-2\phi}+\chi^2 + \tchi^2]^2}\,, \nn \\
\psi &  \longmapsto & -\frac{\psi}{ 4\psi^2+[e^{-2\phi}+\chi^2 + \tchi^2]^2 }\,.
\eeqa
It is straightforward to check that the required compensating transformation
in this case indeed belongs to the maximal compact subgroup $SU(2) \times
U(1)$.

\subsection{The Picard Modular Group}
\label{Section:PicardModularGroup}

We finally discuss the Picard modular group $SU(2, 1;\mbb{Z}[i])$. This group
can be defined as the intersection \cite{FrancsicsLax}
\beq
SU(2, 1;\mbb{Z}[i]) := SU(2,1)\cap SL(3,\mbb{Z}[i]),
\eeq
where $\mbb{Z}[i]$ denotes the Gaussian integers
\beq
\mbb{Z}[i]=\{ z \in \mbb{C} \,:\, \Re(z), \Im(z)\in \mbb{Z}\} = \left\{ m_1+ i m_2 \,:\,m_1,m_2\in\,\mbb{Z}\right\}.
\eeq
This definition implies that any element $g \in SU(2,1)$ which has only
Gaussian integer matrix entries belongs to $SU(2, 1;\mbb{Z}[i]) $. In view of
the discussion of $PU(2,1)\cong SU(2,1)$ the Picard modular group can also be
called $PU(2,1;\Zn[i])$.

Let us now examine the particular $SU(2,1)$-transformations of the previous
subsection to check whether they belong to the Picard group. The Heisenberg group $ N  \subset SU(2,1)$ contains a subgroup
$ N  (\mbb{Z}):= N  \cap \pg $. By inspection of
Eq. (\ref{GeneralHeisenberg}) we see that $ N  (\mbb{Z})$ must be of the
form
\beq
 N  (\mbb{Z})=\big\{ e^{a X_{1}+b\tilde{X}_{1}+cX_{2}}  \hs :\hs a,b,c
\in\mbb{Z} \big\}\,.
\eeq
In view of (\ref{GeneralHeisenberg}), a natural set of generators for
$ N  (\mbb{Z})$ is given by the three matrices in (\ref{heisgen}) $T_1$,
$\tilde{T}_1$ and $T_2$. The action of these discrete shifts are then as given in
(\ref{heisshiftsscalars}) with parameters $a,b,c\in\Zn$. The translations
(\ref{heisgen}) are of infinite order in the Picard modular group.

The rotation $R$ defined in (\ref{rot}) is an element of
order $4$ in the Picard modular group, and $R^{\sigma}$ belongs to $SU(2,1;\mbb{Z}[i])$ for the
discrete values of the exponent $\sigma=0,1,2,3$. The action of $R$ on the scalar fields
is
\beq\label{emdual}
R\,:\quad (\chi, \tilde{\chi}) \longmapsto (-\tilde\chi, \chi),
\eeq
while leaving $\phi$ and $\psi$ invariant. Physically speaking, this corresponds to electric-magnetic duality, which is
expected to be preserved in the quantum theory \cite{BeckerBecker}.

Finally, we will examine the involution $S$ in Eq. (\ref{inv}). Clearly, this
involution is an element (of order $2$) in the Picard modular group. As already noted above, the involution (\ref{inv}) corresponds to the Weyl
reflection of the restricted root system $BC_1$ of the non-split real form
$\mf{su}(2,1)$. This reflection is associated with the long root
$2\alpha$. In this context, we can also give an interpretation of the rotation $R$. This is a
transformation that rotates within the degenerate, two-dimensional $\alpha$
root space, spanned by the generators $X_1$ and $\tilde{X}_1$.

The Picard modular group acts discontinuously on the complex hyperbolic
space
$\cxh^2$. A fundamental domain for its action has been given by
Francsics and
Lax in~\cite{FrancsicsLax}. Recently, together with Falbel and Parker, they have also proven
that the Picard modular group $SU(2,1;\mbb{Z}[i])$ is
generated by the translations ${T}_1$ and $T_2$, together with the
rotation $R$ and the involution $S$ \cite{FaFrLaPa}.\footnote{We are very grateful to G. 
Francsics and P. Lax for communicating this result to us prior to
publication.}

 Since the two translations $T_1$ and $\tilde{T}_1$ are related
through ``electric-magnetic duality'' by $\tilde{T}_1=RT_1R^{-1}$, one
may
equivalently choose either of the translations $T_1$ or $\tilde{T}_1$
associated with the $\alpha$ root space in the theorem. Since all three
translations $T_1$, $\tilde{T}_1$ and $T_2$ will turn out to have a
clear
physical interpretation we present the Picard modular group as generated
(non-minimally) by the following five elements:
 \beqa
 T_1=\left(\begin{array}{ccc}
1 & -1+i & i \\
 0& 1 & 1-i \\
0 &0 & 1‚Äö√Ñ√∂\\
 \end{array}\right), &
  \tilde{T}_1=\left(\begin{array}{ccc}
1 & 1+i & i \\
 0& 1 & 1+i \\
0 & 0& 1‚Äö√Ñ√∂\\
 \end{array}\right),  &
 T_2=\left(\begin{array}{ccc}
1 &  0& 1 \\
 0& 1 &0 \\
 0& 0& 1‚Äö√Ñ√∂\\
 \end{array}\right),
 \nn \\
 R=\left(\begin{array}{ccc}
i & 0& 0\\
0 & -1 &0 \\
0 &0 & i \\
 \end{array}\right), &
 S=\left(\begin{array}{ccc}
0 &0 & i \\
 0 & -1 & 0\\
  -i &0 & 0\\
  \end{array}\right).
  \eeqa

\section{Eisenstein Series for the Picard Modular Group}
\label{Section:Eisenstein}

In this section we shall construct Eisenstein series for the Picard modular
group in the principal continuous series representation of $SU(2,1)$. We 
shall give three different constructions, which, despite being equivalent, mutually enlighten each other. 
In Section \ref{LatticeConstruction} we construct a manifestly $SU(2,1;\mbb{Z}[i])$-invariant function on $SU(2,1)/(SU(2)\times U(1))$ by summing over points in  
the three-dimensional Gaussian lattice $\mbb{Z}[i]^3$. This produces a non-holomorphic Eisenstein series $\mc{E}_s$, parametrized by $s$, which will be the central object of study in the remainder of this paper. In Section \ref{Section:PoincareSeries},  we use the isomorphism between the coset space $SU(2,1)/(SU(2)\times U(1))$ and the complex upper half plane $\mbb{CH}^2$ to construct a Poincar\'e series $\mc{P}_s$ on $\mbb{CH}^2$. This turns out to be identical to $\mc{E}_s$ up to an $s$-dependent Dedekind zeta function
factor. For completeness, in Appendix \ref{pAdicConstruction} we give a third construction using standard 
adelic techniques, which illuminates the representation-theoretic nature of  $\mc{E}_s$.

\subsection{Lattice Construction and Quadratic Constraint}
\label{LatticeConstruction}

Following \cite{ObersPioline}, a non-holomorphic function on the double quotient
\beq
\pg \bas SU(2,1)/ (SU(2)\times U(1))
\eeq
can be constructed  as the Eisenstein series\footnote{We note that the same summand and constraint appear in the analysis of~\cite{Yasaki}.}
\beqa
\mc{E}_s(\mc{K})&:=& \sideset{}{'}\sum_{\vec{\om} \in \mbb{Z}[i]^3\atop 
\vec{\om}^{\dagger}\cdot \eta \cdot \vec{\om}=0}
\Big[\vec{\om}^{\dagger}\cdot \mc{K} \cdot \vec{\om}\Big]^{-s} 
\nn \\
&=&\sideset{}{'}\sum_{\vec{\om}\in\mbb{Z}[i]^3\atop \vec{\om}^{\dagger}\cdot \eta \cdot \vec{\om}=0}
e^{-2s\phi}\Big[|\om_1+\om_2\lambda+\om_3{\ga}|^2
+e^{-2\phi}|\om_2+i{\om}_3{\bar\lambda}|^2+e^{-4\phi}|\om_3|^2\Big]^{-s},
\nn \\
\label{EisensteinPicard}
\eeqa
where $\mc{K}=\mc{V}\mc{V}^{\dagger}$ is the ``generalized metric''  (\ref{generalizedmetric}), 
and the variables $\lambda$ and $\gamma$ were defined as functions of $\cZ=(z_1,z_2)$ in (\ref{eisenvars}). In \eqref{EisensteinPicard} the sum runs over 3-vectors of Gaussian integers 
$\vec{\om}\neq(0,0,0)$ subject to the quadratic constraint 
\beq
 \vec{\om}^{\dagger}\cdot \eta \cdot \vec{\om}=|\om_2|^2-2\Im(\om_1\bar{\om}_3)=0\ ,
 \qquad
\vec{\omega} := \left(\begin{array}{c} \bar{\om}_3\\\bar{\om}_2\\\bar{\om}_1\end{array}\right) \ .
\label{constraint}
\eeq
Setting 
\beq
\om_1=m_1+im_2, \qquad \om_2=n_1+in_2, \qquad \om_3=p_1+ip_2,
\eeq 
this may be rewritten as a sum over six integers $m_i,n_i,p_i$, not all vanishing, subject to the 
constraint 
\beq
\vec{\om}^{\dagger}\cdot \eta\cdot \vec{\om}
=n_1^2+n_2^2+2m_1p_2-2m_2p_1=0 .
\label{App:Constraint}
\eeq
The Eisenstein series defined in (\ref{EisensteinPicard}) converges absolutely for $\Re(s)>2$.

To  explain the role of the quadratic constraint (\ref{constraint}), it is convenient to
utilize the isomorphism between the coset space $SU(2,1)/(SU(2)\times U(1))$ and the complex hyperbolic space $\mbb{CH}^2$, as discussed in Section~\ref{sec:su21coset}. We recall from (\ref{ktildematrix}) that in terms of the variable $\mc{Z}=(z_1, z_2)\in \mbb{CH}^2$, the matrix $\mc{K}$ reads
\beq
\label{KKeta}
\mc{K}=\tilde{\mc{K}}+\eta,
\eeq
where $\eta$ is the $SU(2,1)$-invariant metric, Eq. (\ref{invariantmetric}), and the matrix $\tilde{\mc{K}}$ is given by
\beq
\tcosm =  e^{2\phi}\left(\begin{array}{ccc}
|z_1|^2 & z_1\bar{z}_2 & z_1\\
\bar{z}_1 z_2 & |z_2|^2&z_2 \\
\bar{z}_1& \bar{z}_2 & 1
\end{array}\right) = \tilde{\mathcal{V}}\tilde{\mathcal{V}}^\dagger
\quad \text{for }\,\, \tilde{\mathcal{V}} = e^{\phi}\left(\begin{array}{ccc}
0& 0 & z_1\\
0& 0 & z_2 \\
0&0 & 1
\end{array}\right) .
\eeq
In this new parametrization, the Eisenstein series becomes
\beq
 \mc{E}_s(\mc{Z})= \sideset{}{'}\sum_{\vec{\om} \in \mbb{Z}[i]^3\atop \vec{\om}^{\dagger}\cdot \eta \cdot \vec{\om}=0}
 \Big[\vec{\om}^{\dagger}\cdot \tilde{\mc{K}} \cdot
 \vec{\om}+\vec{\om}^{\dagger}\cdot \eta \cdot \vec{\om}\Big]^{-s}
 = \sideset{}{'}\sum_{\vec{\om} \in \mbb{Z}[i]^3\atop \vec{\om}^{\dagger}\cdot \eta \cdot \vec{\om}=0} 
 e^{-2s \phi} |\om_1+\om_2 z_2+\om_3 {z}_1|^{-2s}\ .
 \label{EisensteinPicard3}
 \eeq
The constraint \eqref{constraint} can now be motivated as follows\cite{ObersPioline}. Since  the coset
representative $\mc{V}\in SU(2,1)/(SU(2)\times U(1))$ transforms in the fundamental
representation $\mc{R}$ of $SU(2,1)$, the generalized metric
$\mc{K}=\mc{V}\mc{V}^{\dagger}$ transforms in the symmetric tensor product
$\mc{R}\otimes_s\mc{R}$. As reflected in \eqref{KKeta}, this tensor product is
not irreducible. In order for $ \mc{E}_s$ to be an eigenfunction of the Laplacian on $\mbb{CH}^2$, 
it is necessary to project out the singlet component in \eqref{KKeta},
hence to enforce the constraint \eqref{constraint} in the sum.
To be specific, the Laplacian on the coset space $\mbb{CH}^2$, written in terms of the real variables $\{y=e^{-2\phi}, \chi, \tilde\chi, \psi\}$, is given by
\beq\label{Laplacian}
\Delta_{\mbb{CH}^2} =\frac14 y (\partial_\chi^2 + \partial_{\tilde{\chi}}^2)
+ \frac14 (y^2 + y(\chi^2+\tilde{\chi}^2))\partial_\psi^2
+ \frac12 y (\tilde{\chi} \partial_\chi - \chi \partial_{\tilde{\chi}})\partial_\psi
+ y^2 \partial_y^2 - y\partial_y.
\eeq
Taking into account the quadratic constraint (\ref{constraint}), it is straightforward to check
that  $\mc{E}_s$ is an eigenvector of the Laplacian with eigenvalue $s(s-2)$, as stated
in \eqref{EigenvalueEquation} above.

Since $SU(2,1)$ admits two Casimir operators of degree 2 and 3, and since 
$\Delta_{\mbb{CH}^2}$ represents the action of the quadratic Casimir on
the space of (square-integrable) functions on $SU(2,1)/(SU(2)\times U(1))$, 
one may ask whether $ \mc{E}_s$ is also an eigenvector of an invariant
differential operator of degree 3. It turns out however, as already noticed in
\cite{PiolineGunaydin}, that the principal representation of the cubic Casimir 
in the space of functions on $SU(2,1)/(SU(2)\times U(1))$ vanishes identically.
In terms of the parametrization of the  Casimir eigenvalues by the
complex variables $(p,q)$ used in  \cite{Bars:1989bb,PiolineGunaydin},
the Eisenstein series $ \mc{E}_s$ is attached to the principal spherical representation 
with $p=q=s-2$ (see Appendix \ref{pAdicConstruction} for some details on the principal series of $SU(2,1)$).

Let us also comment on the functional dimension of the representation associated to $ \mc{E}_s$.
The summation ranges over six (real) integers coordinating the lattice $\IZ[i]^3 \sim\IZ^6$. Since both the summand and the constraint are homogeneous in $\vec{\omega}$ one can factor out an overall common Gaussian integer.
Among the remaining four real integers the (real) quadratic constraint $|\om_2|^2-2\Im(\om_1\bar{\om}_3)=0$
eliminates one of the summation variables, leaving effectively a sum
over 3 integers only. This is consistent with the functional dimension 3 of the
principal continuous series representation of $SU(2,1)$ and the number of expected different instanton contributions.

\subsection{Poincar\'e Series on the Complex Upper Half Plane}
\label{Section:PoincareSeries}

In the mathematical literature, a standard way of constructing non-holomorphic Eisenstein series on a symmetric space $G/K$ is in terms of {Poincar\'e series}. For the case of the coset space $SL(2, \mbb{R})/SO(2)$, parametrized by a complex coordinate $\tau$, such a Poincar\'e series is obtained by summing the function $\Im(\ga\cdot \tau)^s$ over the orbit $\ga\in \Gamma_{\infty}\bas SL(2, \mbb{Z})$, where $\Gamma_{\infty}$ is generated by $T : \tau \mapsto \tau+1$. This indeed produces a non-holomorphic Eisenstein series on the double quotient $SL(2, \mbb{Z})\bas SL(2, \mbb{R})/SO(2)$ with eigenvalue $s(s-1)$ under the Laplacian on $SL(2,\mbb{R})/SO(2)$ (for a very nice treatment, see \cite{Borel}).

Here we  generalize this construction to the case of the complex upper half plane $\mbb{CH}^2$, parametrized by the variable $\mc{Z}=(z_1, z_2)$. The generalization of $\Im(\tau)$ is then given by the $N(\mbb{Z})$-invariant function $\mc{F}(\mc{Z})$, constructed in (\ref{KaehlerArgument}) \cite{Zhang}.\footnote{We are grateful to Genkai Zhang for helpful discussions on this construction.} The invariance of $\mc{F}(\mc{Z})$ under $N(\mbb{Z})$ can be checked by direct substitution of the Heisenberg translations in Eq. (\ref{HeisenbergAction}). As we have seen in Section \ref{Section:SU(2,1)}, the Picard modular group $\pg $ acts by
fractional transformations on $\mc{Z}\in \mbb{CH}^2$ such that the function $\mc{F}(\mc{Z})$
transforms as
\beq
\mc{F}(\ga\cdot \mc{Z})=\f{\mc{F}(\mc{Z})}{|C\mc{Z}+D|^2}, \qquad \ga=\left(\begin{array}{cc}
A & B \\
C & D \\
\end{array}\right)\in \pg .
\eeq
A Poincar\'e series for the Picard group may now be constructed as follows
\beqa\label{PGPoincare}
\mc{P}_s(\mc{Z}) := \sum_{\ga\in N(\mbb{Z})\bas \pg }\mc{F}(\ga\cdot \mc{Z})^{s}
= \sum_{\ga\in N(\mbb{Z})\bas \pg }\Big(\f{\mc{F}(\mc{Z})}{|C\mc{Z}+D|^2}\Big)^s\,.
\eeqa
Taking $C:= (\om_3, \om_2)\in \mbb{Z}[i]^2$ and $D:= \om_1\in \mbb{Z}[i]$, and recalling that $\mc{F}(\mc{Z})=e^{-2\phi}$, then reproduces the same form of the Eisenstein series as in Eq. (\ref{EisensteinPicard3}), i.e.
\beq
\mc{P}_s(\mc{Z})=\sum_{\ga\in N(\mbb{Z})\bas \pg }e^{-2s\phi}|\om_1+\om_2 z_2+\om_3 z_1|^{-2s}.
\eeq
The sum over orbits in $N(\mbb{Z})\bas \pg$ is equivalent to the sum over the Gaussian lattice $\mbb{Z}[i]^3$ modulo the constraint $\vec\om^{\dagger}\cdot \eta\cdot \vec\om=0$, together with a coprime condition on the summation variables $\vec{\om}$ \cite{Orloff}:
\beq
\mc{P}_s(\mc{Z})=\sideset{}{'}\sum_{\vec{\om}\in \mbb{Z}[i]^3,\, \gcd(\om_1^{\prime},\om_2^{\prime},\om_3^{\prime})=1\atop \vec\om^{\prime \dagger }\cdot \eta\cdot \vec\om^{ \prime}=0}e^{-2s\phi}
|\om_1^{\prime}+\om_2^{\prime} z_2+\om_3^{\prime} z_1|^{-2s}.
\eeq
Defining $\vec{\om}=\vec{\om}^{\prime} \be$ with $\be= \gcd(\om_1,\om_2, \om_3)\in\mbb{Z}[i]$ and inserting this into (\ref{EisensteinPicard}) we then have the relation
\beq\label{EisenPoin}
\mc{E}_s(\phi, \lambda, \ga) = 4\zeta_{\mbb{Q}(i)}(s) \mc{P}_s(\mc{Z}),
\eeq
where $\zeta_{\mbb{Q}(i)}(s)$ is the Dedekind zeta function for the quadratic extension $\mbb{Q}(i)$ of the rational numbers, and the overall factor of $4$ originates from the four units in $\mbb{Z}[i]$. This will be discussed in more detail in Section \ref{Section:FirstConstantTerm} (see Eq. (\ref{dedezeta})).

\section{Fourier Expansion of $\mc{E}_s(\phi, \chi, \tilde\chi, \psi)$}
\label{Section:FourierExpansion}
In this section we compute the Fourier expansion of the Eisenstein series 
\eqref{EisensteinPicard}.
We begin by recalling the general decomposition with respect to the action of the Heisenberg subgroup $N\subset SU(2,1)$.

\subsection{General Structure of the Non-Abelian Fourier Expansion}
\label{section:generalFourierexpansion}

The main complication of the Fourier expansion stems from the non-Abelian nature of the nilpotent group $ N \subset SU(2,1)$.  $N$ is isomorphic to a three-dimensional Heisenberg group, where the center $Z=[N,N]$ is parametrized by $\psi$. The Fourier expansion therefore splits into an 
Abelian part and a non-Abelian part. The Abelian term corresponds to an expansion with respect to the abelianized group $N/Z$, while the non-Abelian terms represent the expansion with respect to the center $Z$. This general structure of the Fourier expansion of automorphic forms for the Picard modular group is discussed in detail by Ishikawa \cite{Ishikawa}, to which we refer the interested reader. A similar discussion may also be found in the mathematics \cite{Vinogradov,Proskurin} and physics \cite{AutomorphicNS5} literature for the case of automorphic forms on $SL(3, \mbb{R})/SO(3)$. 

We have seen in Section \ref{Section:SU(2,1)} that the action of an arbitrary Heisenberg shift $\mc{U}_{a,b;c}\in N(\mbb{Z})=N\cap SU(2,1;\mbb{Z}[i])$ on $\chi, \tilde{\chi}$ and $\psi$ is given by:
 \beqa
\mc{U}_{a,b;c} &: & \chi\hs \longmapsto\hs  \chi + a,
 \nn \\
  & & \tilde\chi \hs \longmapsto\hs \tilde\chi +b,
  \nn \\
  & & \psi \hs \longmapsto \psi+ \f{1}{2}c-a\tilde\chi+b\chi
  \eeqa
for $a,b,c\in\mbb{Z}$. Since the Eisenstein series (\ref{EisensteinPicard}) is in particular invariant under $N(\mbb{Z})$ we can organize the Fourier expansion by diagonalizing different subgroups of the non-Abelian Heisenberg group $N(\mbb{Z})$.

Explicitly, we  write the general form of the Fourier expansion as
 \beq
 \mc{E}_s(\phi, \chi, \tilde\chi, \psi)=\mc{E}_s^{(\text{const})}(\phi)+\mc{E}^{(\text{A})}_s(\phi, \chi, \tilde\chi)+\mc{E}_s^{(\text{NA})}(\phi, \chi, \tilde\chi, \psi),
 \eeq
 where $\mc{E}_s^{(\text{const})}(\phi)$ is the constant term and 
 \beqa
 \mc{E}^{(\text{A})}_s(\phi, \chi, \tilde\chi)&=& \sideset{}{'}\sum_{(\ell_1,\ell_2)\in\mbb{Z}^2} \mf{C}_{\ell_1,\ell_2}^{\text{(A)}}(\phi;s) e^{-2\pi i (\ell_1\chi +\ell_2\tilde\chi)},
 \nn \\
 \mc{E}_s^{(\text{NA})}(\phi, \chi, \tilde\chi, \psi)&=&\phantom{kk} \sideset{}{'}\sum_{k\in \mbb{Z}}\mf{C}_{k}^{\text{(NA)}}(\phi, \chi, \tilde\chi;s) e^{-4\pi i k \psi}
 \label{NonAbelianFourierExpansionPicard}
 \eeqa 
are called the Abelian and non-Abelian terms, respectively. Following \cite{Ishikawa,AutomorphicNS5}, we proceed to extract an additional phase factor in the non-Abelian term which accounts for the shifts of $\psi$ along the non-central directions. This yields the following structure of the non-Abelian term
\beq
  \mc{E}_s^{(\text{NA})}(\phi, \chi, \tilde\chi, \psi)=\sideset{}{'}\sum_{k\in \mbb{Z}}\sum_{\ell=0}^{4|k|-1}\sum_{n\in\mbb{Z}+\f{\ell}{4|k|}} \mf{C}_{k, \ell}^{\text{(NA)}}(\phi, \tilde{\chi}-n;s) e^{8\pi i k n \chi-4\pi i k(\psi+\chi\tilde\chi)}.
  \label{NewNonAbelianTerm}
  \eeq
The Abelian term is manifestly invariant under shifts of the form $\mc{U}_{a,b;0}\in N(\mbb{Z})/Z$. For the non-Abelian term, invariance under 
\beqa
\mc{U}_{1,0;0} &:& \chi\hs  \longmapsto\hs \chi +1
\nn \\
&:& \psi\hs \longmapsto\hs  \psi-\tilde\chi
\eeqa
 is manifest since $4kn\in \mbb{Z}$. On the other hand, the transformation
 \beqa
\mc{U}_{0,1;0} &:& \tilde\chi\hs  \longmapsto\hs \tilde\chi +1
\nn \\
&:& \psi\hs \longmapsto\hs  \psi+\chi
\eeqa
requires a compensating shift $n\mapsto n+1$ on the summation, under which the variation of the total phase cancels. Note also the restricted dependence on $\tilde\chi$ in the Fourier coefficient; upon shifting $\tilde\chi \mapsto \tilde\chi+1$ and compensating $n\mapsto n+1$ the coefficient is indeed invariant. Finally, invariance under $\mc{U}_{0,0;1}$ is manifest since this gives an overall phase $e^{-4\pi i k/2}=1$.

Note that in writing the non-Abelian term (\ref{NewNonAbelianTerm}) we have made an explicit choice of {polarization}, in the sense that we have manifestly diagonalized the action of Heisenberg shifts of the restricted form $\mc{U}_{a,0;c}$. We could have chosen the opposite polarization in which we instead diagonalize the action of $\mc{U}_{0,b;c}$. In this case, the non-Abelian term reads
\beq
  \mc{E}_s^{(\text{NA})}(\phi, \chi, \tilde\chi, \psi)=\sideset{}{'}\sum_{k\in\mbb{Z}}\sum_{\ell^{\prime}=0}^{4|k|-1}\sum_{\tilde{n} \in\mbb{Z}+\f{\ell^{\prime}}{4|k|}} \tilde{\mf{C}}_{k, \ell^{\prime}}^{\text{(NA)}}(\phi, {\chi}-\tilde n;s) e^{-8\pi i k \tilde{n} \tilde\chi-4\pi i k(\psi-\chi\tilde\chi)}.
  \label{NewNonAbelianTermSecondPolarization}
  \eeq
The Fourier coefficients $\mf{C}_{k, \ell}^{\text{(NA)}}$ and $\tilde{\mf{C}}_{k, \ell^{\prime}}^{\text{(NA)}}$  in the two different polarizations are related via a Fourier transform (see \cite{AutomorphicNS5}). In the sequel we work for definiteness with the first polarization defined by (\ref{NewNonAbelianTerm}). 

Besides invariance under the Heisenberg group we can also use invariance under the electric-magnetic duality transformation $R: (\chi,\tilde\chi)\mapsto(-\tilde\chi,\chi)$ of (\ref{emdual}). On the Abelian term this implies that the coefficient $\mf{C}_{\ell_1,\ell_2}^{\text{(A)}}$ is invariant under $\pi/2$ rotations of $(\ell_1,\ell_2)$. On the non-Abelian term (\ref{NewNonAbelianTerm}) application of $R$ leads to 
\beq
\mc{E}_s^{(\text{NA})}(\phi, \chi, \tilde\chi, \psi)=\sideset{}{'}\sum_{k\in \mbb{Z}}\sum_{\ell=0}^{4|k|-1}\sum_{n\in\mbb{Z}+\f{\ell}{4|k|}} \mf{C}_{k, \ell}^{\text{(NA)}}(\phi,\chi-n;s) e^{-8\pi i k n \tilde\chi-4\pi i k(\psi-\chi\tilde\chi)}
\eeq
and hence we have $\mf{C}_{k, \ell}^{\text{(NA)}}=\tilde{\mf{C}}_{k, \ell}^{\text{(NA)}}$, relating the two choices of polarization as to be expected from electric-magnetic duality. Applying $R$ again leads to relations among the coefficients $\mf{C}_{k, \ell}^{\text{(NA)}}$ and $\mf{C}_{k, \ell'}^{\text{(NA)}}$ for different $\ell$ and $\ell'$.

Finally, we can use the Laplacian condition on the Eisenstein series $\mc{E}_s$ (see Eq. (\ref{EigenvalueEquation})) to further constrain the Fourier coefficients $\mf{C}_{\ell_1, \ell_2}^{\text{(A)}}$ and $\mf{C}_{k, \ell}^{\text{(NA)}}$ and determine their functional dependence on the moduli. In all cases, we require normalizability of the solution, which physically means a well-behaved `weak-coupling' limit $e^\phi\to 0$. Plugging in the Abelian term $\mc{E}_s^{(\text{A})}$ into the eigenvalue equation (\ref{EigenvalueEquation}) yields an equation for the $\phi$-dependence of the coefficients which is solved by a modified Bessel function. More precisely, we find that the Abelian term in the expansion takes the form
\beq
\mc{E}_s^{(\text{A})}(\phi, \chi, \tilde\chi, \psi)=e^{-2\phi}\sideset{}{'}\sum_{(\ell_1, \ell_2)\in\mbb{Z}^2}C^{\text{(A)}}_{\ell_1, \ell_2}(s) K_{2s-2}\Big(2\pi e^{-\phi}\sqrt{\ell_1^2+\ell_2^2}\Big) e^{-2\pi i (\ell_1\chi+\ell_2\tilde\chi)},
\label{GeneralAbelian}
\eeq
where the remaining coefficients $C^{\text{(A)}}_{\ell_1, \ell_2}(s)$ are now independent of $\phi$ and encode the arithmetic information of the group $SU(2,1;\mbb{Z}[i])$. The precise form of these numerical coefficients will be computed in Section \ref{AbelianCoefficients} below. 

Turning to the non-Abelian term (\ref{NewNonAbelianTerm}), the Laplacian condition on the coefficient separates into a harmonic oscillator equation in the variable $x=\tilde\chi-n$, with solution given by a Hermite polynomial $H$, as well as a hypergeometric equation in the variable $y=e^{-2\phi}$ whose solution can be written in terms of a Whittaker function $W$. The separation of variables induces a sum over the eigenvalues of the harmonic oscillator, leading to the following structure for the non-Abelian term:
\beqa
\mc{E}_s^{(\text{NA})}(\phi, \chi, \tilde\chi, \psi)&=&e^{-\phi} \sideset{}{'}\sum_{k\in\mbb{Z}} \sum_{\ell=0}^{4|k|-1}\sum_{n\in\mbb{Z}+\f{\ell}{4|k|}}\sum_{r=0}^{\infty} 
C^{\text{(NA)}}_{r,k, \ell}(s) \hs |k|^{1/2-s} e^{-4\pi |k| (\tilde\chi-n)^2}
\nn \\
& & \times \,H_{r}\Big(\sqrt{8\pi |k|} (\tilde\chi-n)\Big) W_{-r-\f{1}{2}, s-1} \Big(4\pi |k| e^{-2\phi}\Big) e^{8\pi i k n \chi-4\pi i k(\psi+\chi\tilde\chi)},
\nn \\
\label{GeneralNonAbelianTerm}
\eeqa
where the numerical coefficients $C^{\text{(NA)}}_{r,k, \ell}(s)$ will be further discussed in Section \ref{NonAbelianCoefficients}. 

We shall now proceed to compute the explicit form of the Fourier expansion; that is, determine the constant term $\mc{E}_s^{(\text{const})}$ as well as the Abelian and non-Abelian numerical Fourier coefficients $C^{\text{(A)}}_{\ell_1, \ell_2}(s)$ and $C^{\text{(NA)}}_{r,k, \ell}(s)$.

\subsection{First Constant Term}
\label{Section:FirstConstantTerm}
The constant term\footnote{The terminology constant term is derived from holomorphic Eisenstein series where these terms are truly constant and independent of the scalar fields. For non-holomorphic Eisenstein series, as the one studied here, the constant terms retain a dependence on the fields corresponding to Cartan generators.} is defined generally as
\beq
\mc{E}_s^{(\text{const})}(\phi)=\int_{0}^{1}d\chi \int_0^{1}d\tilde\chi \int_0^{1/2}d\psi \ \mc{E}_s(\phi, \chi, \tilde\chi, \psi),
\label{constanttermformula}
\eeq
where the integral over the coordinate $\psi$ (physically, the NS-axion modulus) runs from $0$ to $1/2$ because of the extra factor of 2 in front of $\psi$ in our parametrization of $N$ in Eq.~(\ref{cosetel}). Since the Cartan subgroup $A$ appearing in the Iwasawa decomposition of $SU(2,1)$ is one-dimensional, the constant term only depends on the dilatonic scalar $\phi$. Moreover, recall from the discussion in Section \ref{subsec:CosetTransf} that the Weyl group of $\mf{su}(2,1)$ is the Weyl group of the restricted root system $BC_1$, which is isomorphic with $\mbb{Z}_2$. Hence, the constant term $\mc{E}_s^{(\text{const})}(\phi)$ consists of two contributions, $\mc{E}_s^{(0)}$ and $\mc{E}_s^{(1)}$, which are permuted by $\mbb{Z}_2$~\cite{Langlands}.\footnote{We are grateful to Pierre Vanhove for helpful discussions on the constant terms.} 

The powers of $e^{\phi}$ in $\mc{E}_s^{(\text{const})}(\phi)$ may be determined by the Laplacian condition on $\mc{E}_s$. In Section \ref{Section:PoincareSeries} we have seen that the Eisenstein series is an eigenfunction of the Laplacian $\Delta_{\mbb{CH}^2}$ with eigenvalue $s(s-2)$. This implies that all the constant terms must individually be eigenfunctions of $\Delta_{\mbb{CH}^2}$ with the same eigenvalue. It turns out that there is a unique solution to this, and we find that $\mc{E}_s^{(0)}$ must be of the form
\beq
\mc{E}_s^{(\text{const})}(\phi)=\mc{E}_s^{(0)}+\mc{E}_s^{(1)}=A(s)e^{-2s\phi}+B(s) e^{-2(2-s)\phi}.
\label{ConstantTerm}
\eeq
Below we will compute the coefficients $A(s)$ and $B(s)$. The first constant term $\mc{E}_s^{(0)}$ corresponds to the leading order term in an expansion at the cusp $e^{\phi}\rightarrow 0$, which physically corresponds to the regime of weak coupling. 

Our strategy for performing the Fourier expansion is as follows: we first consider the
term $\omega_3=0$, which by virtue of the constraint (\ref{App:Constraint}) also requires
 $\omega_2=0$. The remaining sum over $\omega_1\neq 0$ yields the first constant term $\mc{E}_s^{(0)}$. We then consider the case $\omega_3\neq 0$ and solve the constraint (\ref{App:Constraint}) explicitly using the Euclidean algorithm which reduces the remaining sum to one over three integers. On these integers we will perform Poisson resummations to uncover the second constant 
 term, as well the Abelian and non-Abelian Fourier coefficients.  

Accordingly we start by extracting the $\om_3=0$ (implying $\om_2=0$) part of the sum in the Eisenstein series, $\mc{E}_s^{(0)}$, leaving a remainder $\mc{A}^{(s)}$
\beq\label{firstconstant}
\mc{E}_s(\phi, \lambda, \ga)=\mc{E}_s^{(0)}+\mc{A}^{(s)}.
\eeq
The first term $\mc{E}_s^{(0)}$ is the leading order contribution in the limit $e^{\phi}\rightarrow 0$ and corresponds to a sum over $\omega_1=m_1+i m_2$
\beq
\mc{E}_s^{(0)}=e^{-2s\phi}\sideset{}{'}\sum_{(m_1,m_2)\in\mbb{Z}^2} \f{1}{(m_1^2+m_2^2)^s}= 4 \zeta_{\mbb{Q}(i)}(s) e^{-2s\phi}\, ,
\label{LeadingTerm}
\eeq
where $\zeta_{\mbb{Q}(i)}(s)$ is the Dedekind zeta function over the Gaussian integers
\beq\label{dedezeta}
\zeta_{\mbb{Q}(i)}(s)= \frac14 \sideset{}{'}\sum_{ \omega\in \mbb{Z}[i]} |\omega|^{-2s} = \frac14 \sideset{}{'}\sum_{(m,n)\in\mbb{Z}^2} \f{1}{(m^2+n^2)^s}.
\eeq
The factor of $4$ is related to the units of the Gaussian integers (see Appendix \ref{Gaussians}).

The Dedekind zeta function $\zeta_{\mbb{Q}(i)}(s)$ satisfies a functional equation
which is most conveniently written in terms of the ``completed Dedekind zeta function"
\beq\label{completedDedekind}
\zeta_{\mbb{Q}(i)*} (s) := \pi^{-s} \Gamma(s)\zeta_{\mbb{Q}(i)}(s)\,,
\eeq
in terms of which one has
\beq
\zeta_{\mbb{Q}(i)*} (1-s) = \zeta_{\mbb{Q}(i)*} (s)\,.
\label{functionalrelationcompletedDedekind}
\eeq
It is known that the Dedekind function over a quadratic number field can be written as a Dirichlet L-function times the standard Riemann zeta function. In our case this reads (see, e.g., \cite{Cartier} for a proof)
\beq
\label{zeQ}
\zeta_{\mbb{Q}(i)}(s)=\be(s)\zeta(s),
\eeq
where the standard Riemann zeta function is defined as
\beq\label{zetaEuler}
\zeta(s) := \sum_{n=1}^{\infty} n^{-s} = \prod_{p\,\text{prime}} \frac{1}{1-p^{-s}}\quad\quad \text{for $\Re(s)>1$}
\eeq
and $\be(s)$ is the Dirichlet beta function,\footnote{The Dirichlet beta function is also known as $L(\chi_{-4},s)$, i.e. it is the $L$-function associated with the alternating character modulo 4.}
\beq
\be(s):=\sum_{n=0}^{\infty}(-1)^n (2n+1)^{-s}\quad\quad\text{for $\Re(s)>0$}.
\eeq
We also note that $\beta(s)$ has an Euler product representation of the form 
\beq\label{betaEuler}
\beta(s)= \prod_{p\, :\, p=1\, \text{mod} \,4} \frac{1}{1-p^{-s}}
\prod_{p\, :\, p=3\, \text{mod} \, 4} \frac{1}{1+p^{-s}},
\eeq
which together with the Euler product form of the Riemann zeta function $\zeta(s)$ above will be useful later. The functional relation for $\beta(s)$ is again best stated using its completion
\beq\label{completedBeta}
\beta_*(s) := \left(\frac{\pi}{4}\right)^{-\frac{s+1}{2}} \Gamma\left(\frac{s+1}{2}\right)\, \beta(s),
\eeq
for which the functional relation takes the simple form
\beq
\beta_*(s) = \beta_*(1-s).
\label{functionalrelationcompletedbeta}
\eeq

In conclusion, we have found that the first coefficient $A(s)$ in (\ref{ConstantTerm}) is given by the Dedekind function $\zeta_{\mbb{Q}(i)}(s)$ and that it is related to the term $\omega_3 =0$ in the sum over the Gaussian integers. We will now proceed to evaluate the terms with $\omega_3\neq 0$, contained in $\mc{A}^{(s)}$ of (\ref{firstconstant}). We emphasize that the term with $\omega_3=0$ and $\omega_2\neq 0$ vanishes identically because of the quadratic constraint (\ref{App:Constraint}). Thus, $\mc{A}^{(s)}$ only contains terms for which $\om_3\neq 0$.

\subsection{Solution of Constraint and Poisson Resummation}

To solve the constraint (\ref{App:Constraint}) we shall make use of the Euclidean algorithm, which implies that for integers $p_1$ and $p_2$ the equation
\beq
\label{bezout}
q_1p_2-q_2p_1=d
\eeq
has integer solutions for $q_1$ and $q_2$ if and only if $d$ divides $\gcd(p_1,p_2)$. The most general solution is the sum of a particular solution $(q_1,q_2)$ plus an integer times $(p_1,p_2)$.
More precisely, in the case of our constraint (\ref{App:Constraint}) we find that for $\omega_3=p_1+ip_2\neq 0$ there are solutions in $\Zn[i]^3$ if and only if 
\beq
\f{|\om_2|^2}{2d}\in\mbb{Z}\ , \quad\quad \text{ where $d=\gcd(p_1,p_2)$}
\label{problem}
\eeq
and the most general solution for $\omega_1=m_1+i m_2$ is then
\beqa
{} m_1&=&-\f{|\om_2|^2}{2d}q_1+m\f{p_1}{d},
\nn \\
{} m_2 &=& -\f{|\om_2|^2}{2d} q_2 +m\f{p_2}{d}.
\label{SolutionConstraint}
\eeqa
Here, $q_1$ and $q_2$ is any particular solution of  $q_1p_2-q_2p_1=d$ and $m\in\Zn$ is an {\em unconstrained} integer. Therefore, we can rewrite the constrained sum as
\beq
\sum_{\omega_3\neq 0} \sum_{\omega_2\in \mbb{Z}[i]\atop   2d\,\vline\, |\omega_2|^2} \sum_{m\in\Zn}\big(\cdots \big)
\eeq
where in the summand, $\omega_1=m_1+i m_2$ has to be replaced by the expression from (\ref{SolutionConstraint}).

Let us implement this procedure on our Eisenstein series. 
After solving the constraint, the first term in the bracket of (\ref{EisensteinPicard}) becomes
\beqa
|\om_1+\om_2\lambda+\om_3 \ga|^2&=&\f{|\om_3|^2}{d^2}\bigg[\Big(m-\f{|\om_2|^2}{2|\om_3|^2}(q_1p_1+q_2p_2)+\tilde\ell_1\chi+\tilde\ell_2\tilde\chi+2d\psi\Big)^2
\nn \\
& &\phantom{++} +\f{1}{16d^2}\Big((\tilde\ell_1+2d\tilde\chi)^2+(\tilde\ell_2-2d\chi)^2\Big)^2\bigg],
\label{ConstraintSolved}
\eeqa
where we defined
\beqa
{}\tilde{\ell}_1&:=& \f{d}{|\om_3|^2}\Big[(p_1-p_2)n_1+(p_1+p_2)n_2\Big],
\nn \\
{}\tilde\ell_2&:=& \f{d}{|\om_3|^2}\Big[ (p_1+p_2)n_1-(p_1-p_2)n_2\Big].
\eeqa
Extracting an overall factor of $|\om_3|^2/d^2$, the total summand may be written as 
\beqa
\f{d^2}{|\om_3|^2}\, \om^{\dagger}\cdot \mc{K}\cdot \om &=& \bigg[m-\f{|\om_2|^2}{2|\om_3|^2}(q_1p_1+q_2p_2)+\tilde\ell_1\chi+\tilde\ell_2\tilde\chi+2d\psi\bigg]^2
\nn \\
& & +\f{e^{-4\phi}}{d^2}\bigg[ d^2+\f{e^{2\phi}}{4}\Big((\tilde\ell_1+2d\tilde\chi)^2+(\tilde\ell_2-2d\chi)^2\Big)\bigg]^2.
\eeqa
Using an integral representation for the summand in the remainder $\mc{A}^{(s)}$ defined in (\ref{firstconstant}),
\beq \Big[\vec{\om}^{\dagger}\cdot \mc{K} \cdot \vec{\om}\Big]^{-s} =\f{\pi^s}{\Gamma(s)} \int \f{dt}{t^{s+1}} e^{-\f{\pi}{t} \vec{\om}^{\dagger}\cdot \mc{K} \cdot \vec{\om}},
\eeq
and performing a Poisson resummation on $m$ using the standard formula
\beq
\sum_{m\in\mbb{Z}}e^{-\pi x (m+a)^2+2\pi i m b}=\f{1}{\sqrt{x}}\sum_{\tm\in\mbb{Z}}e^{-\f{\pi}{x}(\tm+b)^2-2\pi i (\tm +b)a}\ ,
\eeq
we obtain
\beqa
{}\mc{A}^{(s)}&=&\f{\pi^s}{\Gamma(s)}e^{-2s\phi}\sum_{\tm \in\mbb{Z}}
\sideset{}{'}\sum_{(p_1,p_2)\in\mbb{Z}^2} \sum_{(n_1,n_2)\in\mbb{Z}^2\atop  2d | n_1^2+n_2^2}\f{d}{|\om_3|} e^{-2\pi i \tm \big(-\f{|\om_2|^2}{2|\om_3|^2}(q_1p_1+q_2p_2)+\tilde\ell_1\chi+\tilde\ell_2\tilde\chi+2d\psi\big)}
\nn \\
{}& &\phantom{+}\times  \int_0^{\infty}\f{dt}{t^{s+1/2}} e^{-\pi t \f{d^2}{|\om_3|^2} \tm^2-\f{\pi}{t} \f{|\om_3|^2}{d^4}e^{-4\phi} \Big[ d^2+\f{e^{2\phi}}{4}\big((\tilde\ell_1+2d\tilde\chi)^2+(\tilde\ell_2-2d\chi)^2\big)\Big]^2},
\label{C2}
\eeqa
where we have indicated explicitly the constraint from (\ref{problem}) that $2d$ must divide $|\om_2|^2$.

As we will see in Section \ref{Section:Instantons}, the Abelian terms in the Fourier expansion correspond physically to
instantons with zero NS5-brane charge, independent of the NS-NS scalar $\psi$. 
We therefore split off the Abelian contribution with $\tm= 0$:
\beq
\mc{A}^{(s)} =\mc{D}^{(s)}+\mc{E}_s^{(\text{NA})}, 
\label{AbelianNonAbelianSplit}
\eeq
where $\mc{E}_s^{(\text{NA})}$ denotes the non-Abelian term with $\tm\neq 0$, to be considered later. From $\mc{D}^{(s)}$ we will be able to extract the second constant term $\mc{E}_s^{(1)}$ as well as the Abelian Fourier coefficients $\mf{C}^{\text{(A)}}_{\ell_1,\ell_2}(s)$. Explicitly we have
\beq
 \mc{D}^{(s)}=\f{\pi^s}{\Gamma(s)}e^{-2s\phi}\sideset{}{'}\sum_{(p_1,p_2)\in\mbb{Z}^2} \sum_{(n_1,n_2)\in\mbb{Z}^2\atop  2d | n_1^2+n_2^2}\f{d}{|\om_3|} \int_0^{\infty}\f{dt}{t^{s+1/2}} e^{-\f{\pi}{t} \f{e^{-4\phi}|\om_3|^2}{d^4} \Big[ d^2+\f{e^{2\phi}}{4}\big((\tilde\ell_1+2d\tilde\chi)^2+(\tilde\ell_2-2d\chi)^2\big)\Big]^2}.
 \eeq
To get rid of the square in the exponent, we shall perform the integration over $t$ and then choose a new integral representation of the summand. The current form of the exponent will be convenient for the evaluation of the non-Abelian terms in Section \ref{NonAbelianCoefficients}, but for our present purposes we shall rewrite it in the following way
 \beq
 \f{e^{-4\phi}|\om_3|^2}{d^4}\bigg[d^2+\f{e^{2\phi}}{4}\big((\tilde\ell_1+2d\tilde\chi)^2+(\tilde\ell_2-2d\chi)^2\big)\bigg]^2=\f{1}{4|\om_3|^2}\Big[ |\mc{Y}|^2+2e^{-2\phi} |\om_3|^2\Big]^2,
 \eeq
where we defined the new variable $\mc{Y}=\mc{Y}_1+i\mc{Y}_2$, with
\beqa
\mc{Y}_1 &:=& n_1+(p_1-p_2) \tilde\chi -(p_1+p_2)\chi,
\nn \\
\mc{Y}_2 &:=& n_2 +(p_1+p_2) \tilde\chi+(p_1-p_2) \chi .
\eeqa
Evaluating the integral over $t$ then yields
 \beq
\mc{D}^{(s)}=\f{2^{2s-1}\sqrt{\pi}\Gamma(s-1/2)}{\Gamma(s)} e^{-2s\phi} \sideset{}{'}\sum_{(p_1,p_2)\in\mbb{Z}^2}\sum_{(n_1,n_2)\in\mbb{Z}^2\atop  2d | n_1^2+n_2^2} \f{d}{|\om_3|^{2-2s}}\bigg\{\Big[ |\mc{Y}|^2+2e^{-2\phi} |\om_3|^2\Big]\bigg\}^{1-2s}.
  \eeq
  After replacing the term within brackets by its integral representation we obtain
  \beq
    \mc{D}^{(s)}=\f{(2\pi)^{2s-1}\sqrt{\pi}\Gamma(s-1/2)}{\Gamma(s)\Gamma(2s-1)} e^{-2s\phi} \sideset{}{'}\sum_{(p_1,p_2)\in\mbb{Z}^2}\sum_{(n_1,n_2)\in\mbb{Z}^2\atop  2d | n_1^2+n_2^2} \f{d}{|\om_3|^{2-2s}}\int_0^{\infty} \f{dt}{t^{2s}} e^{-\f{\pi}{t} \big[ |\mc{Y}|^2+2e^{-2\phi} |\om_3|^2\big]}.
    \label{D}
    \eeq
 Since all values of $n_1$ and $n_2$ are almost degenerate we shall perform a further Poisson resummation on these variables. Here we must take into account the remaining constraint that $2d$ divides $n_1^2+n_2^2$. The set of solutions to this constraint can be written as
\beq
n_1 = n_1^0 + \delta n_1 \ ,\qquad n_2 = n_2^0 + \delta n_2\ ,
\label{n1n2solution}
\eeq
where $\delta n_1+i \delta n_2$ runs over the lattice $L$
\beq
L =\left\{d [ (k_1 + k_2)+ i (k_1 - k_2 )]\,:\, (k_1,k_2)\in \mbb{Z}^2 \right\}
\label{LatticeL}
\eeq
and $(n_1^0,n_2^0)$ runs over all solutions
of the quadratic equation $n_1^2+n_2^2=0 \mod 2d$ in a fundamental domain
of $\mbb{Z}[i]/L$, which we take to be $0\leq n_1^0<d$ and $0\leq n_2^0<2d$, with 
area $2d^2$. We denote the set of such solutions as
\beq\label{fundsolutions}
\mc{F}(d) := \left\{ n_1^0 + i  n_2^0\,:\,n_1^2+n_2^2=0 \mod 2d, \ 0\leq n_1^0<d\,,\ 0\leq n_2^0<2d\right\}\,.
\eeq
and its cardinality by
\beq
N(d) := \sharp\mc{F}(d)\ .
\label{MultiplicativeCharacter}
\eeq
As we shall discuss below and in Appendix~\ref{AbelianMeasureDetails}, 
the series $N(d)$ is multiplicative though not completely multiplicative 
(see \cite{SloaneA086933} for the first few values).

After inserting (\ref{n1n2solution}) into (\ref{D}) and performing a Poisson resummation on $\delta n_1$ and $\delta n_2$, we obtain
\beq
\begin{split}
\mc{D}^{(s)}=\f{(2\pi)^{2s-1}\sqrt{\pi}\Gamma(s-1/2)}{2\Gamma(s)\Gamma(2s-1)} e^{-2s\phi} 
\sideset{}{'}\sum_{(p_1,p_2)\in\mbb{Z}^2}\sum_{\tilde{\omega}_2\in L^*}
\sum_{f\in \mc{F}(d)}
\f{1}{d |\om_3|^{2-2s}}  \\
    \times e^{2\pi i \Re(\tilde{\omega}_2 f)}\,  
    \int_0^{\infty} \f{dt}{t^{2s-1}} 
     e^{-\pi t (\tn_1^2+\tn_2^2)-\f{2\pi}{t} e^{-2\phi} |\om_3|^2 + 2\pi i (\ell_1 \chi+\ell_2 \tilde\chi)},
\end{split}
\label{Ds}
\eeq
where $L^*$ is the lattice dual to $L$, 
\beq\label{duallattice}
L^* = \left\{\tilde{\omega}_2 = \tilde{n}_1+i\tilde{n}_2 =\frac{1}{2d} [( \tilde k_1 +\tilde  k_2)
+i  ( \tilde k_1 - \tilde k_2 )]\,:\,(\tilde k_1,\tilde k_2)\in \mbb{Z}^2 \right\}\,,
\eeq
and we defined the new charges
\beqa
\ell_1 &:=& \tn_1 (p_1+p_2)-\tn_2 (p_1-p_2)\,, 
\nn \\
\ell_2 &:=& \tn_1 (p_2-p_1) -\tn_2 (p_1+p_2)\,.
\label{Abeliancharges}
\eeqa     

\subsection{Second Constant Term}
\label{Section:SecondConstantTerm}
We may now extract the second constant term from the $\ell_1=\ell_2=0$ part of the sum, and accordingly we split $\mc{D}^{(s)}$ as
\beq
\mc{D}^{(s)}=\mc{E}_s^{(1)}+\mc{E}_s^{(\text{A})},
\eeq
where $\mc{E}_s^{(\text{A})}$ is the Abelian term in the Fourier expansion to be considered in the next subsection. The $\ell_1=\ell_2=0$ part arises from the $\tilde{\omega}_2=0$ term which reads
\beq
\mc{E}_s^{(1)}=\f{(2\pi)^{2s-1}\Gamma(s-1/2)}{2\sqrt{\pi}\Gamma(s)\Gamma(2s-1)} e^{-2s\phi} \sideset{}{'}\sum_{(p_1,p_2)\in\mbb{Z}^2}\sum_{f\in\mc{F}(d)}  \f{1}{d |\om_3|^{2-2s}}\int_0^{\infty} \f{dt}{t^{2s-1}} e^{-\f{2\pi}{t} e^{-2\phi} |\om_3|^2}.
\eeq
The sum over $f\in\mc{F}(d)$ produces the multiplicative function $N(d)$ in (\ref{MultiplicativeCharacter}). The integral can be explicitly evaluated with the result
\beq
\mc{E}_s^{(1)}= \f{\pi^{3/2}\Gamma(s-1/2)\Gamma(2s-2)}{\Gamma(s)\Gamma(2s-1)} e^{-2(2-s)\phi} \sideset{}{'}\sum_{(p_1,p_2)\in\mbb{Z}^2}N(d)\frac{1}{d |\om_3|^{2s-2}}.
\eeq
The sum can now be expressed in terms of the Riemann zeta function and Dedekind zeta 
function \eqref{dedezeta} as follows. Extract the greatest common divisor of $p_1$ and $p_2$, defining $p_1= d p_1^{\prime}$ and $p_2= d p_2^{\prime}$, with $d=\gcd(p_1, p_2)$ and $\gcd(p_1^{\prime}, p_2^{\prime})=1$. This yields a sum over $d$ and coprime $(p_1',p_2')$
\beq
\sideset{}{'}\sum_{(p_1,p_2)\in\mbb{Z}^2} N(d) d^{-1} |p|^{2-2s}=
\left( \sum_{d>0}  N(d) d^{1-2s} \right) \, 
\left( \sum_{(p_1^{\prime}, p_2^{\prime})=1} \f{1}{(p_1^{\prime2}+p_2^{\prime 2})^{s-1}} \right).
\label{secondconstantterm}
\eeq
The second sum may be rewritten as a ratio of Riemann and Dedekind zeta functions as follows (see Section \ref{Section:FirstConstantTerm})
\beq
 \sum_{(p_1^{\prime}, p_2^{\prime})=1} \f{1}{(p_1^{\prime2}+p_2^{\prime 2})^{s-1}} =\f{4\zeta_{\mbb{Q}(i)}(s-1)}{\zeta(2s-2)}.
 \eeq
Let us now consider the first sum on the right hand side of (\ref{secondconstantterm}) which involves the combinatorial function $N(d)$ defined in (\ref{MultiplicativeCharacter}) (see also~\cite{SloaneA086933}). Given $N(d)$ we may construct the Dirichlet series
\beq\label{dirichletseries}
L(N, s): = \sum_{d=1}^{\infty} N(d) d^{-s}\ ,
\eeq 
that converges for $\Re(s)>2$. Since $N(d)$ is multiplicative, we may evaluate $L(N,s)$
using Euler products. To this end we note that the multiplicative series exhibits the following properties  (mentioned in \cite{SloaneA086933}, and derived in greater generality in Appendix C),
\beq
N(2^m)=2^m\ ,\qquad N(p^m) = 
\left\{ \begin{matrix} 
(m(p-1)+p) p^{m-1}\ , & p=1\, \text{mod}\, 4 \\
p^{2\, \lfloor m/2 \rfloor}\ , &\, \, p=3\, \text{mod} \, 4 .
\end{matrix}  \right.
\label{MultiplicativityN}
\eeq
Therefore, the Dirichlet series (\ref{dirichletseries}) has an Euler product representation given by (see Appendix \ref{Gaussians} for the derivation)
\beq
L(N, s)=\frac{1}{1-2^{1-s}} \prod_{p\, : \, p=1\, \text{mod} \, 4} \frac{1-p^{-s}}{(1-p^{1-s})^2}
\prod_{p\, :\, p=3\, \text{mod} \, 4} \frac{1+p^{-s}}{(1-p^{1-s})(1+p^{1-s})},
\label{EulerProductN(d)}
\eeq
where the product runs over all primes $p> 2$. 
Comparing to (\ref{zetaEuler}) and  (\ref{betaEuler}), we deduce that 
\beq
L(N,s) = \frac{\beta(s-1) \zeta(s-1)}{\beta(s)}.
\eeq
Putting everything together we then find the following expression for the constant term 
\beq
\mc{E}_s^{(1)}= 4\f{\pi^{3/2}\Gamma(s-1/2)\Gamma(2s-2)}{\Gamma(s)\Gamma(2s-1)} \f{L(N,2s-1)}{\zeta(2s-2)} \zeta_{\mbb{Q}(i)}(s-1) \, e^{-2(2-s)\phi} .
\eeq
Referring back to the completed Dedekind zeta function (\ref{completedDedekind}) and Dirichlet beta function (\ref{completedBeta}) we define a completed ``Picard Zeta function'' by
\beq
\label{defxi}
\mathfrak{Z}(s) := \zeta_{\mbb{Q}(i)*}(s) \,  \beta_*(2s-1)\ ,
\eeq 
in terms of which the two constant terms can be neatly summarized by
\beq\label{constantTermstogether}
\mc{E}_s^{(\text{const})}=\mc{E}_s^{(0)}+ \mc{E}_s^{(1)} = 4\zeta_{\mbb{Q}(i)}(s) \left\{ e^{-2s\phi} + \frac{\mathfrak{Z}(2-s)}{\mathfrak{Z}(s)} e^{-2(2-s)\phi}\right\}.
\eeq
Eq. (\ref{constantTermstogether}) can be viewed as an extension of Langlands's constant term formula \cite{Langlands} for Eisenstein series associated to special linear groups to the case of the unitary group $SU(2,1)$. The completed Picard zeta function $\mathfrak{Z}(s)$ 
plays the same role as the  completed Riemann zeta function $\xi(s)=\pi^{-s/2}\Gamma(s/2)\zeta(s)$ 
in Langlands' formula. 

\subsection{Abelian Fourier Coefficients}
\label{AbelianCoefficients}
We now turn to the Abelian Fourier coefficients, corresponding to the terms $(\tn_1,\tn_2)\neq 0$
in \eqref{Ds}. The integral over $t$ leads to a modified Bessel function,
\beqa
\mc{E}_s^{(\text{A})}&=&\f{2\pi^{2s-1/2}\Gamma(s-1/2)}{\Gamma(s)\Gamma(2s-1)} 
e^{-2\phi} \sideset{}{'}\sum_{(p_1,p_2)\in\mbb{Z}^2} 
\sideset{}{'}\sum_{(\tilde{k}_1,\tilde{k}_2)\in\mbb{Z}^2}
\sum_{f\in\mc{F}(d)}
\frac1{d^{2s-1}} \, |u|^{2s-2}\nn\\
&&\quad\times
e^{\f{\pi i}{d}\Re\left[ u f (1-i)\right]}\,
 \, K_{2s-2}\Big(2\pi e^{-\phi}|\Lambda|\Big)\, e^{2\pi i (\ell_1\chi+\ell_2\tilde\chi)}\,,
\eeqa
where we have introduced the following additional notation
\beq
u=\tilde{k}_1+i\tilde{k}_2\,,\quad \Lambda = \ell_2-i \ell_1
\eeq
for
\beq
\ell_1 =\f{1}{d} (\tilde{k}_1p_2+\tilde{k}_2p_1)\,,\quad
\ell_2 = \f{1}{d} (\tilde{k}_2p_2-\tilde{k}_1p_1).
\label{Abeliancharges2}
\eeq
These charges are manifestly integral since $d$ divides $p_1$ and $p_2$. This last relation can also be written as
\beq
\Lambda=\frac{u\, \om_3}{d} = u \, \omega_3',
\eeq
where $\omega_3'=\omega_3/d$ is a primitive Gaussian number (i.e. a Gaussian number whose real and imaginary parts are coprime).
To extract the Abelian Fourier coefficients $\mf{C}^{\text{(A)}}_{\ell_1, \ell_2}(\phi)$ we therefore replace the sum over $\omega_3$ and $u$ by a sum over $d$, $\Lambda$ and $\omega_3'$ where the primitive Gaussian integer $\omega_3'$ has to be a Gaussian divisor of $\Lambda$, to wit
\beqa\label{abterm}
\mc{E}_s^{(\text{A})}&=&C^{\text{(A)}}_s e^{-2\phi}  \sideset{}{'}\sum_{\Lambda\in\mbb{Z}[i]}\left\{\sum_{\om_3' | \Lambda} \left|\f{\Lambda}{\omega_3'}\right|^{2s-2}
\left(\sum_{d>0}\f{1}{d^{2s-1}} \sum_{f\in\mc{F}(d)} e^{\f{\pi i}{d} \Re\big[\f{\Lambda}{\om_3'}f (1-i)\big]}\right)\right\}  \nn\\
& & \phantom{++++}\times K_{2s-2}\Big(2\pi e^{-\phi} |\Lambda|\Big) e^{ 2\pi i (\ell_1 \chi+\ell_2\tilde\chi)}\,,
\eeqa
where the coefficient is given by
\beq\label{coeffab}
C^{\text{(A)}}_s=\f{2\pi^{2s-1/2}\Gamma(s-1/2)}{\Gamma(s)\Gamma(2s-1)} 
= \frac{8\zeta_{\mbb{Q}(i)}(s)\beta(2s-1)}{\mathfrak{Z}(s)}\,.
\eeq
To make contact with the general discussion of Section \ref{section:generalFourierexpansion}, we rewrite this result as a sum over the real variables $\ell_1$ and $\ell_2$:
\beq
\mc{E}_s^{(\text{A})}=2\zeta_{\mbb{Q}(i)}(s) \frac{e^{-2\phi}}{\mathfrak{Z}(s)}\sideset{}{'}\sum_{(\ell_1, \ell_2)\in\mbb{Z}^2}\mu_s(\ell_1, \ell_2) \big[\ell_1^2+\ell_2^2\big]^{s-1} K_{2s-2}\Big(2\pi e^{-\phi}\sqrt{\ell_1^2+\ell_2^2}\Big) e^{2\pi i (\ell_1\chi+\ell_2\tilde\chi)},
\label{FinalAbelianTerm}
\eeq
where we defined the summation measure
\beq
\label{abinstmes}
\mu_s(\ell_1, \ell_2):= 4\beta(2s-1) \sum_{\om_3' | \Lambda} |\omega_3'|^{2-2s} \left(\sum_{d>0} d^{1-2s} \sum_{f\in\mc{F}(d)} e^{\f{\pi i}{d} \Re\big[\f{\Lambda}{\om_3'}f(1-i)\big]}\right),
\eeq
containing the sum over primitive Gaussian divisors of $\Lambda=\ell_2-i\ell_1$. The sum over $d$ in the parenthesis may be carried out for fixed $\Lambda$ and $\omega_3'$ to give the Gaussian divisor function (see Appendix \ref{AbelianMeasureDetails} for the derivation)
\beq
\sum_{d>0} d^{1-2s} \sum_{f\in\mc{F}(d)} e^{\f{\pi i}{d} \Re\big[\f{\Lambda}{\om_3'}f (1-i)\big]} = \frac{1}{4\beta(2s-1)}
\sum_{z|\frac{\Lambda}{\omega_3'}} |z|^{4-4s}\,,
\label{equality}
\eeq
whence the instanton measure (\ref{abinstmes}) simplifies to
\beq\label{FinalAbelianMeasure}
\mu_s(\ell_1,\ell_2) = \sum_{\omega_3'|\Lambda} |\omega_3'|^{2-2s} 
\sum_{z|\frac{\Lambda}{\omega_3'}} |z|^{4-4s}\,.
\eeq
Thus, the Abelian summation measure (\ref{FinalAbelianMeasure}) involves both a sum over primitive divisors of $\Lambda$ and a sum over all divisors of $\Lambda/\omega_3^{\prime}$. By comparing (\ref{FinalAbelianTerm}) to (\ref{GeneralAbelian}) we may now extract the numerical Abelian Fourier coefficients:
\beq
C^{(\text{A})}_{\ell_1, \ell_2}(s)= \frac{2\zeta_{\mbb{Q}(i)}(s)}{\mathfrak{Z}(s)}\mu_s(\ell_1, \ell_2) \big[\ell_1^2+\ell_2^2\big]^{s-1}.
\label{NumericalAbelianFourierCoefficients}
\eeq
We note that the Abelian instanton measure (\ref{FinalAbelianMeasure}) is 
multiplicative in a restricted sense: The relation
\beq\label{eqn:abelianmult}
\mu_s(\Lambda_1) \mu_s(\Lambda_2) = \mu_s(\Lambda_1\Lambda_2)
\eeq
holds if (and only if)  the Gaussian integers $\Lambda_1$ and $\Lambda_2$ 
admit no common prime factor {\em up to  complex conjugation} (this caveat is relevant for
split primes, see appendix~\ref{Gaussians}).

\subsection{Non-Abelian Fourier Coefficients}
\label{NonAbelianCoefficients}
Finally we consider 
the non-Abelian term $\mc{E}_s^{(\text{NA})}$ in (\ref{AbelianNonAbelianSplit}). This term reads
\beqa
{}\mc{E}_s^{(\text{NA})}&=&\f{\pi^s}{\Gamma(s)}e^{-2s\phi}\sideset{}{'}\sum_{\tm \in\mbb{Z} }\sideset{}{'}\sum_{(p_1,p_2)\in\mbb{Z}^2}\sum_{(n_1,n_2)\in\mbb{Z}^2\atop  2d | n_1^2+n_2^2} \f{d}{|\om_3|} e^{-2\pi i \tm \big(-\f{|\om_2|^2}{2|\om_3|^2}(q_1p_1+q_2p_2)+\tilde\ell_1\chi+\tilde\ell_2\tilde\chi+2d\psi\big)}
\nn \\
{}& &\phantom{+}\times  \int_0^{\infty}\f{dt}{t^{s+1/2}} e^{-\pi t \f{d^2}{|\om_3|^2} \tm^2-\f{\pi}{t} \f{|\om_3|^2}{d^4}e^{-4\phi} \Big[ d^2+\f{e^{2\phi}}{4}\big((\tilde\ell_1+2d\tilde\chi)^2+(\tilde\ell_2-2d\chi)^2\big)\Big]^2}.
\label{NonAbelianTerm}
\eeqa
The integral is of Bessel type and yields
\beqa
\mc{E}_s^{(\text{NA})}&=& \f{2\pi^s}{\Gamma(s)}e^{-2s\phi} \sideset{}{'}\sum_{\tm \in\mbb{Z} } \sideset{}{'}\sum_{(p_1,p_2)\in\mbb{Z}^2}\sum_{(n_1,n_2)\in\mbb{Z}^2\atop  2d | n_1^2+n_2^2} \bigg[\f{d}{|\om_3|}\bigg]^{s+1/2} \bigg[\f{|\tm|^2}{\Re(S_{\ell_1,\ell_2,k})}\bigg]^{s-1/2}
\nn \\
& & \phantom{+++}\times K_{s-1/2}\Big(2\pi \Re(S_{\ell_1, \ell_2, k})\Big)e^{-2\pi i \Im(S_{\ell_1, \ell_2, k})}e^{- \f{\pi i}{2kd} (\ell_1^2+\ell_2^2)(q_1p_1+q_2p_2)} ,
\label{NonAbelianTermBessel}
\eeqa
where the real and imaginary parts of $S_{\ell_1,\ell_2,k}$ are given by
\beqa
\Re(S_{\ell_1,\ell_2,k})&=& |k| e^{-2\phi} +\f{1}{4|k|}\Big[ (\ell_1+2k\tilde\chi)^2+(\ell_2-2k\chi)^2\Big],
\nn \\
\Im(S_{\ell_1,\ell_2,k})&=&\ell_1\chi+\ell_2\tilde\chi+2k\psi,
\label{defSl12k}
\eeqa
and we also defined\footnote{The non-Abelian charges ${\ell}_i$ defined in \eqref{defl122} should
not be confused with the Abelian charges ${\ell}_i$ in \eqref{Abeliancharges}.}
\beqa
{}k&:=& \tm d,
\nn \\
{}\ell_1&:=&\f{k}{|\om_3|^2}\Big[(p_1-p_2)n_1+(p_1+p_2)n_2\Big],
\nn \\
{}\ell_2&:=&\f{k}{|\om_3|^2}\Big[ (p_1+p_2)n_1-(p_1-p_2)n_2\Big].
\label{defl122}
\eeqa
In Gaussian notation, $\Lambda=\ell_2-i\ell_1$, the last two relations amount to
\beq
\bar\Lambda=\f{(1+i)k\om_2}{\om_3}.
\eeq
Comparing the expression (\ref{NonAbelianTermBessel}) with the general form of the non-Abelian term (\ref{GeneralNonAbelianTerm}), we see that the former involves a sum of nearly Gaussian wave-functions peaked around $(\ell_2,-\ell_1)/(2k)$ in the $(\chi, \tilde\chi)$ plane, while the latter 
is written in terms of a basis of
Landau-type wave functions which are eigenmodes of $\pa_\chi$ and $\pa_{\psi+\chi\tilde\chi}$,
with quantized charges $4kn$ and $k$. 
To extract the non-Abelian Fourier coefficients $C^{(\text{NA})}_{r,k, \ell}(s)$ we must therefore transform (\ref{NonAbelianTermBessel}) into the correct basis. This can be achieved via Fourier transform 
along the variable $\chi$ (or $\tilde\chi$ in the other polarization). 

To perform the Fourier transform we go back to the integral representation in (\ref{NonAbelianTerm}). The integrand is quartic in $\chi$ and therefore 
inconvenient for Fourier transform. To remedy this we make the following change of integration variables:
\beq
t= \f{t' |\om_3|^2 A}{k^2},
\eeq
where
\beq
A(y, \chi, \tilde\chi)=k \Big[y  + \Big(\tilde\chi+\f{\ell_1}{2k}\Big)^2+\Big(\chi-\f{\ell_2}{2k}\Big)^2\Big],
\eeq
and we recall that $y=e^{-2\phi}$. Implementing this in (\ref{NonAbelianTerm}),
and denoting $ \tilde\ell_i=\ell_i/\tm$, we obtain 
\beqa
\mc{E}_s^{(\text{NA})}&=&\f{\pi^s}{\Gamma(s)} y^s\sideset{}{'}\sum_{\tm \in\mbb{Z}}\sideset{}{'}\sum_{(p_1,p_2)\in\mbb{Z}^2}\sum_{(n_1,n_2)\in\mbb{Z}^2\atop  2d | n_1^2+n_2^2}\f{d|k|^{2s-1}}{|\om_3|^{2s}}  
\nn \\
& & \times e^{-2\pi i \tm \big(-\f{|\om_2|^2}{2|\om_3|^2}(q_1p_1+q_2p_2)+\tilde\ell_1\chi+\tilde\ell_2\tilde\chi+2d\psi\big)}  \int_0^{\infty} \f{dt'}{t'^{s+1/2}} A^{1/2-s} e^{-\pi \big(t'+\f{1}{t'}\big) A},
\eeqa
where the exponent is now quadratic in both $\chi$ and $\tilde\chi$. Using an integral representation for the factor $A^{1/2-s}$, and dropping the prime on $t'$, we may rewrite this expression as follows
\beq
\begin{split}
\mc{E}_s^{(\text{NA})}&=\f{\pi^{2s-1/2}}{\Gamma(s)\Gamma(s-1/2)} y^s\sideset{}{'}\sum_{\tm \in\mbb{Z}}\sideset{}{'}\sum_{(p_1,p_2)\in\mbb{Z}^2}\sum_{(n_1,n_2)\in\mbb{Z}^2\atop  2d | n_1^2+n_2^2}\f{d |k|^{2s-1}}{|\om_3|^{2s}} e^{ \f{\pi i \tm|\om_2|^2}{2|\om_3|^2}(q_1p_1+q_2p_2)}
 \\
&  \times \, e^{-4\pi i k\psi-2\pi i \ell_2\tilde\chi} \int_0^{\infty} \f{dt du}{t^{s+1/2}u^{3/2-s}} e^{-\pi k \big(u+t+\f{1}{t}\big) \big[y+\big(\tilde\chi+\f{\ell_1}{2k}\big)^2\big]} f(y,\chi,\tilde\chi; t,u)\ ,
\label{NonAbelianTerm2}
\end{split}
\eeq
where all the $\chi$-dependence is contained in the function 
\beq
f(y, \chi ;t,u)=e^{-\pi k \big(u+t+\f{1}{t}\big) \big(\chi-\f{\ell_2}{2k}\big)^2-2\pi i \ell_1\chi}.
\label{functionf}
\eeq
The Fourier transform over $\chi$ is now implemented by substituting
\beq
 f(y, \chi ;t,u)=e^{-4\pi i k\chi\tilde\chi}\int dn \hat{f}(y, n, \tilde\chi ;t,u) e^{8\pi i kn \chi},
 \eeq
 with 
 \beqa
 \hat{f}(y, n, \tilde\chi ;t,u) &=& 4|k| \int d\xi \ e^{-8\pi i kn \xi+4\pi i k \xi \tilde\chi} f(y, \xi, \tilde\chi;t,u)
 \nn \\
 &=& 4|k|\ e^{-\f{\pi}{k} \f{(\ell_1-2k\tilde\chi+8\pi kn)^2}{u+t+1/t}}.
 \eeqa
After Fourier transform, the non-Abelian term thus becomes 
\beqa
\mc{E}_s^{(\text{NA})}&=&\f{4\pi^{2s-1/2}}{\Gamma(s)\Gamma(s-1/2)} y^s\sideset{}{'}\sum_{\tm \in\mbb{Z}}\sideset{}{'}\sum_{(p_1,p_2)\in\mbb{Z}^2}\sum_{(n_1,n_2)\in\mbb{Z}^2\atop  2d | n_1^2+n_2^2}\f{d |k|^{2s}}{|\om_3|^{2s}} e^{ \f{\pi i \tm|\om_2|^2}{2|\om_3|^2}(q_1p_1+q_2p_2)}
\nn \\
{}& & \times \int \f{dt du}{t^{s+1/2}u^{3/2-s}}  e^{-\pi k \big(u+t+\f{1}{t}\big) \big[y+\big(\tilde\chi+\f{\ell_1}{2k}\big)^2\big]} 
\nn \\
& & \times \int dn\ e^{-\f{\pi i}{k}\ell_2(\ell_1-4kn)}e^{-\f{\pi}{k} \f{(\ell_1-2k\tilde\chi+8\pi kn)^2}{u+t+1/t}}\ e^{8\pi i kn \chi-4\pi i k(\psi+\chi\tilde\chi)}.
\label{NonAbelianTerm3}
\eeqa

Let us now comment on the structure of Eq. (\ref{NonAbelianTerm3}). After Fourier transforming we see that the non-Abelian term indeed corresponds to an expansion in terms of the invariant wavefunctions on the twisted torus as in Eq. (\ref{GeneralNonAbelianTerm}). However, we have not been able to further manipulate Eq. (\ref{NonAbelianTerm3}) into the form displayed in (\ref{GeneralNonAbelianTerm}) and therefore we cannot extract the numerical Fourier coefficients $C^{(\text{NA})}_{k, \ell}(s)$ in as compact a form as the Abelian coefficients (\ref{NumericalAbelianFourierCoefficients}). Nevertheless, as a consistency check we shall show that the leading order exponential behaviour of (\ref{NonAbelianTerm3}) near the cusp $y\rightarrow \infty$ coincides with that of Eq. (\ref{GeneralNonAbelianTerm}). To this end we may take the saddle point approximation for the integrals over $t$ and $u$ in (\ref{NonAbelianTerm3}) for which the saddle points are located at $t=1$ and $u=0$. We thus find that the leading exponential dependence of (\ref{NonAbelianTerm3}) at the saddle point is given by $e^{-S}$ with 
\beq
\Re(S)=2\pi |k|\Big[y+\Big(\tilde\chi+\f{\ell_1}{2|k|}\Big)^2\Big]+\f{\pi}{2|k|}\big(\ell_1-2|k|\tilde\chi+4|k|n\big)^2.
\eeq
Rearranging terms, this can be written as
\beq
\Re(S)=2\pi |k| y+4\pi |k|\big(\tilde\chi-n\big)^2+4\pi |k|\Big(n+\f{\ell_1}{2|k|}\Big)^2.
\label{saddlepointresult}
\eeq
Using the asymptotic behaviour of the Whittaker function $W_{k,m}(x)\sim e^{-x/2}$ one may indeed verify that the first two terms in (\ref{saddlepointresult}) exactly coincide with the leading behaviour of the general expression (\ref{GeneralNonAbelianTerm}) in the limit $y\rightarrow\infty$. We further expect that the summation over $\ell_1$ and $\ell_2$ (or, more precisely, over $\om_2$ and $\om_3$) will restrict the integral over $n$ such that it localizes on the points in $\mbb{Z}+\ell/(4|k|)$ as is expected from the general expression (\ref{GeneralNonAbelianTerm}). We stress that the result (\ref{saddlepointresult}) is valid in the polarization (\ref{NewNonAbelianTerm}) we have chosen. There is an analogous result for the other polarization. 

Besides the representation of the non-Abelian coefficients in the form (\ref{NonAbelianTerm3}) one could also try to extract the coefficients by other means. One possibility would be to manipulate the expression (\ref{NonAbelianTermBessel}) by expanding out the Bessel function and binomially expanding the resulting power series in $A(y,\chi,\tilde\chi)$ to make contact with the power series expansions of the Hermite polynomials and Whittaker functions of (\ref{GeneralNonAbelianTerm}). Alternatively, one could try to compute the coefficients by going to a suitably chosen point in moduli space (e.g. the cusp $y=\infty$) or by using $\pg$ symmetry or Hecke operators to relate the non-Abelian coefficients to the Abelian ones. We hope to present a complete investigation of the non-Abelian coefficients in a future publication.

\subsection{Functional Relation}

The expression (\ref{constantTermstogether}) for the constant terms of the Eisentein series
$\mc{E}_s$ is suggestive of  a functional relation most conveniently written in terms of the Poincar\'e series \eqref{PGPoincare} and the 
Picard Zeta function \eqref{defxi}, 
\beq
\label{functional}
\mathfrak{Z}(s) \mc{P}_s = \mathfrak{Z}(2-s) \mc{P}_{2-s}\ .
\eeq
Indeed, it is easily checked that both the
constant terms (\ref{constantTermstogether})  and the Abelian Fourier coefficients 
(\ref{FinalAbelianTerm}), (\ref{FinalAbelianMeasure}) satisfy this relation, taking into
account the symmetry of the modified Bessel function $K_{2s-2}(x)=K_{2-2s}(x)$.
Unfortunately, due the unwieldy form of the non-Abelian terms we are unable to 
present a full proof of  \eqref{functional}, which would constitute 
an analog of the familiar functional relation for Eisenstein series associated to special linear groups \cite{Langlands}. We note that a different functional relation for $\mc{P}_s$ has been proposed in \cite{Orloff} but this appears to contradict the constant term formula (\ref{constantTermstogether}).

\section{Instanton Corrections to the Universal Hypermultiplet}
\label{Section:Instantons}

\label{Section:InstantonCorrections}

In this section we propose that the Eisenstein series for the Picard modular group constructed in Section \ref{Section:Eisenstein}, and further analyzed in Section \ref{Section:FourierExpansion}, controls the exact metric
on the universal hypermultiplet moduli space $\mc{M}_{\mathrm{UH}}$, including
the D2- and NS5-brane instanton corrections. We start by recalling some aspects of 
quantum corrections to hypermultiplet moduli spaces in type II Calabi-Yau compactifications, with particular emphasis on recent developments involving twistor techniques.

\subsection{Twistor Techniques for Quaternionic-K\"ahler Spaces}
Quantum corrections to the hypermultiplet moduli space are most conveniently described using
twistor techniques \cite{Salamon,LeBrun,Alexandrov:2008nk,
Alexandrov:2008gh}. Given a quaternion-K\"ahler space $\mc{M}$, one
may construct its twistor space $\mc{Z}_{\mc{M}}$,  a $\mbb{C}P^1$ bundle over 
$\mc{M}$ which admits a canonical complex structure $\mc{J}$, a
complex contact structure $\mc{C}$, a compatible real structure $\tau$ 
and a K\"ahler-Einstein metric $ds^2_{\mc{Z}_{\mc{M}}}$
with K\"ahler potential $K_{\mc{Z}_{\mc{M}}}$.
The contact one-form $\mc{C}^{[i]}$
is proportional to the $(1,0)$-form $Dz := dz + p_+ - i p_3 z + p_- z^2 $, where $z$
is a complex coordinate on the $\mbb{C}P^1$ fiber and $(p_3, p_+, p_-)$ 
is the $SU(2)$ part of the Levi-Civita connection on $\mc{M}$. 
The metric on $\mc{M}$ can be recovered from the K\"ahler-Einstein metric 
on $\mc{Z}_{\mc{M}}$ using
\beq
ds^2_{\mc{Z}_{\mc{M}}} = \frac14 \left( e^{-2 K_{\mc{Z}_{\mc{M}}}} | \mc{C} |^2 + \nu\,
ds^2_{\mc{M}} \right) ,
\eeq
where $\nu$ is a numerical constant related to the curvature of the base manifold $\mc{M}$. Locally, any contact structure is trivial, so there exists an open covering $U_i$ of   $\mc{Z}_{\mc{M}}$
and a local complex coordinate system $(\xi_{[i]}^{\Lambda}, \tilde{\xi}^{[i]}_{\Lambda}, \al_{[i]})$ on 
$U_i$ such that the complex contact structure takes the Darboux form
\beq
\label{DarbouxX}
\mc{C}^{[i]} = d\al_{[i]} + \xi_{[i]}^{\Lambda}\, d\tilde{\xi}^{[i]}_{\Lambda}
:= 2 \, e^{\Phi_{[i]}}\, \f{Dz}{z}\ .
\eeq
The second equality defines the ``contact potential'' $\Phi_{[i]}$ in the patch $U_i$, 
a complex function on $\mc{Z}_{\mc{M}}$ holomorphic along the
fiber. The contact potential in the patch $U_i$ is related to the 
K\"ahler potential $K_{\mc{Z}_{\mc{M}}}$ in the same patch via
\beq
\mc{K}^{[i]}_{\mc{Z}_{\mc{M}}}=\log \f{1+z\bar{z}}{|z|}+\Re\big[\Phi_{[i]}(x^{\mu}, z)\big].
\label{generaltwistorpotential}
\eeq
Globally, the complex contact structure on $\mc{Z}_{\mc{M}}$ is determined by the 
complex contact transformations $S^{[ij]}$ between the Darboux coordinate system 
on the overlap $U_i \cap U_j$. These can be described  e.g. 
by providing holomorphic generating functions 
$S_{ij}\big(\xi_{[i]}^{\Lambda}, \tilde{\xi}^{[j]}_{\Lambda}, \al_{[j]}\big)$, subject to 
compatibility conditions on triple overlaps $U_i \cap U_j \cap U_k$, equivalence under 
local contact transformations on $U_i$ and $U_j$, and reality constraints. 
The quaternion-K\"ahler metric on $\mc{M}$ can then be extracted from these holomorphic data,
by determining the contact twistor lines, i.e. expressing  
$\big(\xi^{\Lambda}, \tilde{\xi}_{\Lambda}, \alpha,\Phi \big)$ in some patch $U$
in terms of the coordinates $x^{\mu}\in \mc{M}$ on the base manifold and the complex coordinate $z\in \mbb{C}P^1$ on the fiber. Plugging the solution into \eqref{DarbouxX}
allows to extract the $SU(2)$ connection $p_\pm, p_3$, the quaternionic 2-forms and 
finally the metric on $\mc{M}$. 
More details on this construction can be found in  \cite{Alexandrov:2008nk,Alexandrov:2008gh}.
It should be noted that these twistor techniques for quaternion-K\"ahler manifolds are 
related to the more standard twistor techniques
for hyperk\"ahler manifolds by the superconformal quotient 
construction \cite{Swann,deWitVandoren2,
Alexandrov:2008ds}.

\subsubsection*{On the Twistor Space of the Tree-Level Universal Hypermultiplet}
We now illustrate this construction in the case of the tree-level universal hypermultiplet moduli space. 
The twistor space $\mc{Z}_{\mc{M}_{\mathrm{UH}}}$ of the classical moduli space $\mc{M}_{\mathrm{UH}}$ can be nicely described group-theoretically as follows. 
Viewing the $\mbb{C}P^1$  twistor fiber as $S^2=SU(2)/U(1)$, the
fibration of $SU(2)/U(1)$ over $\mc{M}_{\mathrm{UH}}$ is such that 
the $SU(2)$ cancels: \cite{PiolineGunaydin,Wolf}:
\beq
\mc{Z}_{\mc{M}_{\mathrm{UH}}}= \f{SU(2)}{U(1)}\ltimes \f{SU(2,1)}{SU(2)\times U(1)}=\f{SU(2,1)}{U(1)\times U(1)}.
\label{TwistorSpace}
\eeq
The twistor space $\mc{Z}_{\mc{M}_{\mathrm{UH}}}$ is a complex 3-dimensional contact manifold, with local coordinates $(\xi, \tilde{\xi}, \al)$. These coordinates parametrize the complexified Heisenberg group $ N _{\mbb{C}}$, or, equivalently, coordinates on the complex coset space $P_{\mbb{C}}\bas SL(3,\mbb{C})$, where $P_{\mbb{C}}$ is the complexification of the parabolic subgroup $P\subset SU(2,1)$ discussed in Appendix~\ref{pAdicConstruction} and $SL(3,\mbb{C})$ is the complexification of $SU(2,1)$. In terms of the coordinates $(\xi, \xit, \al)$ on $P_{\mbb{C}}\bas SL(3,\mbb{C})$ the K\"ahler potential of $\mc{Z}_{\mc{M}_{\mathrm{UH}}}$ takes the following form \cite{PiolineGunaydin}
\beq
K_{\mc{Z}_{\mc{M}_{\mathrm{UH}}}}=\f{1}{2}\log\left[\Big(\big(\xi-\xib\big)^2+\big(\xit-\xitb\big)^2\Big)^2+4\Big(\al-\bar{\al}+\xib\xit-\xi\xitb\Big)^2\right].
\label{TwistorPotentialGeneric}
\eeq
As mentioned above, the contact twistor lines for the unperturbed twistor space correspond to the change of variables that relate the coordinates $(\xi, \tilde{\xi}, \al)$ on $\mc{Z}_{\mc{M}_{\mathrm{UH}}}$ to the coordinates $x^{\mu} =\{ e^{\phi}, \chi, \tilde{\chi}, \psi\}$ on the base $\mc{M}_{\mathrm{UH}}$ and the coordinate $z$ on the fiber $\mbb{C}P^1=SU(2)/U(1)$. These twistor lines were obtained in \cite{PiolineGunaydin}. In our notations they read (away from the north pole $z=0$ and south pole $z=\infty$)
\beqa
\xi&=& -\sqrt{2} \chi+\f{1}{\sqrt{2}}e^{-\phi}\big(z-z^{-1}\big),
\nn \\
\tilde{\xi}&=&-\sqrt{2} \tilde{\chi}-\f{i}{\sqrt{2}}e^{-\phi}\big(z+z^{-1}\big),
\nn \\
\al&=& \phantom{-}2\psi -  e^{-\phi}\Big[z(\tilde\chi+i {\chi})-z^{-1}(\tilde\chi-i {\chi})\Big].
\label{twistorlines}
\eeqa
Plugging these into \eqref{DarbouxX} and (\ref{TwistorPotentialGeneric}), 
we find that the contact potential in this patch is simply
\beq
e^{\Phi(x^{\mu}, z)}=e^{-2\phi}\ ,
\label{classicalcontactpotential}
\eeq
in particular independent of $z$, and verify that \eqref{generaltwistorpotential} is satisfied.
We further note that under an action of $SU(2,1)$, the contact potential and contact
one-form transform as
\beq
\label{transPhi}
e^{\Phi}\hs \longmapsto \hs |C+D\mc{Z}|^{-2}\ e^{\Phi},\qquad
\mc{C} \longmapsto \hs (C+D\mc{Z})^{2}\,  \mc{C} \ ,
\eeq
which ensure that the K\"ahler potential $K_{\mc{Z}_{\mc{M}_{\text{UH}}}}$ transforms by a K\"ahler transformation, and that $SU(2,1)$ acts isometrically on both $\mc{Z}_{\mc{M}_{\text{UH}}}$
and $\mc{M}_{\text{UH}}$ itself.

\subsection{Quantum Corrected Hypermultiplet Moduli Spaces in type IIA}
\label{Section:QuantumCorrectedModuliSpaces}

Using these and related techniques, much progress has been achieved recently in understanding
the hypermultiplet moduli space in type IIA string compactifications on a Calabi-Yau manifold $\mc{X}$. At the perturbative level, the metric on $\mc{M}_{\mathrm{H}}$ is believed 
to receive a one-loop correction, but no higher loop corrections \cite{AntoniadisMinasian1,Strominger,Gunther:1998sc,AntoniadisMinasian2,Anguelova:2004sj,RoblesLlana:2006ez}. For the universal hypermultiplet this was rigorously proven in \cite{Anguelova:2004sj}. The general form of the perturbative corrections can be inferred from compactifications of higher derivative couplings in ten dimensions \cite{AntoniadisMinasian1}, 
or via an explicit string theory calculation in $D=4$ \cite{AntoniadisMinasian2}. 
As a consequence, the contact potential on the twistor space must reduce at large volume, small coupling to
\beq
\label{phipert}
e^\Phi = \frac{\tau_2^2 V_{\mc{X}}}{2} +\f{\chi_E}{192 \pi} + \dots  
\eeq
where $\tau_2=1/g_s$ is the ten-dimensional string coupling, $\chi_{E}$ is the Euler
number of $\mc{X}$ and $V_{\mc{X}}$ is the volume
of  $\mc{X}$ in string units. The complete perturbatively corrected metric 
corresponding to the contact potential \eqref{phipert} can be found in \cite{RoblesLlana:2006ez, Alexandrov:2008nk}. We note that in the corresponding expression in the type IIB hypermultiplet sector there are additional contributions arising from $\alpha^{\prime}$-corrections and worldsheet instantons. However, due to the fact that the metric on the complex structure moduli space of $\mc{X}$ is insensitive to $\alpha^{\prime}$-effects, these corrections are absent in the type IIA expression (\ref{phipert}). The corrections to the contact potential which are non-perturbative in $g_s$ are due to 
D2-brane and NS5-brane instantons \cite{BeckerBeckerStrominger}. Using S-duality and mirror symmetry, the form of the D2-brane instanton corrections was 
obtained in a series of works \cite{Alexandrov:2008nk,Alexandrov:2008gh,Vandoren1,Vandoren2,Alexandrov:2008ds,VandorenAlexandrov}. 
To first order away from the one-loop corrected metric, their contribution to the contact potential
reads
\beq
\label{phiD}
e^{\Phi_{(\text D2 )}} = \frac{1}{4\pi^2} \sum_\gamma n_\gamma 
\sum_{m>0} \frac{e^{-\phi} |Z_\gamma|}{m}\, K_1\left( 8\pi m  e^{-\phi} |Z_\gamma|\right)
e^{2\pi i m \int_{\ga} C_{(3)}}
\eeq
where $\gamma$ runs over the homology classes in $H_{3}(\mc{X},\mbb{Z})$, $Z_\gamma$ is the central charge associated to the cycle $\gamma$, $e^{\phi}$
is the 4D string coupling such that $e^{-2\phi} = \tau_2^2 V_{\mc{X}} / 2$, $C_{(3)}$ is the Ramond-Ramond 3-form and $n_\gamma$ is a
numerical factor which counts the number of BPS states in the homology class $\gamma$. 
NS5-brane contributions have been discussed in \cite{VandorenAlexandrov,AutomorphicNS5}, 
but remain largely mysterious in general.

\subsection{On the Contact Potential and the Picard Eisenstein Series}
\label{Section:TwistorialInterpretation}

We now restrict to the case of type IIA string theory compactified on a rigid Calabi-Yau threefold, 
and propose that the Picard Eisenstein Series $\mc{E}_{s}(\phi, \chi,\tilde\chi,\psi)$, 
for a suitable value of the parameter $s$, controls the exact, quantum corrected metric 
on the universal hypermultiplet moduli space. As in\cite{AutomorphicNS5} (see in particular Section~3.1),
we shall restrict our attention to the contact potential $\Phi(x^\mu,z)$ 
on a certain holomorphic section\footnote{In the presence of NS5-brane corrections, the contact potential
is no longer constant along the fiber. The quaternion-K\"ahler metric on $\mc{M}=\mc{M}_{\rm UH}$
can nevertheless be described, in many different ways, in terms of a single real function $h(x^\mu)$ 
on $\mc{M}$
subject to a non-linear partial differential equation \cite{VandorenAlexandrov,Prza}. 
The latter can be identified with the K\"ahler potential $K_{\mc{Z}_\mc{M}}$ on any holomorphic section $z(x^\mu)$ of $\mc{Z}_\mc{M}$ \cite{Alexandrov:2009vj}. Our proposal refers to 
a specific choice of holomorphic section, which we are not able to specify at this stage. We stress that this technical point plays no role at the level of our present analysis.}
 $z(x^\mu)$ of the twistor space $\mc{Z}_{\mc{M}_{\mathrm{UH}}}$. We also  choose variables such that the action of $SU(2,1)$
on $\mc{M}_{\rm UH}$ is the tree-level action (though it is no longer isometric in general), 
and look for a completion of $\Phi(x^\mu,z(x^\mu))$ which  reproduces the 
expected perturbative contributions.
Determining the specific holomorphic section $z(x^\mu)$ and the exact twistor lines and hypermultiplet metric are  important open problems which lie outside the scope of this work.

Matching the powers of the dilaton, we then propose that, on the holomorphic section $z(x^\mu)$ introduced above, the contact potential for the quantum corrected metric on $\mc{M}_{\rm UH}$ is given by\footnote{Due to the different power of $e^\phi$, the contact potential 
$\Phi_{\text{exact}}$ appears to transform differently from its tree-level 
counterpart \eqref{transPhi}; this is not a fatal flaw however, since the locus $z(x^\mu)$ 
is in general not 
fixed by the action of the Picard modular group.}
 \beq
e^{\Phi_{\text{exact}}(x^{\mu},z(x^\mu))}
= \kappa \, e^{\phi} \mc{E}_{3/2}(\phi, \chi, \tilde\chi, \psi)\ ,
\label{Conjecture}
\eeq
where $\mc{E}_s$ is the Picard Eisenstein series \eqref{EisensteinPicard}, and 
$\kappa$ is an adjustable numerical constant. Using the Fourier expansion
\eqref{FirstFourier}, we see that \eqref{Conjecture} predicts 
\beq
e^{\Phi_{\text{exact}}}=4\zeta_{\mbb{Q}(i)}(3/2)  \kappa 
\left( e^{-2\phi} + \frac{\mathfrak{Z}(1/2)}{\mathfrak{Z}(3/2)} \right) + 
e^{\Phi_{(\text{A})}} +
e^{\Phi_{(\text{NA})}}\ ,
\eeq
where the last two terms correspond to the Abelian and non-Abelian parts of the Fourier expansion, respectively. The two constant terms have the same dependence on the dilaton $e^{\phi}$ as the two perturbative contributions in \eqref{phipert}. We thus want to identify the second constant term at $s=3/2$ with the one-loop coefficient $\chi_E /192 \pi$ in (\ref{phipert}). Here we run into a problem since for $s=3/2$ we find
\beq
\frac{\mathfrak{Z}(1/2)}{\mathfrak{Z}(3/2)} 
 \approx - 2.32607\ ,
 \label{OneLoopMess}
\eeq
implying that matching with the physical one-loop term requires $\chi_E\sim -1403.05$, 
a negative, non-integer number. This contradicts the fact that $\chi_E=2h_{1,1}\in 2\mbb{N}$ 
for a  rigid Calabi-Yau threefold.  Hence, the value of the one-loop coefficient predicted by the second constant term in the Eisenstein series is not physically viable. While this invalidates the proposal that the principal Eisenstein series $\mc{E}_s$ describes the exact universal hypermultiplet metric,
it does not necessarily ruin the idea that the Picard modular group controls that metric.
In the concluding Section \ref{Conclusions}, we speculate that automorphic forms attached to  the quaternionic discrete series
of $SU(2,1,\IZ[i])$ may be relevant. We proceed with our current proposal however, as the
form of the non-Abelian Fourier expansion is largely independent of the details of the automorphic form under consideration. In particular, we show next that the form of the Abelian and non-Abelian contributions to the Fourier expansion of $\mc{E}_{3/2}(\phi, \chi, \tilde\chi, \psi)$ agrees
with the expected form of D2-brane and NS5-brane instanton contributions, respectively. 

\subsubsection*{D2-Brane Instantons}

The Abelian contribution \eqref{FinalAbelianTerm} at $s=3/2$ becomes 
\beq
e^{\Phi_{(\text{A})}} = \frac{2\kappa\, \zeta_{\mbb{Q}(i)}(3/2)\,e^{-\phi}}{\mathfrak{Z}(3/2)}\sideset{}{'}\sum_{(\ell_1, \ell_2)\in\mbb{Z}^2}\mu_{3/2}(\ell_1, \ell_2) \big[\ell_1^2+\ell_2^2\big]^{1/2} K_{1}\Big(2\pi e^{-\phi}\sqrt{\ell_1^2+\ell_2^2}\Big) e^{-2\pi i (\ell_1\chi+\ell_2\tilde\chi)},
\label{FinalAbelianTerm32}
\eeq
where the summation measure $\mu_{3/2}(\ell_1, \ell_2)$ is given in \eqref{FinalAbelianMeasure}.  
In the weak-coupling limit $e^{\phi}\rightarrow 0$ we may use the asymptotic expansion of the modified Bessel function at large~$x$,
\beq
K_t(x) \sim \sqrt{\f{\pi}{2x}} e^{-x}\sum_{n\geq 0} \frac{\Gamma\left(t+n+\frac12\right)}{\Gamma(n+1)\Gamma\left(t-n+\frac12\right)}(2x)^{-n}\ ,
\eeq
to approximate 
\beq 
e^{\Phi_{(\text{A})}}\sim \frac{\kappa \, \zeta_{\mbb{Q}(i)}(3/2)}{\mf{Z}(3/2)}e^{-\phi/2}\sideset{}{'}\sum_{(\ell_1, \ell_2)\in\mbb{Z}^2}\mu_{3/2}(\ell_1, \ell_2)(\ell_1^2+\ell_2^2)^{1/4} e^{-2\pi S_{\ell_1, \ell_2}}\Big[1+\mc{O}(e^\phi)\Big].
\label{D2Term}
\eeq
We thus find that $e^{\Phi_{(\text{A})}}$ exhibits exponentially suppressed corrections in the limit $e^\phi\rightarrow 0$, weighted by the instanton action
\beq
S_{\ell_1, \ell_2}= e^{-\phi} \sqrt{\ell_1^2+\ell_2^2} + i (\ell_1\chi+\ell_2\tilde\chi)\ .
\label{D2action}
\eeq 
This is recognized as the action for Euclidean D2-branes wrapping special Lagrangian 3-cycles in the homology class $\ell_1\mc{A}+\ell_2\mc{B}\in H_3(\mc{X}, \mbb{Z})$, where $(\mc{A},\mc{B})$ provides an integral symplectic basis of $H_3(\mc{X}, \mbb{Z})$. To see this, we note that generally the instanton action for D2-branes wrapping a special Lagrangian submanifold in the homology class $\ga\in H_3(X,\mbb{Z})$ inside a Calabi-Yau threefold $\mc{X}$ is given by 
\beq
S_{\ga}=\f{1}{g_s} \Big| \int_{\ga} \Omega\Big| + i\int_{\ga} C_{(3)},
\label{generalD2action}
\eeq
where ${g}_s$ is the ten-dimensional string coupling, $\Omega\in H_{3,0}(X)$ is the holomorphic 3-form and $C_{(3)}\in H^{3}(X,\mbb{R})/H^{3}(X,\mbb{Z})$ is the RR 3-form. The real  part of the action can further be written in terms of the central charge $Z_{\ga}=e^{K/2} \int_{\ga}\Omega$ as  $\Re(S_{\ga})= e^{-K/2} |Z_{\ga}| / g_s$, where $K=-\log \int_X \Omega\wedge \bar{\Omega} $ is the K\"ahler potential of the complex structure moduli space. Noting that $K=-\log V_{\mc{X}}$ we then find 
\beq
S_{\ga}= e^{-\phi} |Z_{\ga}| + i\int_{\ga} C_{(3)},
\eeq
where we defined the four-dimensional dilaton by $e^{\phi} := V_{\mc{X}}^{-1/2} g_s$. Restricting to a rigid Calabi-Yau threefold $\mc{X}$, we recall from Section \ref{Section:Introduction} that the prepotential is $F=\tau X /2$ with $\tau$ being the period ``matrix'' $\int_{\mc{B}} \Omega /\int_{\mc{A}}\Omega$. In this case the D2-brane wraps a 3-cycle in the homology class $\ga=\ell_1\mc{A}+\ell_2\mc{B}\in H_3(\mc{X}, \mbb{Z})$, which gives $Z_{\ga}=(\ell_1+\tau\ell_2)/\sqrt{\Im\tau}$, so the instanton action reduces to
\beq
S_{\ell_1,\ell_2}(\tau)=e^{-\phi} \frac{|\ell_1+\tau \ell_2|}{\sqrt{\Im\tau}}  + i \int_{\ell_1\mc{A}+\ell_2\mc{B}}C_{(3)}.
\eeq
Further setting $\tau=i$, which is the relevant value for our analysis, and using Eq. (\ref{RRperiods}) for the periods of the Ramond-Ramond 3-form $C_{(3)}$, this action indeed coincides with the instanton action (\ref{D2action}) predicted from $SU(2,1;\mbb{Z}[i])$-invariance. Thus, we may conclude that the Abelian term (\ref{FinalAbelianTerm32}) in the Fourier expansion agrees with the general form of D-instanton corrections in \eqref{phiD} upon restricting to a rigid Calabi-Yau threefold which admits complex multiplication by $\mbb{Z}[i]$. 

The infinite series within the brackets in (\ref{D2Term}) should, in the spirit of \cite{Green:1997tv}, arise from perturbative contributions around  the instanton background. The summation measure is given by 
specifying \eqref{FinalAbelianMeasure} to $s=3/2$,
\beq
\mu_{3/2}(\ell_1,\ell_2) = \sum_{\omega_3'|\Lambda} |\omega_3'|^{-1} \sum_{z|\frac{\Lambda}{\omega_3'}}|z|^{-2}\ ,
\label{D2measure}
\eeq
where we recall that $\Lambda=\ell_2-i\ell_1$ is a complex combination of the electric and magnetic charges $(\ell_1, \ell_2)$. The instanton measure $\mu_{3/2}(\ell_1, \ell_2)$ should count the degeneracy of Euclidean D2-branes in the homology class $\ell_1\mc{A}+\ell_2\mc{B}\in H_3(\mc{X}, \mbb{Z})$. For 
D2-instantons with $A$-type charge only,  i.e. $\ell_2=0$, and such that $\ell:=\ell_1$ 
is a product of inert primes  (those of the form $p=4n+3$, see Appendix~\ref{Gaussians}), the first sum collapses to $\omega_3'=1$ and the instanton measure reduces to 
\beq\label{D2AMeasure}
\mu_{3/2}(\ell, 0) = \sum_{z|\ell} |z|^{-2}\ .
\eeq
This reproduces the instanton measure found on the basis of $SL(2,\IZ)$ invariance 
in \cite{Alexandrov:2008gh,Vandoren1,Vandoren2,Pioline:2009ia},  which 
by analogy with \cite{Green:1997tv,Kostov:1998pg,Moore:1998et} 
should count ways of splitting a marginal bound state into smaller constituents.  However, it is possible
that $\ell$  be prime over the integers but
factorizable over the Gaussian integers, e.g. $2=-i(1+i)^2$ or $5=(2+i)(2-i)$, in which case
the measure \eqref{D2measure} involves additional contributions compared to \cite{Alexandrov:2008gh,Vandoren1,Vandoren2,Pioline:2009ia}. 
This novel feature of compactifications on rigid Calabi-Yau manifolds is an interesting
prediction of $SU(2,1;\mbb{Z}[i])$-invariance which deserves further investigation.

\subsubsection*{NS5-Brane Instantons}

As mentioned above, the non-Abelian term $e^{\Phi_{(\text{NA})}}$
may be interpreted as NS5-brane instanton contributions.
 Although we have not been able to extract the coefficients  $C^{\text{(NA)}}_{r,k, \ell}(s)$
 in the non-Abelian Fourier expansion \eqref{GeneralNonAbelianTerm}, 
 we can still extract the instanton action by taking the semiclassical limit. This corresponds to the asymptotic behaviour of (\ref{GeneralNonAbelianTerm}) in the limit $y\rightarrow\infty$, or, equivalently, to the saddle point approximation of the $t$-integral in (\ref{NonAbelianTerm3}) as analyzed in Section \ref{NonAbelianCoefficients}. Expanding the Whittaker function around $x=\infty$ yields
\beqa
W_{k,m}(x) &\sim&e^{-x/2} x^k \sum_{n\geq 0}\frac{\Gamma\left(m-k+n+\frac12\right)\Gamma\left(m+k+\frac12\right)}{\Gamma(n+1)\Gamma\left(m-k+\frac12\right)\Gamma\left(m+k-n+\frac12\right)} x^{-n}
\nn\\
&\sim& 
e^{-x/2} x^{k} \Big[1+\mc{O}\big(1/x\big)\Big].
\eeqa
Implementing this in (\ref{GeneralNonAbelianTerm}) and extracting the leading $r=0$ term, we deduce that the leading order contribution to $e^{\Phi_{(\text{NA})}}$ is given by
\beq
e^{\Phi_{(\text{NA})}}\sim e^{\phi} \sideset{}{'}\sum_{k\in\mbb{Z}} \sum_{\ell=0}^{4|k|-1} \sum_{n\in\mbb{Z}+\f{\ell}{4|k|}} C_{r,k, \ell} |k|^{-s} e^{-2\pi S_{k, n}} \Big[1+\mc{O}\big(e^{2\phi}\big)\Big],
\label{NS5D2contribution}
\eeq
where we have defined
\beq S_{k, n}=|k| e^{-2\phi}+2 |k|\big(\tilde\chi-n\big)^2-4i k n  \chi+2ik(\psi+\chi\tilde\chi).
\label{D2NS5action}
\eeq
This reproduces the Euclidean action of $4k$ NS5-branes  bound to $4 k n \in \mathbb{Z}$
 D2-branes wrapping a 3-cycle $\mathcal{A}\in H_3(\mathcal{X},\mathbb{Z})$. The fact that the number of NS5-brane states for fixed $k$ is a multiple of 4 is a consequence of demanding invariance under $SU(2,1;\mathbb{Z}[i])$. Note that even in the absence of D2-brane instanton contributions ($n=0$), the real part of the action receives a contribution from the background Ramond-Ramond flux $\tilde\chi$,
as found  previously in \cite{VandorenAlexandrov}. For vanishing $\tilde\chi$, this reduces to the  pure NS5-brane instanton action of \cite{BeckerBeckerStrominger}:
\beq
S_{k}= |k| e^{-2\phi}+2ik\psi.
\eeq
It should be emphasized that the result (\ref{NS5D2contribution}) displays the contribution from $A$-type D2-brane instantons only. The $B$-type D2-branes could be exposed by choosing the alternative polarization displayed in (\ref{NewNonAbelianTermSecondPolarization}), but then the $A$-type D2-brane effects are not visible. This is in contrast to the situation in \cite{AutomorphicNS5}, where the 
appearance of an extra summation in the non-Abelian term made it possible to expose the 
D$(-1)$, D5 and NS5-brane effects simultaneously.\footnote{We note that the presence of an extra ``theta-angle'' in the NS5-brane instanton action of \cite{AutomorphicNS5}, compared to our result (\ref{D2NS5action}), is related to the fact that the spherical vector $f_K$ in the principal series of $SL(3,\mbb{R})$ displays a cubic phase factor \cite{AutomorphicMembrane} which is absent in the corresponding spherical vector for $SU(2,1)$ \cite{PiolineGunaydin}.} 

Finally, we observe that  the asymptotic expansion of the Whittaker function predicts an 
infinite series of perturbative corrections around the NS5-brane instanton background. This is in marked contrast to the case of type IIA Euclidean NS5-branes wrapping $K3\times T^2$, where
the perturbative corrections around the instanton background truncate at one loop \cite{Obers:2001sw}.

\section{Conclusions}
\label{Conclusions}

In this work we postulated that quantum corrections to the hypermultiplet moduli space in 
type IIA string theory compactified on a rigid Calabi-Yau threefold with complex multiplication by $\mbb{Z}[i]$ are controlled by the Picard modular group $SU(2,1;\mbb{Z}[i])$. We investigated the consequences of this assumption
for the simplest automorphic form, the Eisenstein series \eqref{eisenintro}. Despite a serious
discrepancy with the sign of the one-loop term, the fact that we were led to D2- and NS5-brane 
instanton corrections with the correct classical action provides support for our postulate. In
the case of D2-brane instantons, the prediction of the Eisenstein series   \eqref{FinalAbelianTerm32}
is in fact in full agreement with the general form predicted in~\cite{Alexandrov:2008gh,Vandoren2,Pioline:2009ia}, though one could argue that it is largely
a consequence of the Laplace equation  \eqref{EigenvalueEquation}.
The instanton measure (\ref{D2AMeasure}) is also similar to the dilogarithm sum
found in \cite{Alexandrov:2008gh}, with additional refinements
when the charges include non-inert prime factors. It would be interesting 
to compare the instanton summation measure with the generalized Donaldson-Thomas 
invariants of rigid Calabi-Yau manifolds.

While the sign of the one-loop term invalidates our proposal that the Eisenstein series \eqref{eisenintro} governs the exact metric on the hypermultiplet moduli space, and so 
forbids us to expect a detailed agreement between our summation measure and 
the  generalized Donaldson-Thomas invariants,  we do not think that
it ruins the basic postulate that the Picard modular group $SU(2,1;\mbb{Z}[i])$ should
act isometrically on the exact universal hypermultiplet moduli space. Rather, we take it
as an incentive to construct a more sophisticated automorphic form which would produce
the correct one-loop term, as well as produce a non-trivial dependence on the coordinate $z$ on the twistor fiber $\mbb{C}P^1$, which is generally expected when all isometries are broken. In fact, since the twistor space
is known to be described by holomorphic contact transformations, it is natural to expect
that automorphic forms attached to the quaternionic discrete series of $SU(2,1)$ should be
relevant. Indeed, these forms can be lifted to sections of a certain complex line bundle 
on the twistor space $\mc{Z}_{\mc{M}_{\text{UH}}}=SU(2,1)/(U(1)\times U(1))$ \cite{PiolineGunaydin,GrossWallach}. 
It is challenging to construct such automorphic forms explicitly, and adapt the analysis in \cite{Alexandrov:2009qq} to produce a manifestly
$SU(2,1,\mbb{Z}[i])$-invariant description of the twistor space.  We anticipate, 
however, that the resulting
instanton corrections will be qualitatively similar to the ones considered here,
although the summation measure will certainly be quite different.

In this work we have concentrated exclusively on rigid Calabi-Yau threefolds whose
intermediate Jacobian $J(\mc{X})$ admits complex multiplication by $\mbb{Z}[i]$. 
It is interesting to ask\footnote{Note added: the extension to other CM types has been 
analyzed more recently in \cite{Bao:2010cc}.}
 how our construction may generalize to other values 
of the period matrix $\tau$. When $J(\mc{X})$ admits complex multiplication by the ring of integers $\mc{O}_d$ in the imaginary quadratic number field $\mbb{Q}(\sqrt{-d}), \, d> 0$, it is natural to conjecture
that the relevant arithmetic subgroup of $SU(2,1)$ would be $SU(2,1;\mc{O}_d)$. For example, choosing $\tau=(1+i\sqrt{3})/2:= \omega$  should correspond to the ``Picard-Eisenstein" modular group $SU(2,1;\mbb{Z}[\omega])$, where $\mbb{Z}[\omega]$ are the Eisenstein integers, corresponding to the ring of integers $\mc{O}_3=\mbb{Z}[\om]$ in $\mbb{Q}(\sqrt{-3})$~\cite{Holzapfel}. In contrast to the $\tau=i$ case, it is interesting to note that $SU(2,1;\mbb{Z}[\omega])$ does not contain the ``rotation'' generator $R$ in (\ref{emduality}) \cite{FalbelParker}. Indeed, one does not generally
expect the full electric-magnetic duality group to be a quantum symmetry, but rather its
subgroup generated by monodromies in the moduli space of complex structures (which
is non-existent in the case of rigid CY threefolds). 

{}From a purely mathematical point of view, we have provided several explicit 
constructions of an automorphic form for the Picard modular group attached to the principal continuous series of $SU(2,1)$. In addition, we analyzed its Abelian and
non-Abelian Fourier expansion in detail and found evidence for its functional equation. Sums over Gaussian divisors and Dirichlet $L$-series for the Gauss
field play central roles in the analysis. We expect that our results will be useful
in subsequent investigations of automorphic forms for various types of Picard groups.

\section*{Acknowledgements}

We are grateful to Sergei Alexandrov, Guillaume Bossard, Claudia Colonnello, Gabor Francsics, Ulf Gran, Nick Halmagyi, Peter D. Lax, Jakob Palmkvist, Ulf Persson, Christoffer Petersson, Per Salberger, Frank Saueressig, Jan Stienstra, Stefan Theisen, Stefan Vandoren, Pierre Vanhove, Niclas Wyllard, Don Zagier and Genkai Zhang for helpful discussions and correspondence. We also thank Jakob Palmkvist and Christoffer Petersson for many useful comments on an early version of this manuscript. 

A.K. is a Research Associate of the Fonds de la Recherche Scientifique-FNRS, Belgium. This work has been supported in part by  IISN-Belgium (conventions 4.4511.06, 4.4505.86 and 4.4514.08) and by the Belgian Federal Science Policy Office through the Interuniversity Attraction Pole P6/11. 
\newpage
\appendix

\section{Dirichlet Series and Gaussian Integers}
\label{Gaussians}

In this appendix, we collect for the reader's convenience some standard facts about Dirichlet series and Gaussian integers.

\subsection{Euler Products and Dirichlet Series}

A series $a(n)$ for $n\in\mbb{N}$ is called multiplicative if and only if $a(n_1 n_2)=a(n_1)a(n_2)$ whenever $n_1$ and $n_2$ are coprime~\cite{Apostol}. The associated Dirichlet series
\beq
L(a,s) = \sum_{n>0} a(n) n^{-s}
\eeq 
constructed from a multiplicative $a(n)$ can be recast as an Euler product over the primes ($p>1$)
\beq
L(a,s) = \prod_{p\text{ prime}} P(p,s)\,,
\eeq
where 
\beq
P(p,s) = \sum_{k\geq 0} a(p^k) p^{-ks}\,.
\eeq
As an example consider the multiplicative series (\ref{MultiplicativeCharacter}). One finds
\beq
P(2,s) = \sum_{k\geq 0} (2^{1-s})^k = \frac1{1-2^{1-s}}
\eeq
and for Pythagorean primes $p=1\mod4$
\beq
P(p,s) =\sum_{k\geq 0} (p^{1-s})^k + \frac{p-1}{1-s} \partial_p \sum_{k\geq 0} (p^{1-s})^k = \frac{1-p^{-s}}{(1-p^{1-s})^2}\,.
\eeq
For primes of the form $p=3\mod 4$ one has
\beq
P(p,s) = \sum_{k\geq 0} (p^{2-2s})^k +p^{-s} \sum_{k \geq 0} (p^{2-2s})^k = \frac{1+p^{-s}}{(1-p^{1-s})(1+p^{1-s})}\,,
\eeq
whence one recovers (\ref{EulerProductN(d)}).

\subsection{Structure of Gaussian Primes}
The ring of Gaussian integers $\mbb{Z}[i]$ forms a principal ideal domain~\cite{Neukrich,Cartier}, i.e.
every element admits a unique prime factorization up to migration of the four units $\pm 1,\pm i$.
We will use the notation $g$ for Gaussian primes and $p$ for standard (rational) primes. 
Gaussian prime numbers $g=a+ib\in \mbb{Z}[i]$ fall under three different cases, 
called {\em ramified}, {\em inert} and {\em split}.
\begin{itemize}
\item[(i)] The first case consists solely of $g=1+i$. Since $2=-i(1+i)^2$,
this implies that $p=2$, despite being prime in $\mbb{Z}$, is no longer prime in $\mbb{Z}[i]$.
The rational prime $p=2$ is said to be ramified over the Gaussian integers.
\item[(ii)] Inert primes are of the form $g=4n+3$ for some $n\in\mbb{N}$ such
that $4n+3$ is prime in $\mbb{Z}$. The name inert indicates that such integers 
are prime both over $\mbb{Z}$ and $\mbb{Z}[i]$.
\item[(iii)] Split primes come in complex conjugate pairs $g$ and $\bar{g}$. Such $g=a+ib, b\neq 0$ are prime if and only if $p:=g\bar{g}=a^2+b^2$ is a standard prime in  $\mbb{Z}$ and $p>2$. By Fermat's theorem on the sums of squares $p$ must be of the form $p=4n+1$ for some $n\in\mbb{N}$, i.e. a Pythagorean prime.
\end{itemize}

\section{Spherical Vector and $p$-Adic Eisenstein Series}
\label{pAdicConstruction}

Automorphic forms can be constructed quite generally using adelic methods,
as explained for the layman e.g. in 
\cite{PiolineWaldron1,Kazhdan,PiolineWaldron2}. In this Appendix, we apply
this method to recover the Eisenstein series
$\mc{E}_s(\phi, \lambda, \ga)$ for the Picard modular group. This
alternative approach also sheds light on the relation between the quadratic
constraint (\ref{constraint}) and the representation-theoretic structure 
underlying the Eisenstein series. This Appendix may be viewed as an 
automorphic extension of the results in Section 2 of \cite{PiolineGunaydin}.

\subsection{Formal Construction}

In general, to construct an automorphic form $\Psi$ on
$G/K$, invariant under a discrete subgroup
$G(\mbb{Z})\subset G$, we require three ingredients: $(1)$ a
$K$-invariant {spherical vector} $f_{K}\in \mc{H}\hs $
($\mc{H}$ being a Hilbert space of square integrable functions), $(2)$ a
linear representation $\rho$ of $G$ acting on $\mc{H}$, and $(3)$ a
$G(\mbb{Z})$-invariant {distribution} $f_{\mbb{Z}}\in
\mc{H}^{\star}$ in the dual space of $\mc{H}$. Using the natural pairing
$\left< \ , \ \right>$ between $\mc{H}$ and $\mc{H}^{\star}$, the automorphic
form $\Psi$ can then be defined formally as
\beq
\Psi(g) :=  \left<f_{\mbb{Z}}, \rho(g)\cdot f_{K}\right>,
\eeq
with $g\in G$. By virtue of the Iwasawa decomposition,
\beq
G=NAK,
\eeq
an arbitrary group element $g\in G$ splits as $g=nak:=\mc{V}k$, and, since $f_{K}$ is $K$-invariant, $\Psi$ simplifies to
\beq
\Psi(\mc{V}) =  \left<f_{\mbb{Z}}, \rho(\mc{V})\cdot
  f_{K}\right>.
\eeq
The coset representative $\mc{V}\in G/K$ transforms by $k^{-1}\in
K$ from the right and $\ga\in G(\mbb{Z})$ from the left,
\beq
\mc{V} \longmapsto \ga\mc{V}k^{-1}.
\eeq
From the point of view of $\Psi(\mc{V})$ the right action by $k^{-1}$ on $\rho(\mc{V})$ becomes a left action on $f_{K}$, which is invariant by definition, and the left action of $\ga$ becomes a right action on
$f_{\mbb{Z}}$, which is also invariant. Hence, $\Psi(\mc{V})$ is by
construction a function on the double quotient $G(\mbb{Z})\bas G /
K$ as desired.

Although very appealing, this method is often unpractical due to the 
difficulty of obtaining the invariant distribution. Adelic methods offer
a powerful way to obtain $f_{\mbb{Z}}$, by reducing this problem to that
of finding the $p$-adic spherical vector $f_p$ for all primes $p$
(see \cite{Koblitz,BrekkeFreund} for an introduction to $p$-adic numbers, and\cite{PiolineWaldron1,Kazhdan,PiolineWaldron2} for illustrations
of the adelic method).
The distribution $f_{\mbb{Z}}(x)$ is then obtained as the product
of $f_p(x)$ over all prime numbers of the given number field. For our purposes, $p$ runs over
the Gaussian primes and we will denote it by $g$ in accordance with Appendix~\ref{Gaussians}. Nevertheless, we will refer to the approach as the $p$-adic approach. The function $\Psi$ then can be rewritten formally as  
\beq
\Psi(\mc{V})=\sideset{}{'}\sum_{\vec{x}\in \mbb{Q}(i)^n} 
\rho(\mc{V})\cdot \Big[\prod_{g\, \text{prime}} f_g(\vec{x})\Big] ,
\label{pAdicAutomorphicForm}
\eeq
where $\vec{x}$ is a vector of rational Gaussian numbers in $\mbb{Q}(i)^n$,  the
product runs over all Gaussian prime numbers $g$ including the ``place at infinity" $g=\infty$,
and we defined $f_\infty=f_K$. Note that we restrict to the case where $\mc{V} \in G(\mbb{R})$,
such that  $\rho(\mc{V})$ acts only on $f_K$, but it is natural to
extend \eqref{pAdicAutomorphicForm} to the case where $\cV$ is an element of
the adele group $G(\mbb{A})$, in which case $\rho(\mc{V})$ acts on all the $f_g$'s as well. 
We shall now see that the Eisenstein
series $\mc{E}_s(\phi, \lambda, \ga)$, constructed in Section
\ref{LatticeConstruction}, can indeed be obtained from this adelic point of view. 

\subsection{Real and $p$-Adic Spherical Vector}

To reproduce the Picard Eisenstein series \eqref{EisensteinPicard} by this method, 
we consider the principal continuous series representation of $SU(2,1)$,
induced from the Heisenberg parabolic $P$ whose Lie algebra 
consists of the non-positive grade part of the 5-grading (\ref{5grading}):
\beq
\mf{p}=\mf{g}_{-2}\oplus \mf{g}_{-1}\oplus \mf{g}_0 \subset \mf{su}(2,1).
\eeq
The parabolic group $P$ thus corresponds to the subgroup of lower-triangular matrices,
\beq
P = \left\{ \left(\begin{array}{ccc}
t_1 &  & \\
 *  & t_2 & \\
 *  &  *  & t_3 \\
\end{array} \right) \in SU(2,1)\hs :\hs  ~ t_1t_2t_3=1 \right\}.
\eeq
The coset space $P\bas SU(2,1)$ is isomorphic to
the Heisenberg group ${N}$,  and can be parameterized as follows:
\beq\label{HeisenbergGroup}
n=e^{xX_{1}+\tilde{x}\tilde X_{1}+2yX_{2}}=\left(\begin{array}{ccc}
1 & i\bar C_2 & C_1 \\
  & 1 & {C}_2\\
  &   & 1 \\
\end{array} \right):= \left(\begin{array}{c}
\vec{r}_1  \\
\vec{r}_2 \\
\vec{r}_3 \\
\end{array}\right) \in N,
\eeq
where
\beq
\label{C12}
C_1:=2y+\f{i}{2}|C_2|^2\,,\qquad C_2:= x+\tilde{x}+i(\tilde x-x)
\eeq
satisfy the quadratic relation
\beq
 |C_2|^2-2\Im (C_1) = 0\,,
\label{QuadraticRelation}
\eeq
and the last equality in (\ref{HeisenbergGroup}) defines the row vectors $\vec r_i$ of the Heisenberg group element. 

The coset space $N=P\bas SU(2,1)$ admits an action of $g\in SU(2,1)$ by multiplication from the right,
followed by a compensating action by $p(g)\in P$ from the left so as to restore the upper triangular gauge \eqref{HeisenbergGroup}. The principal continuous series representation consists
of functions $f(x,\tilde{x},y)$ on $N$ 
transforming by the character $\chi_s(p(g))$ under the action of $g$, where
\beq
\chi_s(p):=t_1^{-2s}, \qquad p=\left(\begin{array}{ccc}
t_1 &  & \\
 *  & t_2 & \\
 *  &  *  & t_3 \\
\end{array} \right)\in P.
\eeq
The spherical vector $f_K$ can be obtained straightforwardly as follows \cite{PiolineGunaydin}: 
while the compensating left-action of $P$ on the second and third rows, $\vec{r}_2$ and $\vec{r}_3$, of $n$ is quite complicated,  the action on the first row $\vec{r}_1$ is very simple: $p\in P$ simply modifies $\vec{r}_1$ by an overall factor of $t_1$. Moreover, the action of $k\in SU(2)\times U(1)$ leaves invariant the (complex) norms of the rows $\vec{r}_i$. The spherical vector $f_K$ 
can therefore be obtained by raising the norm of the first row $\vec{r}_1$ of $n$ to the appropriate power of $s$ \cite{PiolineGunaydin}\footnote{See also \cite{AutomorphicMembrane} for a similar construction in the context of $SL(3, \mbb{R})$.}:
\beq
f_{K}(x, \tilde{x}, y):=|\vec{r}_1|^{-2s}=\big(1+|C_1|^2+|C_2|^2\big)^{-s}=\Big(1+2(x^2+\tilde{x}^2)+4y^2+(x^2+\tilde{x}^2)^2\Big)^{-s}.
\nn \\
\eeq
This object is indeed invariant under $SU(2)\times U(1)$, since the right action of $k$ on $n$ is a ``rotation'' that preserves the norm, while the compensating left action of $p$ merely modifies $f_{K}$ by an overall factor $t_1^{2s}$, which in turn is canceled against the character $\chi_s(p)=t_1^{-2s}$ which is present since $f_{K}$ is in the principal series. 

The next step is to compute the action of $\rho(\mc{V})$ on $f_{K}$. Following the prescription above, this can be done by first computing $n\cdot \mc{V}=p_0 \cdot n^{\prime}$, with
\beq
p_0=\left(\begin{array}{ccc}
e^{-\phi} & & \\
& 1 & \\
& & e^{\phi}\\
\end{array}\right)\in P, \qquad n^{\prime}=\left(\begin{array}{ccc}
1 & ie^{\phi}(\bar\lambda+\bar{C}_2) & e^{2\phi}(\ga+i\bar{C}_2\lambda+C_1) \\
 & 1 & e^{\phi}(\lambda+C_2)\\
 &  &  1\\
 \end{array}\right)\in P\bas SU(2,1).
 \eeq
Applying this to the spherical vector $f_K(x, \tilde x, y)=f_K(n)$ yields
\beq
\rho(\mc{V})\cdot f_K(n)=f_K(n\mc{V})=f_K(p_0 n^{\prime})=\chi_s(p_0)f_K(n^{\prime})=e^{2s\phi}|{\vec{r}_1}^{\hs \prime}|^{-2s},
\eeq
which may be written explicitly in the form
\beq
\rho(\mc{V})\cdot f_{K}(C_1, C_2)=e^{-2s\phi}\Big(|\bar{C_1}-i{C_2}\bar\lambda+\bar\ga|^2+e^{-2\phi}|{C_2}+{\lambda}|^2+e^{-4\phi}\Big)^{-s}.
\eeq
The $p$-adic spherical vector $f_p(C_1,C_2)$ can now be found by
replacing the Euclidean norm $|\cdot |$
appearing in the real spherical vector $f_{K}$ by its counterpart over the  $p$-adic 
Gaussian numbers\footnote{Note that with this definition, $|z|_g^{\mbb{Q}(i)}$ is not invariant under complex conjugation, as can be seen easily by taking $z$ to be a split prime. The definition of the $p$-adic norm in~\cite{BrekkeFreund} differs from the one we use. It is invariant under complex conjugation but 
misses other desirable properties; in particular it does not reproduce Eq. \eqref{primeproduct} correctly.} (see, e.g., \cite{Shintani})
\beq\label{Gaussianpadic}
|z|_g^{\mbb{Q}(i)}:=|g|^{-k}, \qquad z\in \mbb{Q}(i),
\eeq
for any Gaussian prime $g$, with $k\in \mbb{Z}$ being the maximum power of $g$ appearing in the prime factorization of $z$ in Gaussian primes.
 The $p$-adic spherical vector for a Gaussian prime $g$ is then given by
\beq
f_g(C_1,C_2):=\Big[\big|\vec{r}_1\big|^{\mbb{Q}(i)}_g\Big]^{-2s}=\mathrm{max}\Big(1,\big|\bar{C}_1\big|_g^{\mbb{Q}(i)},\big|C_2\big|_g^{\mbb{Q}(i)}\Big)^{-2s}.
\eeq

\subsection{Product over Primes}

The automorphic form $\Psi(\mc{V})$ in this representation now reads
\beq
\Psi(\mc{V})=\sideset{}{'}\sum_{(C_1,C_2)\in\mbb{Q}(i)^2\atop |C_2|^2-2\Im(C_1)=0}
\rho(\mc{V})\cdot  \Big[\prod_{g<\infty}f_g(C_1,C_2)\Big]f_{K}(C_1, C_2).
\label{padicEisensteinSeries}
\eeq
Next we must evaluate the infinite product over Gaussian prime numbers $g$. To this end we split the rational variables $C_1$ and $C_2$ in the following way:
\beq
{C_1}=\f{\om_1}{\om_3}\,,\quad {C_2}= \f{i \bar\om_2}{\bar\om_3}\,,
\label{variablerelation}
\eeq
with $\om_j\in \mbb{Z}[i]$, for $j=1,2,3$, $\gcd(\omega_1,\omega_2,\omega_3)=1$.\footnote{We note that the greatest common divisor in $\mbb{Z}[i]$ is defined up to Gaussian units which are a subgroup of order 4 in the Gaussian integers $\mbb{Z}[i]$. See Appendix \ref{Gaussians} for more details on the Gaussian integers.} Using the definition (\ref{Gaussianpadic}) we can explicitly evaluate the infinite product over primes in (\ref{padicEisensteinSeries}) as
\beq
\prod_{g<\infty}\mathrm{max}\Big(1,\Big|\f{\bar{\om}_1}{\bar{\om}_3}\Big|_g^{\mbb{Q}(i)},\Big|\f{\bar{\om}_2}{\bar{\om}_3}\Big|_g^{\mbb{Q}(i)}\Big)^{-2s}=|\om_3|^{-2s}.
\label{primeproduct}
\eeq
Multiplying the constraint \eqref{QuadraticRelation} further by a factor of $|\om_3|^2$ one obtains
\beq
|\om_3|^2\Big(|C_2|^2-2\Im(C_1)\Big)=  |\om_2|^2-2\Im(\om_1\bar{\om}_3)=\vec{\om}^{\dagger}\cdot \eta \cdot \vec{\om}=0.
\label{NewConstraint}
\eeq
Combining Eqs. (\ref{padicEisensteinSeries}), (\ref{primeproduct}) and (\ref{NewConstraint})
and adding  the contribution at $C_1=C_2=\infty$ (i.e. $\omega_3=0$) 
then yields the final form of $\Psi(\mc{V})$:
\beq
\Psi(\mc{V})=\sideset{}{'}\sum_{\vec{\om}\in \mbb{Z}[i]^3,\  \gcd(\om_1,\om_2,\om_3)=1 \atop   |\om_2|^2-2\Im(\om_1\bar{\om}_3)=0}e^{-2s\phi}\Big[|\bar\om_1+\bar\om_2\bar\lambda+\bar\om_3\bar{\ga}|^2+e^{-2\phi}|\bar\om_2-i\bar\om_3{\lambda}|^2+e^{-4\phi}|\om_3|^2\Big]^{-s},
\label{padicEisensteinSeries2}
\eeq
which we recognize as the Eisenstein series $\mc{P}_s(\phi, \lambda, \ga)$ constructed in Section \ref{Section:PoincareSeries}.

\section{More on the Abelian Measure}
\label{AbelianMeasureDetails}

This appendix contains a detailed analysis of the norm constraint (\ref{problem}) entering the Fourier expansion at various places and the derivation of the Abelian measure (\ref{FinalAbelianMeasure}) as a sum over Gaussian divisors.

\subsection{Analysis of the Norm Constraint}

The norm constraint (\ref{problem}) requires to find, for a fixed integer $d$, all Gaussian integers with norm squared divisible by $2d$. In this appendix we write this constraint as
\beq\label{normconstraint}
|\alpha|^2 := 0 \mod 2d.
\eeq

\subsubsection{Multiplicative Structure}

The solutions to the norm constraint possess a multiplicative structure. Let $d_1$ and $d_2$ be coprime integers and let $\alpha_1$ and $\alpha_2$ be Gaussian integers such that $2d_i$ divides $|\alpha_i|^2$. Then clearly $\alpha_1\alpha_2$ satisfies the norm constraint for $d_1d_2$. Due to prime factorization of Gaussian integers we know that this describes all solutions and it is therefore sufficient to study the solutions to the norm constraint for powers of primes $d=p^k$. There are three qualitatively different cases.
\begin{itemize}
\item[(i)] $p=2$, whence $d=2^k$. The structure of the set of solutions looks different for $k$ even and odd. For $k$ even one has that
\beq \label{sol2even}
\alpha = 2^{k/2} (n_1+ in_2)\quad \quad\text{for } n_1+n_2\in 2\mbb{Z}
\eeq
solves the constraint, whereas for $k$ odd
\beq\label{sol2odd}
\alpha = 2^{(k+1)/2} (n_1+ in_2)
\eeq
solves the constraint without restriction on the integers $n_1$ and $n_2$.

\item[(ii)] $p=4n+3$. Again one has to distinguish $k$ even and $k$ odd in solving (\ref{normconstraint}) for  $d=p^k$. For $k$ even one has
\beq\label{sol3even}
\alpha = p^{k/2} (n_1+ in_2)\quad \quad\text{for } n_1+n_2\in 2\mbb{Z}
\eeq
solves the constraint, whereas for $k$ odd
\beq\label{sol3odd}
\alpha = p^{(k+1)/2} (n_1+ in_2) \quad \quad\text{for } n_1+n_2\in 2\mbb{Z}
\eeq
solves the constraint. Note that there are restrictions on the integers $n_1$ and $n_2$ in both cases.

\item[(iii)] $p=4n+1$. This case is the most complicated one. Any such prime can be written as $p=a^2+b^2$ for some integers $a$ and $b$ and we assume $a>b$ without loss of generality. To describe the set of solutions to (\ref{normconstraint}) for $d=p^k$ we again distinguish even and odd $k$. An important auxiliary definition is furnished by 
\beq
e_k = (a-ib)^k (1+i)\quad\quad \Rightarrow \quad |e_k|^2 = 2p^k\,,
\eeq
providing an elementary solution of the constraint. With the help of the Gaussian integer $e_k$ one can define the following pairs of lattices for $k$ odd and $j=0,\ldots,\frac{k-1}2$
\beqa
\Lambda_{j+1} &=& \left\{ p^j(k_1  e_{k-2j}+ k_2  i e_{k-2j})\,:\, k_1,k_2\in\mbb{Z}\right\}\,,\nn\\
\bar\Lambda_{j+1} &=& \left\{ p^j(k_1  \bar{e}_{k-2j}+ k_2 i \bar{e}_{k-2j})\,:\, k_1,k_2\in\mbb{Z}\right\}\,.
\eeqa
The set of all solutions for $k$ odd is then given by
\beq\label{sol1odd}
\bigcup_{j=0}^{(k-1)/2} (\Lambda_{j+1}\cup \bar{\Lambda}_{j+1})\,.
\eeq
For $k$ even one also requires the lattice
\beq
\Lambda_{k+1} = \left\{ p^{k/2}(k_1 + i k_2 )\,:\, k_1,k_2\in\mbb{Z} \text{ and } k_1+k_2\in 2\mbb{Z}\right\}
\eeq
and then all solutions are given by
\beq\label{sol1even}
\bigcup_{j=0}^{k/2-1} (\Lambda_{j+1}\cup \bar{\Lambda}_{j+1})\cup \Lambda_{k+1}\,.
\eeq
\end{itemize}

Pictures of the three kinds of solution sets will be given momentarily when discussing the restriction to a fundamental domain under the action of a translation group.

\subsubsection{Restriction to a Fundamental Domain}

In the Abelian measure we made use of writing the solution to the constraint in terms of solutions in a fundamental domain in (\ref{n1n2solution}). We denote by
\beq\label{fundd}
\mc{F}(d) = \left\{\alpha \in \mbb{Z}[i]\,:\, |\alpha|^2:= 0 \mod 2d \text{ and } 0\leq \Re(\alpha)<d,\, 0\leq \Im(\alpha)<2d\right\}
\eeq
the set of solutions to (\ref{normconstraint}) in the fundamental domain. From the analysis above we know that for $d_1$ and $d_2$ coprime the following holds
\beq\label{multfund}
\mc{F}(d_1 d_2) \cong \mc{F}(d_1)\times \mc{F}(d_2)
\eeq
where the solutions are of the form $d_2 f_1 +d_1 f_2$ for $f_i\in\mc{F}(d_i)$ up to translation by the lattice $L$ of (\ref{LatticeL}) defining the fundamental domain. Therefore it is sufficient to restrict to $d=p^k$ being a power of a prime. For describing (\ref{fundd}) more explicitly we have to make recourse to the results of the preceding section and distinguish three cases.

\begin{itemize}

\begin{figure}[t!]
\centering
\includegraphics[scale=0.6]{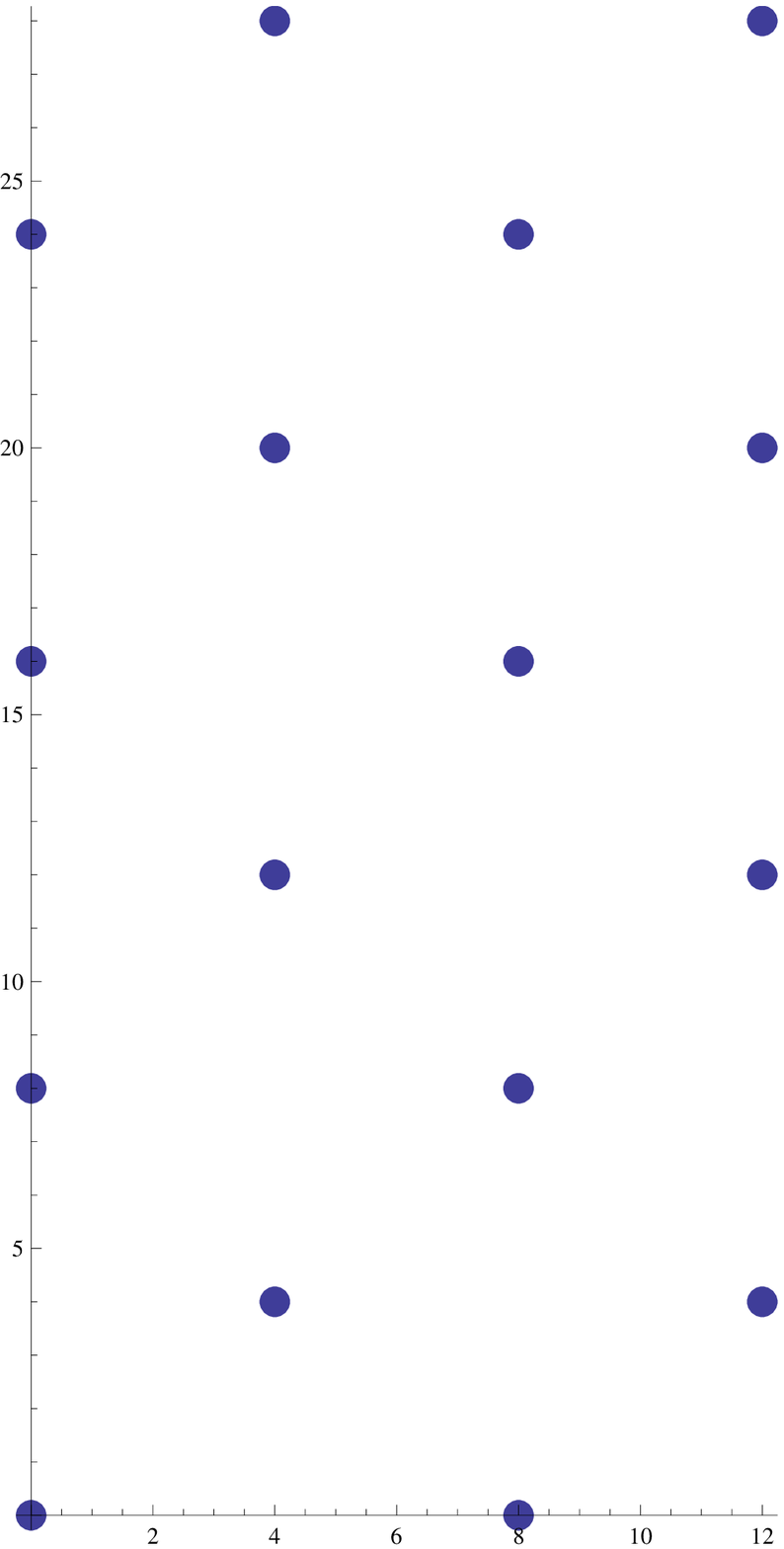}\hspace{8mm}
\includegraphics[scale=0.77]{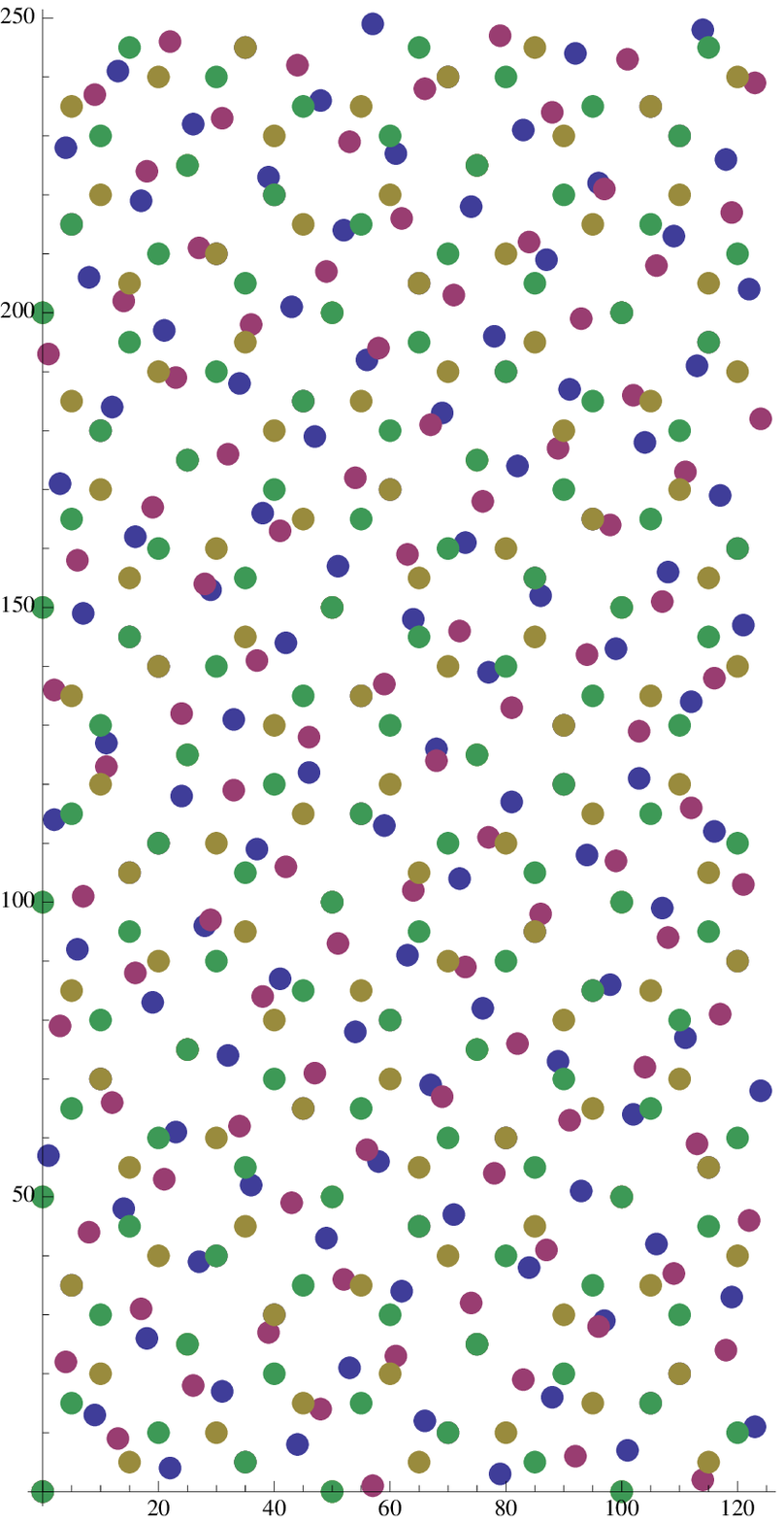}
\caption{\label{fundsols}Left: {\sl The set $\mc{F}(2^4)$ is a $\pi/4$ rotated and rescaled square lattice.} Right: {\sl The set $\mc{F}(5^3)$ as the union of four lattices with common points. The blue and purple points belong to the lattices $\Lambda_1$ and $\bar{\Lambda}_1$, whereas the green and ochre points are those of $\Lambda_2$ and $\bar\Lambda_2$.}}
\end{figure}

\item[(i)] $d=2^k$. Here one simply restricts the integers $n_1$ and $n_2$ in (\ref{sol2even}) and (\ref{sol2odd}). The number of points in the fundamental domain is
\beq
\sharp\, \mc{F}(2^k) = 2^k = N(2^k)
\eeq
in agreement with (\ref{MultiplicativityN}). The easiest way of doing the counting is by computing the sizes of the fundamental cell of the lattices and comparing to the total area of the fundamental domain $2d^2$. A typical lattice is depicted on the left of fig.~\ref{fundsols}. For $k=1$ there are two points in the fundamental domain.

\item[(ii)] $d=p^k$ for $p=4n+3$. The counting works similar but one has to take into account the additional constraint in (\ref{sol3odd}) leading to 
\beq
\sharp\,\mc{F}(p^k) = p^{2 \lfloor k/2 \rfloor} = N(p^k)\,.
\eeq
The form of the lattice is identical to that of the left part of fig.~\ref{fundsols}. For $k=1$ there is only a single point $\alpha=0$ in the fundamental domain.

\item[(iii)] $d=p^k$ for $p=4n+1$. The counting of points in the fundamental domain is now more involved since the individual lattices in (\ref{sol1odd}), or (\ref{sol1even}), have common points that should not be overcounted. Each lattice has $p^k$ points but, for example, there are $p^{2j}$ common points for the lattices $\Lambda_{j+1}$ and $\bar{\Lambda}_{j+1}$ forming the square lattice
\beq
\Lambda_{j+1}\cap \bar{\Lambda}_{j+1} = \left\{ p^{k-j} (k_1+i k_2)\,:\, k_1+ k_2\in 2\mbb{Z}\right\}\,,
\eeq
implying
\beq
\sharp\, \left((\Lambda_{j+1}\cup \bar{\Lambda}_{j+1})\cap\mc{F}(d)\right) = 2p^k -p^{2j}\,.
\eeq
One can also show that 
\beq
\sharp\, \left((\Lambda_{j+1}\cup \bar{\Lambda}_{j+1})\cap(\Lambda_{j+2}\cup \bar{\Lambda}_{j+2})\cap \mc{F}(d)\right) = 2p^{k-1} -p^{2j}\,
\eeq
and that all common points between pairs of lattices whose indices are farther apart are already contained in the intersection above. Putting everything together one arrives at the following count of points in the fundamental domain for $k$ odd
\beq
\sharp\,\mc{F}(p^k) = 2p^k -p^0 +\sum_{j=1}^{(k-1)/2}\big( 2p^k-p^{2j} -(2p^{k-1}-p^{2j-2})\big) = (k+1)p^k - k p^{k-1}= N(p^k)\,.
\eeq
For $k$ even the analysis is similar. An example of lattices with intersection points can be found on the right of fig.~\ref{fundsols}.
\end{itemize}

\subsection{Rewriting the Abelian Measure}

We now turn to deriving (\ref{FinalAbelianMeasure}) from the Abelian measure (\ref{abinstmes}). This involves mainly demonstrating the equality (\ref{equality}). To this end we introduce an additional function on the Gaussian integers
\beqa\label{auxnu}
\nu_s(q) &=& |q|^{2s-2} \beta(2s-1)\left(\sum_{d>0}d^{1-2s}\sum_{f\in\mc{F}(d)}e^{\frac{\pi i}{d}\Re[q f (1-i)]}\right)\nn\\
&=&\left|q\right|^{2s-2} \beta(2s-1)\sum_{d>0}d^{1-2s} a_{q}(d)\,.
\eeqa
The function $\nu_s(q)$ is related to the l.h.s. of (\ref{equality}) in an obvious way.

\subsubsection{Evaluation of the Auxiliary Functions $a_q(d)$ and $\nu_s(q)$}

The first observation is that the series $a_q(d)$ is multiplicative in $d$ for fixed $q$ (but not in $q$). This follows from (\ref{multfund}) and a simple rewriting of the exponent. Therefore it is sufficient to determine $a_q(d)$ for $d=p^k$. This is where the description of the sets $\mc{F}(p^k)$ enters.
The series $\nu_s(q)$ can be shown to be multiplicative so we only require $a_q(d)$ for $q=g^m$ as a power of a Gaussian prime and $d=p^k$ the power of standard prime.

The next observation is that
\beq
a_{g^m}(p^k) = \sum_{f\in\mc{F}(p^k)} e^{\frac{\pi i}{p^k}\Re[g^m f (1-i)]}
= \sum_{f\in g^m \mc{F}(p^k)} e^{\frac{\pi i}{p^k}\Re[ f (1-i)]}
\eeq
by rotating (and rescaling) the set of fundamental solutions. If $g$ does not divide $p$ then the rotated set is an equivalently good fundamental set of solutions. Hence
\beq
a_{g^m}(p^k) = a_1(p^k)\quad\text{if $g$ does not divide $p$.}
\eeq
For this reason we will first evaluate $a_1(p^k)$ and treat the case when $g$ divides $p$ afterwards.

It turns out that it suffices to count the number of times the lattice containing only the point $\alpha=0$ in $\mc{F}(p^k)$ appears in the sum over $\mc{F}(p^k)$. For all other lattices the sum over phases is zero. Hence one finds immediately
\beq
a_1(2^k) = 0 \quad\quad\text{for $k>0$}
\eeq
and for $p=4n+3$ that
\beq
a_1(p^k) = \left\{ \begin{array}{cl}1 &,\, k=1\\ 0 &,\,k>1\end{array}\right.\,.
\eeq
For $p=4n+1$ one has to count more carefully due to the intersection points. For $k=1$ the origin is the only common point in $\Lambda_1$ and $\bar\Lambda_1$ and hence is overcounted once leading to $a_1(p)=-1$. For $k>1$ this is offset by the intersection with $\Lambda_2\cup\bar\Lambda_2$, making $a_1(p^k)$ vanish. In total one has therefore for the Pythagorean primes
\beq
a_1(p^k) = \left\{ \begin{array}{cl}-1 &,\, k=1\\ 0 &,\,k>1\end{array}\right.\,.
\eeq
Constructing the Dirichlet series in (\ref{auxnu}) via its Euler product therefore leads to, after referring back to (\ref{betaEuler})
\beq
\sum_{d>0} d^{1-2s} a_1(d) = \frac{1}{\beta(2s-1)}\quad\Rightarrow\quad \nu_s(1) = 1\,.
\eeq

For $q=g^m$ one can perform a scaling of the lattices involved. Starting with the case of $g=1+i$ the value $p=2$ is important and one has
\beq
(1+i)^m \mc{F}(2^k) \cong [\mc{F}(2^{k-m})]^{2^m}
\eeq
defining the right hand side to consist only of the origin for $k\leq m$. Counting now the number of times the origin appears leads to 
\beq
a_{(1+i)^m} (2^k) = \left\{\begin{array}{cl} 
0 & \text{for $k>m$}\\
2^k=N(2^k) & \text{for $k\le m$}
\end{array}\right.\,. 
\eeq
The corresponding auxiliary series (\ref{auxnu}) is then
\beqa
\nu_s((1+i)^m) &=& 2^{m(s-1)}\beta(2s-1)\left(1+\sum_{k=1}^m 2^{k(2-2s)}\right) \sum_{d>0} d^{1-2s} a_1(d)\nn\\
& =& \frac{1}{4} 2^{m(1-s)} \sum_{ z | (1+i)^m} |z|^{4s-4}
\eeqa
and so is the usual divisor function multiplied by the proper overall factor to make it symmetric under $s\leftrightarrow 2-s$.
 
A similar analysis can be carried out for inert primes leading to 
\beq
a_{p^m}(p^k) = \left\{\begin{array}{cl} 
0 & \text{for $k\geq 2(m+1)$}\\
p^{2 \lfloor k/2 \rfloor}=N(p^k) & \text{for $k< 2(m+1)$}
\end{array}\right.\,. 
\eeq
Therefore the full result for (\ref{auxnu}) for inert primes is
\beqa
\nu_s(p^m) &=& p^{2m(s-1)}\beta(2s-1)\left(1+\sum_{k=1}^{2m+1} p^{2 \lfloor k/2 \rfloor+k(1-2s)}\right) \frac{\sum_{d>0} d^{1-2s} a_1(d)}{1+p^{1-2s}}\nn\\
& =& \frac{1}{4} p^{2m(1-s)} \sum_{ z | p^m} |z|^{4s-4}.
\eeqa
For split primes one finds
\beq
a_{g^m}(p^k) = \left\{\begin{array}{cl}
0 & \text{for $k> m+1$}\\
-p^{m} & \text{for $k= m+1$}\\
p^{k-1}(p-1) & \text{for $k < m+1$}
\end{array}\right.\,. 
\eeq
Therefore the full result for split primes is
\beqa
\nu_s(g^m) &=& \frac{p^{m(s-1)}}{1-p^{1-2s}}\left(1+\sum_{k=1}^{m} (p-1)p^{k-1+k(1-2s)}-p^{m+(m+1)(1-2s)}\right) \nn\\
& =& \frac{1}{4} |g|^{2m(1-s)} \sum_{ z | g^m} |z|^{4s-4}\,.
\eeqa
In summary the function $\nu_s(q)$ defined in (\ref{auxnu}) takes the value
\beq\label{auxnufinal}
\nu_s(q) = \frac{1}{4} |q|^{2-2s} \sum_{ z | q} |z|^{4s-4}\,,
\eeq
for any Gaussian integer $q\neq 0$ and therefore is a Gaussian divisor function in disguise. 

\subsubsection{The Abelian Instanton Measure}

The Abelian instanton measure of (\ref{abinstmes}) is thus given by a sum over primitive divisors of $\Lambda=\ell_2-i\ell_1$
\beq
\mu_s(\ell_1,\ell_2) = 4 \sum_{\omega_3'|\Lambda} |\Lambda|^{2-2s} \nu_s\left(\frac{\Lambda}{\omega_3'}\right)\,,
\eeq
which, together with (\ref{auxnufinal}), leads to (\ref{FinalAbelianMeasure}).



\begin{thebibliography}{99}

\bibitem{Aspinwall}
  P.~S.~Aspinwall,
  ``Compactification, geometry and duality: $\mc{N} = 2$,''
  [arXiv:hep-th/0001001].

\bibitem{Hull:1994ys}
  C.~M.~Hull and P.~K.~Townsend,
  ``Unity of superstring dualities,''
  Nucl.\ Phys.\  B {\bf 438}, 109 (1995)
  [arXiv:hep-th/9410167].

\bibitem{Witten:1995ex}
  E.~Witten,
  ``String theory dynamics in various dimensions,''
  Nucl.\ Phys.\  B {\bf 443} (1995) 85
  [arXiv:hep-th/9503124].

  \bibitem{Obers:1998fb}
  N.~A.~Obers and B.~Pioline,
  ``U-duality and M-theory,''
  Phys.\ Rept.\  {\bf 318} (1999) 113
  [arXiv:hep-th/9809039].

\bibitem{Green:1997tv}
  M.~B.~Green and M.~Gutperle,
  ``Effects of D-instantons,''
  Nucl.\ Phys.\  B {\bf 498} (1997) 195
  [arXiv:hep-th/9701093].

\bibitem{Green:1981yb}
  M.~B.~Green and J.~H.~Schwarz,
  ``Supersymmetrical String Theories,''
  Phys.\ Lett.\  B {\bf 109} (1982) 444.

\bibitem{Gross:1986iv}
  D.~J.~Gross and E.~Witten,
  ``Superstring Modifications Of Einstein's Equations,''
  Nucl.\ Phys.\  B {\bf 277} (1986) 1.

\bibitem{D'Hoker:2005ht}
  E.~D'Hoker, M.~Gutperle and D.~H.~Phong,
  ``Two-loop superstrings and S-duality,''
  Nucl.\ Phys.\  B {\bf 722} (2005) 81
  [arXiv:hep-th/0503180].

\bibitem{Kostov:1998pg}
  I.~K.~Kostov and P.~Vanhove,
  ``Matrix string partition functions,''
  Phys.\ Lett.\  B {\bf 444} (1998) 196
  [arXiv:hep-th/9809130].
  
\bibitem{Moore:1998et}
  G.~W.~Moore, N.~Nekrasov and S.~Shatashvili,
  ``D-particle bound states and generalized instantons,''
  Commun.\ Math.\ Phys.\  {\bf 209} (2000) 77
  [arXiv:hep-th/9803265].

\bibitem{PiolineKiritsis}
E. Kiritsis and B. Pioline, ``On $R^4$ threshold corrections in type IIB string theory and $(p,q)$ string  instantons'',
Nucl. phys. {\bf B508}, 509-534 (1997), [arXiv:hep-th/9707018].

\bibitem{Green:1997di}
  M.~B.~Green and P.~Vanhove,
  ``D-instantons, strings and M-theory,''
  Phys.\ Lett.\  B {\bf 408} (1997) 122
  [arXiv:hep-th/9704145].

\bibitem{ObersPioline}
N.~Obers and B.~Pioline, ``Eisenstein series and string thresholds'', Commun.\ Math.\ Phys.\ {\bf 209}, 275-324 (2000),
[arXiv:hep-th/9903113].

\bibitem{MirrorSymmetry}
C. Vafa and E. Zaslow (Eds.), ``Mirror Symmetry'',
Clay Mathematics Monographs, Volume 1, 929 pp. (2003).

\bibitem{Cecotti:1988qn}
  S.~Cecotti, S.~Ferrara and L.~Girardello,
  ``Geometry of Type II Superstrings and the Moduli of Superconformal Field
  Theories,''
  Int.\ J.\ Mod.\ Phys.\  A {\bf 4} (1989) 2475.



\bibitem{BeckerBeckerStrominger}
K. Becker, M. Becker and A. Strominger, ``Five-branes, membranes and nonperturbative string theory'',
Nucl. Phys. {\bf B456}, 130-152 (1995), [arXiv:hep-th/9507158].

\bibitem{AntoniadisMinasian1}
I. Antoniadis, S. Ferrara, R. Minasian and K. S. Narain, ``$R^4$ couplings in M- and type II theories on Calabi-Yau spaces'', Nucl. Phys. {\bf B507}, 571-588 (1997), [arXiv:hep-th/9707013].

\bibitem{Strominger}
A. Strominger, ``Loop corrections to the universal hypermultiplet'',
Phys. Lett. {\bf B421}, 139-148 (1998), [arXiv:hep-th/9706195].

\bibitem{Gunther:1998sc}
  H.~Gunther, C.~Herrmann and J.~Louis,
  ``Quantum corrections in the hypermultiplet moduli space,''
  Fortsch.\ Phys.\  {\bf 48}, 119 (2000)
  [arXiv:hep-th/9901137].

\bibitem{AntoniadisMinasian2}
I. Antoniadis, R. Minasian, S. Theisen and P. Vanhove, ``String loop corrections to the universal hypermultiplet'', Class. Quant. Grav. {\bf 20}, 5079-5102 (2003), [arXiv:hep-th/0307268].

\bibitem{Nick}
  N.~Halmagyi, I.~V.~Melnikov and S.~Sethi,
  ``Instantons, Hypermultiplets and the Heterotic String,''
  JHEP {\bf 0707} (2007) 086
  [arXiv:0704.3308 [hep-th]].

\bibitem{Salamon}  S. M. Salamon,``Quaternionic K\"ahler manifolds'', Invent. Math. {\bf 67}, no. 1, 143-171 (1982).

\bibitem{Swann}
A. Swann, ``Hyper-K\"ahler and quaternionic K\"ahler geometry'',
Math. Ann. {\bf 289}, no. 3, 421-450 (1991).

\bibitem{LeBrun}
C. LeBrun, ``Fano manifolds, contact structures, and quaternionic geometry'',
Internat. J. Math. {\bf 6}, no 3, 419-437 (1995), [arXiv:dg-ga/9409001].

\bibitem{deWitVandoren1}
B. de Wit, B. Kleijn and S. Vandoren, ``Superconformal hypermultiplets'',
Nucl. Phys. {\bf B568}, 475-502 (2000), [arXiv:hep-th/9909228].

\bibitem{deWitVandoren2}
B. de Wit, M. Rocek and S. Vandoren, ``Hypermultiplets, hyperk\"ahler cones and quaternion-K\"ahler geometry'',
JHEP {\bf 02}, 039 (2001), [arXiv:hep-th/0101161].


\bibitem{Alexandrov:2008ds}
  S.~Alexandrov, B.~Pioline, F.~Saueressig and S.~Vandoren,
  ``Linear perturbations of hyperk\"ahler metrics,''
  Lett.\ Math.\ Phys.\  {\bf 87} (2009) 225
  [arXiv:0806.4620 [hep-th]].

\bibitem{Alexandrov:2008nk}
  S.~Alexandrov, B.~Pioline, F.~Saueressig and S.~Vandoren,
  ``Linear perturbations of quaternionic metrics,''
  Commun.\ Math.\ Phys.\  {\bf 296} (2010) 353
  [arXiv:0810.1675 [hep-th]].


\bibitem{Alexandrov:2008gh}
  S.~Alexandrov, B.~Pioline, F.~Saueressig and S.~Vandoren,
  ``D-instantons and twistors,''
  JHEP {\bf 0903} (2009) 044
  [arXiv:0812.4219 [hep-th]].
  
\bibitem{Vandoren1}
D. Robles-Llana, M. Rocek, F. Saueressig, U. Theis and S. Vandoren, ``Nonperturbative corrections to $4D$ string theory effective actions from $SL(2,\mbb{Z})$ duality and supersymmetry'',
Phys. Rev. Lett. {\bf 98}, 211602 (2007), [arXiv:hep-th/0612027].

\bibitem{Vandoren2}
  D.~Robles-Llana, F.~Saueressig, U.~Theis and S.~Vandoren,
  ``Membrane instantons from mirror symmetry,''
  Commun.\ Num.\ Theor.\ Phys.\  {\bf 1}, 681 (2007)
  [arXiv:0707.0838 [hep-th]].

\bibitem{Alexandrov:2009zh}
  S.~Alexandrov,
  ``D-instantons and twistors: some exact results,''
  J.\ Phys.\ A  {\bf 42}, 335402 (2009)
  [arXiv:0902.2761 [hep-th]].

\bibitem{Alexandrov:2009qq}
  S.~Alexandrov and F.~Saueressig,
  ``Quantum mirror symmetry and twistors,''
  JHEP {\bf 0909} (2009) 108
  [arXiv:0906.3743 [hep-th]].


\bibitem{AutomorphicNS5}
  B.~Pioline and D. Persson,
  ``The automorphic NS5-brane ,'' Comm. Num. Th. Phys.  {\bf 3}, No. 4, 697-754 (2009),
   [arXiv:0902.3274v3 [hep-th]].



\bibitem{Noriko}
N.~Yui, ``Update on the Modularity of Calabi-Yau Varieties,''
Fields Institute Communications, {\bf 38}, 307-362 (2003).

\bibitem{Candelas:1985en}
  P.~Candelas, G.~T.~Horowitz, A.~Strominger and E.~Witten,
  ``Vacuum Configurations For Superstrings,''
  Nucl.\ Phys.\  B {\bf 258} (1985) 46.

\bibitem{Strominger:1985it}
  A.~Strominger and E.~Witten,
  ``New Manifolds For Superstring Compactification,''
  Commun.\ Math.\ Phys.\  {\bf 101} (1985) 341.

\bibitem{Candelas:1993nd}
  P.~Candelas, E.~Derrick and L.~Parkes,
  ``Generalized Calabi-Yau manifolds and the mirror of a rigid manifold,''
  Nucl.\ Phys.\  B {\bf 407}, 115 (1993)
  [arXiv:hep-th/9304045].

\bibitem{Bershadsky:1993cx}
  M.~Bershadsky, S.~Cecotti, H.~Ooguri and C.~Vafa,
  ``Kodaira-Spencer theory of gravity and exact results for quantum string
  amplitudes,''
  Commun.\ Math.\ Phys.\  {\bf 165}, 311 (1994)
  [arXiv:hep-th/9309140].

\bibitem{Aspinwall:1995vk}
  P.~S.~Aspinwall and J.~Louis,
  ``On the Ubiquity of K3 Fibrations in String Duality,''
  Phys.\ Lett.\  B {\bf 369}, 233 (1996)
  [arXiv:hep-th/9510234].

\bibitem{Ferrara:1988ff}
  S.~Ferrara and S.~Sabharwal,
  ``Dimensional reduction of type II superstrings,''
  Class.\ Quant.\ Grav.\  {\bf 6} (1989) L77.

\bibitem{Morrison:1995yi}
  D.~R.~Morrison,
  ``Mirror symmetry and the type II string,''
  Nucl.\ Phys.\ Proc.\ Suppl.\  {\bf 46} (1996) 146
  [arXiv:hep-th/9512016].



\bibitem{Schoen}
C. Schoen, ``Varieties dominated by product varieties,''
Internat. J. Math. {\bf 7}, no. 4, 541-571 (1996).

\bibitem{Moore:1998pn}
  G.~W.~Moore,
  ``Arithmetic and attractors,''
  arXiv:hep-th/9807087.

\bibitem{Gukov:2002nw}
  S.~Gukov and C.~Vafa,
  ``Rational conformal field theories and complex multiplication,''
  Commun.\ Math.\ Phys.\  {\bf 246}, 181 (2004)
  [arXiv:hep-th/0203213].

\bibitem{CompactificationPaper}
L.~Bao, C.~Colonnello, A.~Kleinschmidt, B.~E.~W.~Nilsson and D.~Persson, 
{\it Work in progress.}


\bibitem{BeckerBecker}
K. Becker and M. Becker, ``Instanton action for type II hypermultiplets'',
Nucl. Phys. {\bf B551}, 102-116 (1999), [arXiv:hep-th/9901126].

\bibitem{LambertWest2}
N.~Lambert and P.~West, ``Duality groups, automorphic forms and
higher derivative corrections,'' Phys.\ Rev.\ {\bf D75}, 066002
(2007), [arXiv:hep-th/0611318].

\bibitem{Bao:2007er}
  L.~Bao, M.~Cederwall and B.~E.~W.~Nilsson,
  ``Aspects of higher curvature terms and U-duality,''
  Class.\ Quant.\ Grav.\  {\bf 25}, 095001 (2008)
  [arXiv:0706.1183 [hep-th]].

\bibitem{Michel}
  Y.~Michel and B.~Pioline,
  ``Higher Derivative Corrections, Dimensional Reduction and Ehlers Duality,''
  JHEP {\bf 0709}, 103 (2007)
  [arXiv:0706.1769 [hep-th]].

\bibitem{Colonnello:2007qy}
  C.~Colonnello and A.~Kleinschmidt,
  ``Ehlers symmetry at the next derivative order,''
  JHEP {\bf 0708}, 078 (2007)
  [arXiv:0706.2816 [hep-th]].

\bibitem{Bao:2007fx}
  L.~Bao, J.~Bielecki, M.~Cederwall, B.~E.~W.~Nilsson and D.~Persson,
  ``U-Duality and the Compactified Gauss-Bonnet Term,''
  JHEP {\bf 0807}, 048 (2008)
  [arXiv:0710.4907 [hep-th]].

\bibitem{deWit:1996ix}
  B.~de Wit,
  ``N = 2 electric-magnetic duality in a chiral background,''
  Nucl.\ Phys.\ Proc.\ Suppl.\  {\bf 49} (1996) 191
  [arXiv:hep-th/9602060].

\bibitem{Bohm:1999uk}
  R.~Bohm, H.~Gunther, C.~Herrmann and J.~Louis,
  ``Compactification of type IIB string theory on Calabi-Yau threefolds,''
  Nucl.\ Phys.\  B {\bf 569} (2000) 229
  [arXiv:hep-th/9908007].

\bibitem{Manschot:2009ia}
  J.~Manschot,
  ``Stability and duality in N=2 supergravity,''
  [arXiv:0906.1767 [hep-th]].



\bibitem{FaFrLaPa} 
  E.~Falbel, G.~Francsics, P.~D.~Lax and J.~R.~Parker,
  ``Generators of a Picard modular group in two complex dimensions,''
  arXiv:0911.1104 [math.CV].
  

\bibitem{FrancsicsLax}
G. Francsics and P. D. Lax, ``A Fundamental Domain for the Picard Modular Group in $\mbb{C}^2$'',
Vienna, Preprint ESI 1273 (2003).


\bibitem{Gunaydin:2001bt}
  M.~Gunaydin, K.~Koepsell and H.~Nicolai,
  ``The Minimal Unitary Representation of $E_{8(8)}$,''
  Adv.\ Theor.\ Math.\ Phys.\  {\bf 5} (2002) 923
  [arXiv:hep-th/0109005].
  
\bibitem{Gunaydin:2004md}
  M.~Gunaydin and O.~Pavlyk,
  ``Minimal unitary realizations of exceptional U-duality groups and their
  subgroups as quasiconformal groups,''
  JHEP {\bf 0501} (2005) 019
  [arXiv:hep-th/0409272].


\bibitem{PiolineWaldron1}
D. Kazhdan, B. Pioline and A. Waldron, ``Minimal representations, spherical vectors, and exceptional theta  series. I'',
Commun. Math, Phys. {\bf 226}, 1-40 (2002) [arXiv:hep-th/0107222].

\bibitem{Kazhdan}
D. Kazhdan and A. Polishchuk, ``Minimal representations: spherical vectors and automorphic functionals'',
[arXiv:math/0209315 [math.RT]].

\bibitem{PiolineWaldron2}
B. Pioline and A. Waldron, ``Automorphic forms: A physicist's survey'',
[arXiv:hep-th/0312068].

\bibitem{PiolineGunaydin}
  M.~Gunaydin, A.~Neitzke, O.~Pavlyk and B.~Pioline,
  ``Quasi-conformal actions, quaternionic discrete series and twistors: $SU(2,1)$
  and $G_{2(2)}$,''
  Commun.\ Math.\ Phys.\  {\bf 283}, 169 (2008)
  [arXiv:0707.1669 [hep-th]].


\bibitem{Ishikawa}
Y. Ishikawa, ``The Generalized Whittaker Functions for $SU(2,1)$ and the Fourier Expansion of Automorphic Forms'',
J. Math. Sci. Univ. Tokyo {\bf 6}, 477-526 (1999).

\bibitem{Langlands}
R. P. Langlands, ``On the Functional Relations Satisfied by Eisenstein Series'',
Springer Lecture Notes in Mathematics {\bf 544}, 1-337 (1976).


\bibitem{Orloff}
T. Orloff, ``Dirichlet Series and Automorphic Forms on Unitary Groups'',
Trans. Am. Math. Soc. {\bf 290}, no. 2, 431--456 (1985).

  \bibitem{LivRel}
  M.~Henneaux, D.~Persson and P.~Spindel,
  ``Spacelike Singularities and Hidden Symmetries of Gravity'',
  Living Rev.\ Rel.\  {\bf 11} (2008) 1
  [arXiv:0710.1818 [hep-th]].

\bibitem{Yasaki}
D. Yasaki, ``Explicit reduction theory for $\pg $'',
[arXiv:math/0601071].

\bibitem{Bars:1989bb}
  I.~Bars and Z.~J.~Teng,
  ``The Unitary Irreducible Representations of $SU(2,1)$,''
  J.\ Math.\ Phys.\  {\bf 31} (1990) 1576.

\bibitem{Borel}
A. Borel, ``Automorphic Form on $SL(2, \mbb{R})$,''
Cambridge University Press, 1st Edition (1997).

\bibitem{Zhang}
G. Zhang, ``Shimura invariant differential operators and their eigenvalues'',
Math. Ann. {\bf 319}, no. 2, 235--265 (2001).




\bibitem{Vinogradov}
A. I. Vinogradov and L. A. Takhtadzhyan, ``Theory of Eisenstein series for the group $SL(3, \mbb{R})$ and its application to a binary problem,''
Journal of Soviet Mathematics, {\bf 18:3}, 293‚Äö√Ñ√∂ (1983).

\bibitem{Proskurin}
N. V. Proskurin, ``Expansions of Automorphic Functions'',
Translated from ``Zapiski Nauchnykh Seminarov Leningradskogo Otdeleniya Matematicheskogo Instituta'' im. V. A. Steklova AN USSR, {\bf 116}, 119-141 (1982).


\bibitem{Cartier} P.~Cartier, ``An introduction to zeta functions'', in: {\sl From Number Theory to Physics}, M. Waldschmidt et al. (eds.), Springer (1995)


\bibitem{SloaneA086933} Series A086933 in the ``Online Encyclopedia of Integer Sequences'', \href{http://www.research.att.com/~njas/sequences/}{\tt http://www.research.att.com/\~{}njas/sequences/}

\bibitem{Wolf}
J. A. Wolf, ``Complex Homogeneous Contact Manifolds and Quaternionic Symmetric Spaces'',
J. Math. Mech. {\bf 14}, No. 6, 1033-1047 (1965).

\bibitem{Anguelova:2004sj}
  L.~Anguelova, M.~Rocek and S.~Vandoren,
  ``Quantum corrections to the universal hypermultiplet and superspace,''
  Phys.\ Rev.\  D {\bf 70} (2004) 066001
  [arXiv:hep-th/0402132].

\bibitem{RoblesLlana:2006ez}
  D.~Robles-Llana, F.~Saueressig and S.~Vandoren,
  ``String loop corrected hypermultiplet moduli spaces,''
  JHEP {\bf 0603} (2006) 081
  [arXiv:hep-th/0602164].

\bibitem{VandorenAlexandrov}
S. Alexandrov, F. Saueressig and S. Vandoren, ``Membrane and fivebrane instantons from quaternionic geometry'',
JHEP, {\bf 09}, 040 (2006), [arXiv:hep-th/0606259].

\bibitem{Prza}
M. Przanowski, 
``Locally Hermite-Einstein, self-dual gravitational instantons," 
Acta Phys. Polon. B 14 (1983), no. 8, 625--627. 


\bibitem{Alexandrov:2009vj}
  S.~Alexandrov, B.~Pioline and S.~Vandoren,
  ``Self-dual Einstein Spaces, Heavenly Metrics and Twistors,''
  [arXiv:0912.3406 [hep-th]].
  
\bibitem{Pioline:2009ia}
  B.~Pioline and S.~Vandoren,
  ``Large D-instanton effects in string theory,''
  JHEP {\bf 0907} (2009) 008
  [arXiv:0904.2303 [hep-th]].

\bibitem{AutomorphicMembrane}
  B.~Pioline and A.~Waldron,
  ``The automorphic membrane,''
  JHEP {\bf 0406} (2004) 009
  [arXiv:hep-th/0404018].

\bibitem{Obers:2001sw}
N.~A.~Obers and B.~Pioline, ``Exact thresholds and instanton effects in string theory'', Fortsch.\ Phys.\  {\bf 49} (2001) 359, [arXiv:hep-th/0101122].

\bibitem{GrossWallach}
B. H. Gross and N. R. Wallach, ``On quaternionic discrete series representations, and their continuations,''
J. Reine. Angew. Math. {\bf 481}, 73-123 (1996).


\bibitem{Holzapfel} R.-P. Holzapfel, ``Geometry and Arithmetic around Euler partial differential equations'', D.Reidel Publishing Company, Springer; 1st edition (1986).

\bibitem{FalbelParker}
E. Falbel and J. R. Parker, ``The geometry of the Eisenstein-Picard modular group,''
Duke Math. J. {\bf 131} No. 2, 249 (2006).

\bibitem{Apostol} T.~.M.~Apostol, ``Introduction to analytic number theory'', Springer (1976).

\bibitem{Neukrich} H.~Neukirch, ``Algebraic number theory'', Springer (2007).

\bibitem{Koblitz}
N.~Koblitz, ``$p$-adic Numbers, $p$-adic Analysis and Zeta functions", Springer (1977).

\bibitem{BrekkeFreund}
L. Brekke and P. G. O. Freund, ``$p$-adic numbers in physics'', Phys. Rept. {\bf 233}, 1-66 (2003).

\bibitem{Shintani}
T.~Shintani, ``On automorphic forms on unitary groups of order 3'', (preprint, 1979).

\bibitem{Bao:2010cc}
  L.~Bao, A.~Kleinschmidt, B.~E.~W.~Nilsson, D.~Persson and B.~Pioline,
  ``Rigid Calabi-Yau threefolds, Picard Eisenstein series and instantons,''
  [arXiv:1005.4848 [hep-th]].




\end{thebibliography}
\end{document}